\documentclass[aps,floats]{revtex4}
\usepackage{amsmath,amssymb}
\usepackage{graphicx,epsfig}
\usepackage[greek,english]{babel}
\usepackage{bbold}
\usepackage{braket}

\begin{document}
\bibliographystyle {plain}

\pdfoutput=1
\def\oppropto{\mathop{\propto}} 
\def\opsimeq{\mathop{\simeq}}
\def\opoverderline{\mathop{\overline}}
\def\operarrow{\mathop{\longrightarrow}}
\def\opsim{\mathop{\sim}}

\def\fig#1#2{\includegraphics[height=#1]{#2}}
\def\figx#1#2{\includegraphics[width=#1]{#2}}


\title{ Explicit properties of the simplest inhomogeneous Matrix-Product-State \\
including the Riemann metric of the MPS manifold  }


\author{ C\'ecile Monthus }
 \affiliation{Institut de Physique Th\'{e}orique, 
Universit\'e Paris Saclay, CNRS, CEA,
91191 Gif-sur-Yvette, France}

\begin{abstract}

We consider the simplest inhomogeneous Matrix-Product-State for an open chain of N quantum spins that involves only two angles per site and two angles per bond with the following direct physical meanings. The two angles associated to the site $k$ are the two Bloch angles that parametrize the two orthonormal eigenvectors of the reduced density matrix $\rho_k$ of the spin $k$ alone. The two angles associated to the bond $(k,k+1)$ parametrize the entanglement properties of the Schmidt decomposition across the bond $(k,k+1)$. Explicit results are given for the reduced density matrix $\rho_{k,k+1}$ of two consecutive sites that is needed to evaluate the energy of two-body Hamiltonians, and for the reduced density matrix $\rho_{k,k+r}$ of two sites at distance $r$ that is needed to evaluate the spin-spin correlations at distance $r$. The global structure of the MPS manifold as parametrized by these $(4N-2)$ angles is then characterized by its explicit Riemann metric. Finally, the generalizations to any tree-like structure without loops and to the chain with periodic boundary conditions are discussed.

\end{abstract}

\maketitle

\section{ Introduction }

The difficulty in studying systems with a large number $N$ of quantum spins
comes from the exponential growth of the size of the Hilbert space ${\cal N}=2^N$.
The exact representation of a ket in the tensor Pauli basis 
\begin{eqnarray}
\ket{ \psi_N} = \sum_{S_1=\pm 1}\sum_{S_2=\pm 1} ... \sum_{S_N=\pm 1} 
c_{S_1,...,S_N} 
\ket {\sigma^z_1=S_1 }\ket {\sigma^z_2=S_2 } ... \ket {\sigma^z_N=S_N }
\label{ketfull}
\end{eqnarray}
involves ${\cal N}=2^N $ complex coefficients $c_{S_1,...,S_N}  $,
i.e. $2{\cal N}$ real parameters. The normalization $ \braket{ \psi_N \vert \psi_N }=1$ and the removal of the global phase with no physical meaning allows to reduce by two the number of parameters,
so that the total number of real parameters needed to parametrize the ket grows
exponentially with $N$ as
\begin{eqnarray}
P^{full}_N =2{\cal N}  - 2 = 2 (2^N-1) 
\label{pfull}
\end{eqnarray}

The key idea of the whole Tensor Network field
(see the reviews \cite{wolf,ver,cirac,vidal_intro,ulrich2011,phd-evenbly,mera-review,hauru,orus14a,orus14b,orus19} and references therein) is that in most physical problems,
the big tensor $c_{S_1,...,S_N} $ may be decomposed in terms of 
elementary small tensors that can be assembled 
in various ways in order to adapt to the geometry and to the entanglement properties of
the problem under focus. Indeed the notion of entanglement between the different regions
has emerged as the central idea and has completely changed the perspective on many-body quantum systems
(see the reviews \cite{amico08,horo,calabrese09,qi,laflorencie16,chiara} and references therein).
 In particular for one-dimensional quantum spin chains,
the Matrix-Products-States (MPS) are well adapted to describe non-critical states
displaying area-law entanglement \cite{wolf,ver,cirac,vidal_intro,ulrich2011,phd-evenbly,mera-review,hauru,orus14a,orus14b,orus19},
i.e. for most ground-states of local Hamiltonians \cite{hastings}.
In the Vidal canonical form of MPS \cite{vidalcanonical},
the coefficients $c_{S_1,...,S_N}  $ of Eq. \ref{ketfull}
are decomposed into
\begin{eqnarray}
 c_{S_1,...,S_N} = \sum_{\alpha_1=1}^D\sum_{\alpha_2=1}^D ... \sum_{\alpha_{N-1}=1}^D 
\Gamma^{[1]S_1}_{1,\alpha_1} \lambda^{[1,2]}_{\alpha_1} 
\Gamma^{[2]S_2}_{\alpha_1,\alpha_2} \lambda^{[2]}_{\alpha_2} 
...
\lambda^{[N-1,N]}_{\alpha_{N-1}} \Gamma^{[N]S_N}_{\alpha_{N-1},1} 
\label{vidal}
\end{eqnarray}
where the $D$ variables $ \lambda^{[k,k+1]}_{\alpha_k=1,..,D} $ associated to the bond $(k,k+1)$ are
 the Schmidt coefficients associated to the Schmidt decomposition 
of the MPS across the bond $(k,k+1)$, while the tensor 
$ \Gamma^{[k]S_k}_{\alpha_{k-1},\alpha_k}$ associated to the site $k$
represents some appropriate transformation within the local Hilbert space of the site $k$.
The number of parameters in the MPS grows only linearly in $N$
(instead of the exponential growth of Eq. \ref{pfull})
and these MPS parameters can be optimized numerically
to obtain the best approximation within the some given MPS manifold.
This variational point of view can be seen as a reformulation of
the Density-Matrix-RG algorithm \cite{white1,white2,ulrich2005}.
More generally, the Tensor Network activity has been mostly
oriented towards the production of extremely powerful numerical algorithms,
based on the variational optimization of the whole Tensor Network 
and where the numerical precision 
can be systematically improved by increasing the bond dimension $D$.

In the field of disordered spin chains, area-law states are even more important
since they appear not only for 
non-critical ground-states of generic local random Hamiltonians
 but also for excited states in Many-Body-Localized phases  \cite{pekker1,pekker2,friesdorf,tensor,sondhi,dmrgx,dmrgxfloquet}.
While in pure systems, the translational invariance allows
to choose the same elementary matrix with a finite number of parameters in the MPS
and to apply the transfer-matrix formalism to compute all physical observables
(see the reviews \cite{wolf,ver,cirac,vidal_intro,ulrich2011,phd-evenbly,mera-review,hauru,orus14a,orus14b,orus19} and references therein),
the presence of disorder breaks the translational invariance
and one needs to consider inhomogeneous MPS, where the parameters of the elementary tensors can adapt to the local disorder variables.
In the present paper, we will thus 
focus on the simplest inhomogeneous Matrix-Product-States 
of bond dimension $D=2$
 that involves only two real parameters per site and two real parameters per bond,
so that the number of real parameters for an open chain of $N$ spins and $(N-1)$ bonds will grow linearly with respect to the size $N$ as
\begin{eqnarray}
P^{MPS}_{N} = 2 N +2 (N-1) = 4N-2
\label{pmps}
\end{eqnarray}
 In addition, these parameters will have a very direct physical meaning,
since the bond variables will be related to the entanglement across the bond as in the Vidal canonical form of Eq. \ref{vidal} even if we will use a different choice of gauge fixing for the Schmidt decomposition,
while the sites variables will be the Bloch angles that parametrize the eigenvectors of the single-site reduced density matrices. Our goal will be to compute explicitly the properties
of this simplest inhomogeneous MPS.

The paper is organized as follows.
In section \ref{sec_gauge}, we describe the specific choice of gauge fixing 
for the Schmidt decomposition and introduce the useful notations for the whole paper.
In section \ref{sec_vidalmps}, 
we explain the parametrization of the simplest inhomogeneous Matrix-Product-States
and analyze its basic properties.
In section \ref{sec_rhointerval}, we study the reduced density matrix of an interval in the bulk,
whereas its diagonalization is given in Appendix \ref{app_diago}.
In section \ref{sec_energy}, we discuss the optimization of the MPS parameters
to approximate the ground-state of local Hamiltonian with one-body and two-body terms.
In section \ref{sec_corre}, we obtain the reduced density matrix of two sites at distance $r$
and evaluate the spin-spin correlations.
In section \ref{sec_riemann}, we compute the Riemann metric of the MPS manifold.
Finally, we describe the generalization of the simple MPS
to an arbitrary tree-like structure without loops in section \ref{sec_tree},
and the generalization to the chain with periodic boundary conditions in section \ref{sec_ring}.
Section \ref{sec_conclusion} summarizes our conclusions.


\section{ Present choice of gauge fixing for the Schmidt decomposition   }

\label{sec_gauge}

\subsection{ Schmidt decomposition for a state in the tensor product of two Hilbert spaces of dimension $D$ }

In this section, we focus on a quantum state $\ket{ \psi} $
 in the tensor product Hilbert space ${\cal H}_A \otimes {\cal H}_B$, where 
the two Hilbert spaces ${\cal H}_A $ and ${\cal H}_B $ have the same dimension $D$.
The diagonalization of the reduced density matrices
\begin{eqnarray}
\rho_A && = {\rm Tr}_{\{B\}}( \ket{\psi} \bra{\psi} ) = \sum_{\alpha=1}^{D} p_{\alpha} \ket{ A_{\alpha} }  \bra{ A_{\alpha} }
\nonumber \\
\rho_B && = {\rm Tr}_{\{A \}}( \ket{\psi} \bra{\psi} ) =\sum_{\alpha=1}^{D} p_{\alpha} \ket{ B_{\alpha} }  \bra{ B_{\alpha} }
\label{diagorhoa}
\end{eqnarray}
involve the same real positive weights $ p_{\alpha} \geq 0  $
normalized to unity
\begin{eqnarray}
 \sum_{\alpha=1}^{D} p_{\alpha}= 1 
\label{normaweights}
\end{eqnarray}
while the corresponding eigenvectors $\ket{A_{\alpha} }$ and $\ket{B_{\alpha}} $ form 
orthonormal basis of $A$ and $B$ respectively
\begin{eqnarray}
\braket{ A_{\alpha} \vert A_{\alpha'} }  && =\delta_{\alpha \alpha'}
\nonumber \\
\braket{ B_{\alpha} \vert B_{\alpha'} }  && =\delta_{\alpha \alpha'}
\label{orthorhoa}
\end{eqnarray}
The new basis $\ket{A_{\alpha} }$ can be obtained from the initial basis of $A$
by a unitary matrix of size $D \times D$ that contains a priori $D^2$ real parameters. 
However each ket $\ket{ A_{\alpha} } $ for $\alpha=1,..,D$ contains a global phase
that disappears from the projector $  \ket{ A_{\alpha} }  \bra{ A_{\alpha} }$ appearing in the reduced density matrix of Eq. \ref{diagorhoa},
so the $D$ projectors $ \ket{ A_{\alpha} }  \bra{ A_{\alpha} } $ can be parametrized with only $(D^2 - D )$ parameters.
Similarly, the $D$ projectors $\ket{ B_{\alpha} }  \bra{ B_{\alpha} } $ can be parametrized with only $(D^2 - D )$ parameters.

So together,
the two reduced density matrices $\rho_A $ and $\rho_B $ of Eq. \ref{diagorhoa}
involve
$(D-1)$ real parameters for the $D$ weights $p_{\alpha}$ normalized to unity (Eq \ref{normaweights}),
$(D^2 - D )$ real parameters for the $D$ projectors $  \ket{ A_{\alpha} }  \bra{ A_{\alpha} }$ in A
and $(D^2 - D )$ real parameters for the $D$ projectors $  \ket{ B_{\alpha} }  \bra{ B_{\alpha} }$ in B,
so that the total number of real parameters contained in the pair $(\rho_A,\rho_B)$ is
\begin{eqnarray}
P^{(\rho_A,\rho_B)} = (D-1) + 2 (D^2 - D ) = 2 D^2 - D - 1
\label{prho}
\end{eqnarray}
The comparison with the total number of real parameters needed to parametrize 
the ket $\ket{\psi}$ in the Hilbert space of dimension $D\times D$ (Eq \ref{pfull})
\begin{eqnarray}
P^{full} =  2 (D^2 - 1 ) = 2 D^2 - 2
\label{pfullD}
\end{eqnarray}
 yields that the number of missing parameters to reconstruct the ket $\ket{\psi} $ reduces to
\begin{eqnarray}
P^{missing} = P^{full} - P^{(\rho_A,\rho_B)}= D - 1 
\label{pmissing}
\end{eqnarray}
These $(D-1)$ missing parameters are the phases $\phi_{\alpha} \in [0,2 \pi[$ for $\alpha=2,..,D$ 
that are needed to write the Schmidt decomposition 
in the present gauge fixing,
 where we have already chosen the phases of the eigenvectors $\ket{ A_{\alpha} }  $ and $\ket{ B_{\alpha} }  $
as explained above
\begin{eqnarray}
\ket{\psi} = \sum_{\alpha=1}^{D} \sqrt{ p_{\alpha} }  e^{i \phi_{\alpha} } \ket{ A_{\alpha} } \otimes \ket{ B_{\alpha} }
\equiv \sum_{\alpha=1}^{D} \lambda_{\alpha} \ket{ A_{\alpha} } \otimes \ket{ B_{\alpha} }
\label{schmidtcl}
\end{eqnarray}
so in the present paper, the Schmidt values $\lambda_{\alpha}$
in front of the tensor product $ \ket{ A_{\alpha} } \otimes \ket{ B_{\alpha} } $ 
will be complex numbers
\begin{eqnarray}
\lambda_{\alpha} \equiv \sqrt{ p_{\alpha} }  e^{i \phi_{\alpha} } 
\label{lambdacomplex}
\end{eqnarray}
except for the first phase that will be chosen to vanish
\begin{eqnarray}
\phi_{\alpha=1}=0
\label{phi1}
\end{eqnarray}
  in order to fix the global phase of the ket of Eq. \ref{schmidtcl}.

Here it should be stressed that the Schmidt decomposition
 is usually formulated via the Singular Value Decomposition of matrices,
where the Schmidt values are real positive $\lambda^{usual}_{\alpha}=\sqrt{ p_{\alpha} } $,
i.e. the phases $e^{i \phi_{\alpha} }  $ introduced above are actually included 
in the eigenvectors $ \ket{A_{\alpha}}$ and $\ket{ B_{\alpha} }$.
 In the present paper however, it will be more convenient
 to choose the global phases of the eigenvectors as explained above,
and to work with complex Schmidt values as in Eq. \ref{lambdacomplex},
in order to avoid the gauge freedom and the gauge redundancy that exist
in the usual definition of Matrix-Product-States.
Let us now describe in detail what this gauge fixing means for the dimension $D=2$,
since it will be the basic building block of the MPS studied in the further sections.

\subsection{ Example with the dimension $D=2$ }

Here $A$ corresponds to a single spin with the Pauli basis $\ket{\sigma^z_1=\pm } $,
and $B$ corresponds to a single spin with the Pauli basis $\ket{\sigma^z_2=\pm } $.
The Schmidt decomposition of Eq. \ref{schmidtcl} contains only $D=2$ terms,
and it will be more convenient to label them with $\alpha=\pm$ (instead of $\alpha=1,2$).
The two weights normalized to unity $p_+ + p_- =1$ can be parametrized 
by the single angle $\theta \in [0,\frac{\pi}{2}]$ 
(if one chooses the ordering $p_+ \geq p_-$) 
\begin{eqnarray}
p_+ && = \cos^2 \left( \frac{\theta}{2} \right) = \frac{1+\cos \theta }{2}
\nonumber \\
p_- && = \sin^2 \left( \frac{\theta}{2} \right) =  \frac{1-\cos \theta }{2}
\label{normalambda}
\end{eqnarray}
The two complex Schmidt values of Eq. \ref{lambdacomplex}
are then parametrized in terms of two angles $(\theta,\phi)$
\begin{eqnarray}
\lambda_+ && = \cos \left( \frac{\theta}{2} \right)
\nonumber \\
\lambda_- && = \sin \left( \frac{\theta}{2} \right)e^{i \phi } 
\label{lambdathetaphi}
\end{eqnarray}
leading to the following Schmidt decomposition in the present gauge fixing
\begin{eqnarray}
\ket{\psi} = \cos \left( \frac{\theta}{2} \right) \ket {\tau^z_1=+ } \otimes \ket {\tau^z_2=+ } 
+\sin \left( \frac{\theta}{2} \right)   e^{i \phi } \ket {\tau^z_1=- } \otimes \ket {\tau^z_2=- } 
\label{schmidt2spins}
\end{eqnarray}
where $\ket {\tau^z_k=\pm } $ should be the two eigenvectors of the reduced density matrix $\rho_k$ for $k=1,2$.
It is convenient to parametrize the eigenvector $\ket{\tau^z_k=+ } $
by the usual Bloch angles $\theta_k \in [0,\pi]$ and $\phi_k \in [0,2 \pi[$,
while the orthogonal eigenvector $\ket{\tau^z_k=- } $
will correspond to the opposite point on the Bloch sphere with Bloch angles  $(\pi-\theta_k)$ and $( \phi_k+\pi) [2 \pi]$
\begin{eqnarray}
\ket{\tau^z_k=+ } && \equiv \ket {\theta_k, \phi_k }   
= \cos \frac{\theta_k}{2} \ket{  \sigma^z_k=+ } + \sin \frac{\theta_k}{2} e^{i \phi_k} \ket{  \sigma^z_k=- } 
\nonumber \\
 \ket{\tau^z_k=- }   && \equiv \ket {\pi-\theta_k, \phi_k+\pi }
= \sin \frac{\theta_k}{2} \ket{  \sigma^z_k=+ } - \cos \frac{\theta_k}{2} e^{i \phi_k} \ket{  \sigma^z_k=- } 
\label{blochk}
\end{eqnarray}

From the point of view of Pauli operators, the ket change of basis from $\ket{\sigma^z_k=\pm } $ to $\ket{\tau^z_k=\pm } $ 
corresponds to the change from the initial Pauli basis $(1,\sigma_k^z,\sigma_k^x,\sigma_k^y)$
to the new Pauli basis $(1,\tau_k^z,\tau_k^x,\tau_k^y)$
\begin{eqnarray}
\tau_k^z && 
= \cos \theta_k \sigma_k^z + \sin \theta_k \left[ \cos \phi_k \sigma_k^x 
+  \sin \phi_k \sigma_k^y  \right]
\nonumber \\
\tau_k^x 
&&= \sin \theta_k \sigma_k^z - \cos \theta_k \left[ \cos \phi_k \sigma_k^x 
+  \sin \phi_k \sigma_k^y  \right]
\nonumber \\
\tau_k^y 
&&=   \sin \phi_k \sigma_k^x-  \cos \phi_k  \sigma_k^y 
\label{taukzxy}
\end{eqnarray}
It will be convenient to denote by $R_k$ the reciprocal $3 \times 3$ rotation matrix
\begin{eqnarray}
\begin{pmatrix}
\sigma_k^z \\
 \sigma_k^x \\
\sigma_k^y
\end{pmatrix}
= R_k
\begin{pmatrix}
\tau_k^z \\
 \tau_k^x \\
\tau_k^y
\end{pmatrix}
=
\begin{pmatrix}
\cos \theta_k & \sin \theta_k &  0 \\
\sin \theta_k \cos \phi_k& - \cos \theta_k \cos \phi_k & \sin \phi_k\\
\sin \theta_k \sin \phi_k& - \cos \theta_k\sin \phi_k & - \cos \phi_k
\end{pmatrix}
\begin{pmatrix}
\tau_k^z \\
 \tau_k^x \\
\tau_k^y
\end{pmatrix}
\label{operatorsigmatau}
\end{eqnarray}
and to denote its matrix elements by $R_k^{ab}$ with $a=a,y,z$ and $b=x,y,z$
\begin{eqnarray}
\sigma_k^a = \sum_{b=x,y,z} R_k^{ab} \tau_k^b
\label{rotationab}
\end{eqnarray}

In terms of the new Pauli basis $(1,\tau_k^z,\tau_k^x,\tau_k^y)$,
the reduced density matrix for the spin $k$ is diagonal and involves only the two operator $(1,\tau_k^z )$
\begin{eqnarray}
\rho_k && = p_+\ket{\tau^z_k=+ }\bra{\tau^z_k=+ } +p_- \ket{\tau^z_k=- }\bra{\tau^z_k=- }
 = \frac{1}{2} \left( 1+\cos \theta \tau^z_k  \right)
\label{rhoktau}
\end{eqnarray}
while the full density matrix for the two spins computed from Eq \ref{schmidt2spins}
\begin{eqnarray}
\rho_{1,2}  \equiv \ket{\psi} \bra{\psi} 
 && =
\left( \frac{1+\cos \theta }{2} \right)
 \left( \frac{1+\tau^z_1}{2} \right) \left( \frac{1+\tau^z_2}{2} \right) 
+\left( \frac{1-\cos \theta }{2} \right)
\left( \frac{1-\tau^z_1}{2} \right) \left( \frac{1-\tau^z_2}{2} \right) 
\nonumber \\
&& + 
\frac{\sin \theta }{2} 
\left[ e^{-i \phi } \left( \frac{\tau_1^x+i \tau^y_1}{2} \right)\left( \frac{\tau_2^x+i \tau^y_2}{2} \right)
+   e^{i \phi }   \left( \frac{\tau_1^x-i \tau^y_1}{2} \right)\left( \frac{\tau_2^x-i \tau^y_2}{2} \right)
\right]
\nonumber \\
&& 
= \frac{1}{4} \left[ 1 + \cos \theta (\tau^z_1 + \tau^z_2 ) + \tau^z_1\tau^z_2 
+ \sin \theta \cos \phi ( \tau^x_1\tau^x_2 - \tau^y_1\tau^y_2 )
+ \sin \theta \sin \phi ( \tau^x_1\tau^y_2 + \tau^y_1\tau^x_2 )
\right]
\label{rho12}
\end{eqnarray}
involves only the 8 operators that commute with $\tau^z_1\tau^z_2  $
among the 16 operators of the tensor basis $\tau_1^a \otimes \tau_2^b$ with
$a=0,x,y,z$ and $b=0,x,y,z$.
In the next section, the exact decomposition of Eq. \ref{schmidt2spins}
for $N=2$ spins is used as the building block to construct
the simplest approximation for the ket $\ket{\psi_N}$ fo an open chain of $N$ spins
via a Matrix-Product-State of Schmidt dimension $D=2$.


\section{ Simplest Matrix-Product-State for an open chain of $N$ spins }

\label{sec_vidalmps}

In this section, the gauge fixing described in the previous section
is used to build the simplest Matrix-Product-State for an open chain of $N$ spins
in terms of $(4N-2)$ parameters with a clear physical meaning.

\subsection{ Parametrization of the MPS ket $\ket{\psi}  $ in terms of $( 4N-2) $ angles }

For each spin $k=1,..,N$, two 
Bloch angles $\theta_k $ and $\phi_k $ will be used
to parametrize the new appropriate local basis $\ket{\tau^z_k=\pm } $
as in Eq. \ref{blochk}.
For each bond $(k,k+1)$ with $k=1,..,N-1$,
two angles $\theta_{k+\frac{1}{2}} $ and $ \phi_{k+\frac{1}{2}}$
will be used to parametrize the two complex Schmidt values 
 as in Eq \ref{lambdathetaphi}
\begin{eqnarray}
\lambda_{k,k+1}^+ && \equiv \cos \left( \frac{\theta_{k+\frac{1}{2}}}{2} \right)
\nonumber \\
 \lambda_{k,k+1}^- && \equiv \sin \left( \frac{\theta_{k+\frac{1}{2}}}{2} \right) e^{i \phi_{k+\frac{1}{2}} }
\label{lambdacomplexk}
\end{eqnarray}

With these $(4N-2 )$ angles, the following Matrix-Product-State
is constructed
by adapting the Vidal canonical form recalled Eq. \ref{vidal}
to the present gauge fixing
\begin{eqnarray}
\ket{\psi}  && =  \sum_{\alpha_1=\pm} ...  \sum_{\alpha_{N-1}=\pm} 
 \ket{\tau^z_1= \alpha_1} 
 \lambda^{\alpha_1}_{1,2} 
\ket{\tau^z_2= \alpha_1\alpha_2} 
\lambda^{\alpha_2}_{2,3} 
\ket{\tau^z_3= \alpha_2\alpha_3} 
\lambda^{\alpha_3}_{3,4} ....
\nonumber \\
&&
.....
\ket{\tau^z_k= \alpha_{k-1} \alpha_k} 
\lambda^{\alpha_k}_{k,k+1} 
\ket{\tau^z_{k+1}= \alpha_k\alpha_{k+1}} 
......
\ket{\tau^z_{N-1}= \alpha_{N-2}  \alpha_{N-1}} 
\lambda^{\alpha_{N-1}}_{N-1,N} 
\ket{\tau^z_N= \alpha_{N-1}} 
\nonumber \\
&& =  \sum_{\alpha_1=\pm} ...  \sum_{\alpha_{N-1}=\pm} 
\left[ \prod_{k=1}^{N-1}\lambda^{\alpha_k}_{k,k+1}  \right]
\left[ \prod_{k=1}^N \ket{\tau^z_k= \alpha_{k-1} \alpha_k} \right]
\label{mpsfinal}
\end{eqnarray}
with the following convention for the boundary conditions on the last line
\begin{eqnarray}
\alpha_0 && =1
\nonumber \\
\alpha_N && =1
\label{alphaboundary}
\end{eqnarray}
Eq. \ref{mpsfinal} represents the simplest generalization beyond the mean-field product state of the $\ket{\tau^z_k= +} $
(that would corresponds to the Schmidt values $\lambda_{k,k+1}^+=1$ and $\lambda_{k,k+1}^-=0$)
\begin{eqnarray}
\ket{\psi^{MF}} = \ket{\tau^z_1= +} \ket{\tau^z_2= +} ... \ket{\tau^z_N= +}
\label{meanfield}
\end{eqnarray}
where one introduces quantum fluctuations
on each site via the opposite ket $\ket{ \tau^z_k= - } $,
and where one introduces entanglement on each bond via the Schmidt values $\lambda_{k,k+1}^- \ne 0 $ 
of Eq. \ref{lambdacomplexk}.

\subsection{ Corresponding Matrix-Product-Operator form of the full density matrix $\rho$   }

\label{sec_rhompo}

Using the same convention for the boundary conditions as in Eq. \ref{alphaboundary}
\begin{eqnarray}
\beta_0 && =1
\nonumber \\
\beta_N && =1
\label{betaboundary}
\end{eqnarray}
the MPS state of Eq. \ref{mpsfinal} translates into the following Matrix-Product-Operator form for the full density matrix 
\begin{eqnarray}
\rho = \ket{\psi} \bra{\psi}  = 
\sum_{ \substack{\alpha_1=\pm \\ \beta_1=\pm} }
...
\sum_{ \substack{\alpha_{N-1}=\pm \\ \beta_{N-1}=\pm} }
\left[ \prod_{k=1}^{N-1} \Lambda_{k,k+1}^{\alpha_k,\beta_k}  \right]
 \left[ \prod_{k=1}^N  O_k^{\alpha_{k-1} \alpha_k,\beta_{k-1} \beta_k}\right]
\label{mporho}
\end{eqnarray}
with the following notations.
The variables $\Lambda_{k,k+1}^{\alpha,\beta} $ associated to the bond $(k,k+1)$ involve two indices
$\alpha=\pm$ and $\beta=\pm$ and read in terms of the complex Schmidt values of Eq. \ref{lambdacomplexk}
\begin{eqnarray}
\Lambda_{k,k+1}^{\alpha,\beta} \equiv  \lambda^{\alpha}_{k,k+1} \overline{ \lambda^{\beta}_{k,k+1}  } 
\label{lambdabondrho}
\end{eqnarray}
i.e. they correspond to the following four values in terms of the two angles of Eq. \ref{lambdacomplexk}
\begin{eqnarray}
\Lambda_{k,k+1}^{+,+} && = \left \vert \lambda^{+}_{k,k+1} \right \vert^2=
\cos^2 \left( \frac{\theta_{k+\frac{1}{2}}}{2} \right) = \frac{1+\cos \left( \theta_{k+\frac{1}{2}} \right)  }{2}
\nonumber \\
\Lambda_{k,k+1}^{-,-} && =\left \vert  \lambda^{-}_{k,k+1} \right \vert^2=
\sin^2 \left( \frac{\theta_{k+\frac{1}{2}}}{2} \right) = \frac{1-\cos \left( \theta_{k+\frac{1}{2}} \right)  }{2}
\nonumber \\
\Lambda_{k,k+1}^{+,-} && =  \lambda^{+}_{k,k+1} \overline{ \lambda^{-}_{k,k+1}  } =
\cos \left( \frac{\theta_{k+\frac{1}{2}}}{2} \right)\sin \left( \frac{\theta_{k+\frac{1}{2}}}{2} \right) 
e^{-i \phi_{k+\frac{1}{2}} }
 = \frac{\sin \left( \theta_{k+\frac{1}{2}} \right)  }{2}e^{-i \phi_{k+\frac{1}{2}} }
\nonumber \\
\Lambda_{k,k+1}^{-,+} && =  \lambda^{-}_{k,k+1} \overline{ \lambda^{+}_{k,k+1}  } =
\sin \left( \frac{\theta_{k+\frac{1}{2}}}{2} \right) 
e^{i \phi_{k+\frac{1}{2}} }
\cos \left( \frac{\theta_{k+\frac{1}{2}}}{2} \right)
 = \frac{\sin \left( \theta_{k+\frac{1}{2}} \right)  }{2}e^{i \phi_{k+\frac{1}{2}} }
\label{lambdabondrhoexplicit}
\end{eqnarray}
that can be summarized by
\begin{eqnarray}
\Lambda_{k,k+1}^{\alpha,\beta} 
= \delta_{\alpha,\beta } \left( \frac{1+\alpha\cos \left( \theta_{k+\frac{1}{2}} \right)  }{2}\right)
+ \delta_{\alpha,-\beta } \left(\frac{\sin \left( \theta_{k+\frac{1}{2}} \right) 
\left[   \cos \left( \phi_{k+\frac{1}{2}}  \right) -i \alpha\cos \left( \phi_{k+\frac{1}{2}}  \right)
\right]
 }{2}\right)
\label{lambdarhofinal}
\end{eqnarray}

The operators $O_k^{T,T'}$ associated to the site $k$ involve two indices $T=\pm 1$ and $T'=\pm$
\begin{eqnarray}
O_k^{T,T'} \equiv \ket{\tau^z_k= T} \bra{\tau^z_k= T'} 
\label{operateork}
\end{eqnarray}
i.e. these four operators correspond to the two projectors that can be rewritten in terms of 
 $(1,\tau_k^z)$
\begin{eqnarray}
O_k^{+,+}  &&= \ket{\tau^z_k= +} \bra{\tau^z_k= +}   = \frac{ 1 +\tau^z_k }{2}
\nonumber \\
O_k^{-,-}  &&=\ket{\tau^z_k= -} \bra{\tau^z_k= - } = \frac{ 1 - \tau^z_k }{2}
\label{tauproj}
\end{eqnarray}
and to the two ladder operators that can be rewritten in terms of $(\tau_k^x,\tau_k^y)$
\begin{eqnarray}
O_k^{+,-} && = \ket{\tau^z_k= +} \bra{\tau^z_k= - } =\tau_k^{+} = \frac{ \tau^x_k +i \tau^y_k }{2}
\nonumber \\
O_k^{-,+}  && =\ket{\tau^z_k= -} \bra{\tau^z_k= + } = \tau_k^{-} = \frac{ \tau^x_k - i \tau^y_k }{2}
\label{tauladder}
\end{eqnarray}
that can be summarized by
\begin{eqnarray}
O_k^{T,T'}  = \ket{\tau^z_k=T }\bra{\tau^z_k=T' } 
= \delta_{T,T'} \left( \frac{1+T \tau_k^z}{2}\right)
+ \delta_{T,-T'} \left( \frac{ \tau^x_k +i T \tau^y_k }{2}\right)
\label{operatorfinal}
\end{eqnarray}
These operators could be rewritten in terms of the two angles $(\theta_k,\phi_k)$
and of the initial Pauli basis $(1,\sigma_k^z,\sigma_k^x,\sigma_k^y)$ using Eq \ref{taukzxy},
but in the present paper it will be more convenient to do all the calculations in the basis $\tau$.

Plugging the explicit forms of Eq. \ref{lambdarhofinal} and Eq. \ref{operatorfinal}
into Eq. \ref{mporho},
one obtains the more explicit Matrix-Product-Operator 
form
of the full density matrix 
\begin{eqnarray}
\rho && =  
\sum_{ \substack{\alpha_1=\pm \\ \beta_1=\pm} }
...
\sum_{ \substack{\alpha_{N-1}=\pm \\ \beta_{N-1}=\pm} }
\left( \prod_{k=1}^N \left[ 
\delta_{ \alpha_{k-1} \alpha_k,\beta_{k-1} \beta_k} \left( \frac{1+ \alpha_{k-1} \alpha_k \tau_k^z}{2}\right)
+ \delta_{ \alpha_{k-1} \alpha_k,-\beta_{k-1} \beta_k} \frac{ \tau^x_k +i  \alpha_{k-1} \alpha_k \tau^y_k }{2}
 \right]
\right)
 \nonumber \\   &&
 \left( \prod_{k=1}^{N-1} \left[
\delta_{\alpha_k,\beta_k } \left( \frac{1+\alpha_k\cos \left( \theta_{k+\frac{1}{2}} \right)  }{2}\right)
+ \delta_{\alpha_k,-\beta_k } \left(\frac{\sin \left( \theta_{k+\frac{1}{2}} \right) 
\left[   \cos \left( \phi_{k+\frac{1}{2}}  \right) -i \alpha_k\cos \left( \phi_{k+\frac{1}{2}}  \right)
\right]
 }{2}\right)
\right]
  \right)
\label{mporhoexplicit}
\end{eqnarray}

\subsection{ Schmidt decomposition across the bond $[k,k+1]$ }

The Schmidt decomposition of the ket of Eq. \ref{mpsfinal}
with respect to the bond $[k,k+1]$ 
reads
\begin{eqnarray}
\ket{\psi} =  \sum_{\alpha=\pm} \ket{\Phi^{[1,...,k]}_{\alpha} } 
\lambda_{k,k+1}^{\alpha}
\ket{\Phi^{[k+1,...,N]}_{\alpha} } 
\label{schmidtk}
\end{eqnarray}
where the two complex Schmidt values $ \lambda_{k,k+1}^{\pm}$ have been given in Eq \ref{lambdacomplexk},
while the corresponding orthonormalized Schmidt eigenvectors $\ket{\Phi^{[1,...,k]}_{\pm} } $ of the Left part $[1,2,..,k]$
and the corresponding orthonormalized Schmidt eigenvectors $\ket{\Phi^{[k+1,...,N]}_{\pm} } $  of the Right part $[k+1,...,N]$
\begin{eqnarray}
 \braket{ \Phi^{[1,...,k]}_{\alpha} \vert \Phi^{[1,...,k]}_{\beta} } && =\delta_{\alpha \beta}
\nonumber \\
 \braket{ \Phi^{[k+1,...,N]}_{\alpha} \vert \Phi^{[k+1,...,N]}_{\beta} }  && =\delta_{\alpha \beta}
\label{schmidortho}
\end{eqnarray}
read
\begin{eqnarray}
\ket{\Phi^{[1,...,k]}_{\alpha} }  && \equiv 
\sum_{\alpha_1=\pm} ...  \sum_{\alpha_{k-1}=\pm} 
 \ket{\tau^z_1= \alpha_1} 
 \lambda^{\alpha_1}_{1,2} 
\ket{\tau^z_2= \alpha_1\alpha_2} 
\lambda^{\alpha_2}_{2,3} 
\ket{\tau^z_3= \alpha_2\alpha_3} 
\lambda^{\alpha_3}_{3,4} ....
\lambda^{\alpha_{k-1}}_{k-1,k} 
\ket{\tau^z_k= \alpha_{k-1} \alpha} 
\nonumber \\
&& =  \sum_{\alpha_1=\pm} ...  \sum_{\alpha_{k-1}=\pm} 
\left[ \prod_{n=1}^{k-1}\lambda^{\alpha_n}_{n,n+1}  \right]
\left[ \prod_{n=1}^{k-1} \ket{\tau^z_n= \alpha_{n-1} \alpha_n} \right]\ket{\tau^z_k= \alpha_{k-1} \alpha} 
 \nonumber  \\
\ket{\Phi^{[k+1,...,N]}_{\alpha} } && \equiv 
\sum_{\alpha_{k+1}=\pm} ...  \sum_{\alpha_{N-1}=\pm} 
\ket{\tau^z_{k+1}= \alpha\alpha_{k+1}} 
\lambda^{\alpha_{k+1}}_{k+1,k+2} 
.....
\ket{\tau^z_{N-1}= \alpha_{N-2}  \alpha_{N-1}} 
\lambda^{\alpha_{N-1}}_{N-1,N} 
\ket{\tau^z_N= \alpha_{N-1}} 
\nonumber \\
&& =  \sum_{\alpha_{k+1}=\pm} ...  \sum_{\alpha_{N-1}=\pm} 
\left[ \prod_{n=k+1}^{N-1}\lambda^{\alpha_n}_{n,n+1}  \right]
\ket{\tau^z_{k+1}= \alpha \alpha_{k+1}}  \left[ \prod_{n=k+2}^N \ket{\tau^z_n= \alpha_{n-1} \alpha_n} \right]
\label{schmidteigenvectors} 
\end{eqnarray}
and satisfy the recurrences
\begin{eqnarray}
\ket{\Phi^{[1,...,k]}_{\alpha} }  && \equiv 
\sum_{\alpha_{k-1}=\pm} 
\ket{\Phi^{[1,...,k-1]}_{\alpha_{k-1}} } 
\lambda^{\alpha_{k-1}}_{k-1,k} 
\ket{\tau^z_k= \alpha_{k-1} \alpha} 
 \nonumber  \\
\ket{\Phi^{[k,...,N]}_{\alpha} } && 
= \sum_{\alpha_k=\pm} \ket{\tau^z_k= \alpha \alpha_k} \lambda^{\alpha_{k}}_{k,k+1} 
\ket{\Phi^{[k+1,...,N]}_{\alpha_k} } 
\label{receigenvectors} 
\end{eqnarray}

The Schmidt decomposition of Eq. \ref{schmidtk}
for the MPS across the bond $[k,k+1]$ 
translates into the following decomposition for the density matrix 
\begin{eqnarray}
 \rho = \ket{\psi} \bra{\psi}
&& = 
\sum_{ \substack{\alpha=\pm \\ \beta=\pm} }
\left(\ket{\Phi^{[1,...,k]}_{\alpha} } \bra{\Phi^{[1,...,k]}_{\beta} } 
\right)
\Lambda_{k,k+1}^{\alpha,\beta}
\left(\ket{\Phi^{[k+1,...,N]}_{\alpha} } \bra{\Phi^{[k+1,...,N]}_{\beta} } \right)
\label{rho1bond}
\end{eqnarray}

The reduced density matrices of the Left part $[1,..,k] $ and of the Right part $[k+1,..,N] $ read

\begin{eqnarray}
\rho_{[1,..,k]} && 
 =    \sum_{\alpha=\pm} \Lambda_{k,k+1}^{\alpha,\alpha}
 \ket{\Phi^{[1,...,k]}_{\alpha} }  \bra{\Phi^{[1,...,k]}_{\alpha} } 
 \nonumber \\&& 
= 
\left( \frac{1+\cos \left( \theta_{k+\frac{1}{2}} \right)  }{2}\right)
\ket{\Phi^{[1,...,k]}_{+} }  \bra{\Phi^{[1,...,k]}_{+} } 
+ \left( \frac{1-\cos \left( \theta_{k+\frac{1}{2}} \right)  }{2}\right)
\ket{\Phi^{[1,...,k]}_{-} }  \bra{\Phi^{[1,...,k]}_{-} } 
\nonumber \\
\rho_{[k+1,..,N]} && =   \sum_{\alpha=\pm} 
\Lambda_{k,k+1}^{\alpha,\alpha}
\ket{\Phi^{[k+1,..,N]}_{\alpha} }  \bra{\Phi^{[k+1,..,N]}_{\alpha} } 
\nonumber \\
&&=\left( \frac{1+\cos \left( \theta_{k+\frac{1}{2}} \right)  }{2}\right)
\ket{\Phi^{[k+1,..,N]}_{+} }  \bra{\Phi^{[k+1,..,N]}_{+} } 
+ \left( \frac{1+\cos \left( \theta_{k+\frac{1}{2}} \right)  }{2}\right)
\ket{\Phi^{[k+1,..,N]}_{-} }  \bra{\Phi^{[k+1,..,N]}_{-} } 
\label{schmidtkrho}
\end{eqnarray}

For the special case $k=1$ of the Schmidt decomposition across the bond $(1,2)$, 
one obtains that the ket $\ket{\tau_1^z=\pm }  $
corresponds to the Schmidt eigenvectors $\ket{\Phi^{[1]}_{\pm} }$
\begin{eqnarray}
\ket{\Phi^{[1]}_{\pm} }  && \equiv \ket{ \tau_1^z=\pm  } 
\label{psi1}
\end{eqnarray}
that diagonalize the reduced density matrix for the boundary spin $k=1$ 
\begin{eqnarray}
\rho_{1} && 
= \frac{1 }{2} \left[1+ \cos \left( \theta_{\frac{3}{2}} \right) \tau_1^z \right] 
\label{rho1boundary}
\end{eqnarray}

Similarly, the special case $k=N-1$, the Schmidt decomposition across the bond $(k,k+1)=(N-1,N)$
yields that the boundary ket $ \ket{ \tau_1^z=\pm  } $
are directly the Schmidt eigenvectors 
\begin{eqnarray}
\ket{\Phi^{[N]}_{\pm} } && \equiv  \ket{ \tau_N^z=\pm  } 
\label{psin}
\end{eqnarray}
that diagonalize the reduced density matrix
\begin{eqnarray}
\rho_{N} 
&& = \frac{1 }{2} \left[1+\cos \left( \theta_{N-\frac{1}{2}} \right)  \tau_N^z \right] 
\label{rhoNboundary}
\end{eqnarray}


\subsection{ Reduced density matrix $\rho_{k}$ for the site $k$ in the bulk ($k=2,..,N-1$) }

If one wishes to focus on the spin $k$ with the simultaneous Schmidt decomposition with respect to the Left part $[1,...,k-1]$
and to the Right part $[k+1,...,N]$, Eq  \ref{mpsfinal} becomes with Eq. \ref{schmidteigenvectors}
\begin{eqnarray}
\ket{\psi} =  \sum_{\alpha_L=\pm} \sum_{\alpha_R=\pm}
\ket{\Phi^{[1,...,k-1]}_{\alpha_L} }
\lambda_{k-1,k}^{\alpha_L}
 \ket{ \tau_k^z=\alpha_L \alpha_R  }
\lambda_{k,k+1}^{\alpha_R}
\ket{\Phi^{[k+1,...,N]}_{\alpha_R} } 
\label{schmidtkm1}
\end{eqnarray}
leading to the full density matrix
\begin{eqnarray}
&& \rho  = \ket{\psi} \bra{\psi} =  
\sum_{ \substack{\alpha_L=\pm \\ \beta_L=\pm} }
\sum_{ \substack{\alpha_R=\pm \\ \beta_R=\pm} }
 \left(
\ket{\Phi^{[1,...,k-1]}_{\alpha_L} }
\bra{\Phi^{[1,...,k-1]}_{\beta_L} }
\right)
\Lambda_{k-1,k}^{\alpha_L,\beta_L}
O_k^{\alpha_L \alpha_R,\beta_L \beta_R  }
\Lambda_{k,k+1}^{\alpha_R,\beta_R}
\left( 
\ket{\Phi^{[k+1,...,N]}_{\alpha_R} } 
\bra{\Phi^{[k+1,...,N]}_{\beta_R} } 
\right)
\label{schmidtkmrho}
\end{eqnarray}

The orthonormalization of the Schmidt eigenvectors (Eq \ref{schmidortho})
yields that 
the trace over the Left part $[1,..,k-1]$ imposes $\beta_L=\alpha_L$,
while the trace over the Right part $[k+1,..,N]$ imposes $\beta_R=\alpha_R$,
so that the reduced density matrix for the spin $k$ alone reduces to
\begin{eqnarray}
 \rho_k && = {\rm Tr}_{\{1,..,k-1\},\{k+1,..,N\}} (\rho) =
 \sum_{\alpha_L=\pm} \sum_{\alpha_R=\pm}
\Lambda_{k-1,k}^{\alpha_L,\alpha_L}
O_k^{\alpha_L \alpha_R,\alpha_L \alpha_R  }
\Lambda_{k,k+1}^{\alpha_R,\alpha_R}
\label{rhokbulkinter}
\end{eqnarray}
The explicit expressions of Eq. \ref{lambdarhofinal}
and of Eq. \ref{operatorfinal}
lead to the final result
\begin{eqnarray}
 \rho_k && = 
 \sum_{\alpha_L=\pm} \sum_{\alpha_R=\pm}
\left( \frac{1+\alpha_L\cos \left( \theta_{k-\frac{1}{2}} \right)  }{2}\right)
\left( \frac{1+\alpha_L \alpha_R \tau_k^z}{2}\right)
\left( \frac{1+\alpha_R\cos \left( \theta_{k+\frac{1}{2}} \right)  }{2}\right)
 \nonumber \\ &&
= \frac{1 }{2} \left[1+\cos \left( \theta_{k-\frac{1}{2}} \right)  \tau_k^z \cos \left( \theta_{k+\frac{1}{2}} \right)
\right] 
\label{rhokbulkz}
\end{eqnarray}
So the two ket $ \ket{ \tau_k^z=\pm } $ parametrized by the two Bloch angles 
$(\theta_k ,\phi_k )$ (Eq. \ref{blochk})
are the two eigenvectors 
of the reduced density matrix $\rho_k$ with the corresponding eigenvalues
\begin{eqnarray}
 p_k^{\pm} 
&& 
 = \frac{1 }{2} \left[1\pm\cos \left( \theta_{k-\frac{1}{2}} \right) \cos \left( \theta_{k+\frac{1}{2}} \right) \right]
\label{rhokbulkw}
\end{eqnarray}
that involve the angles $\theta_{k \pm \frac{1}{2}} $ of the two neighboring bonds.


\subsection{ Multifractality of the MPS components in the $\{\tau_k^z\}$ basis }

The groundstate wavefunction of manybody quantum systems 
have been found to be generically multifractal, with many studies
concerning the Shannon-R\'enyi entropies in quantum spin models
\cite{jms2009,jms2010,jms2011,moore,grassberger,atas_short,
atas_long,luitz_short,luitz_o3,luitz_spectro,luitz_qmc,jms2014,alcaraz1,alcaraz2,c_renyi,jms2017,c_treetensor},
while multifractal properties have been also much studied recently in the field of Many-Body-Localization
\cite{toulouse,mace1,mace2,lev}.
Of course this multifractal analysis depends on the basis, and a natural question
is thus to determine what is the more appropriate basis in each context.
Here since the ket $ \ket{ \tau_k^z=\pm }  $ are the two eigenvectors of the one-site density matrix $\rho_k$
 (Eq \ref{rhokbulkz}), it is interesting to consider the expansion of the MPS of Eq \ref{mpsfinal}
in this basis $\tau_k^z=\pm 1$ (instead of the expansion of Eq. \ref{ketfull}
in the initial basis $\sigma_k^z=\pm 1 $)
\begin{eqnarray}
\ket{ \psi} = \sum_{T_1=\pm 1}\sum_{T_2=\pm 1} ... \sum_{T_N=\pm 1} 
\psi_{T_1,...,T_N} 
\ket {\tau^z_1=T_1 }\ket {\tau^z_2=T_2 } ... \ket {\tau^z_N=T_N }
\label{kettau}
\end{eqnarray}
in order to characterize the statistics of the coefficients
\begin{eqnarray}
\psi_{T_1,...,T_N} 
= && \left( \bra {\tau^z_1=T_1 }\bra {\tau^z_2=T_2 } ... \bra {\tau^z_N=T_N }  \right) 
\ket{\psi} 
\nonumber \\
&& = 
 \sum_{\alpha_1=\pm} ...  \sum_{\alpha_{N-1}=\pm} 
\delta_{ T_1,\alpha_1}
 \lambda^{\alpha_1}_{1,2}
\delta_{T_2,\alpha_1 \alpha_2}
\lambda^{\alpha_2}_{2,3}
\delta_{T_3,\alpha_2 \alpha_3}
... 
 \lambda^{\alpha_{N-2}}_{N-2,N-1}
\delta_{T_{N-1},\alpha_{N-2} \alpha_{N-1}}
 \lambda^{\alpha_{N-1}}_{N-1,N}
\delta_{T_N,\alpha_{N-1}}
\nonumber \\
&& = \delta_{1,(T_1 T_2 ... T_N)}
 \lambda^{T_1}_{1,2}
\lambda^{(T_1T_2)}_{2,3}
\lambda^{(T_1T_2 T_3)}_{3,4}
... 
 \lambda^{(T_1 T_2 ... T_{N-2} ) }_{N-2,N-1}
 \lambda^{(T_1 T_2 ... T_{N-1} ) }_{N-1,N}
=  \delta_{1,(T_1 T_2 ... T_N)} \prod_{k=1}^{N-1} \lambda^{(T_1 T_2 ... T_k ) }_{k,k+1}
\label{psitaufull}
\end{eqnarray}
Besides the global constraint $1=T_1 T_2 ... T_N$,
these coefficients are thus simply given by the product of the complex Schmidt values $\lambda_{k,k+1}^{\alpha_k}$ 
(Eq \ref{lambdacomplexk}) along the whole chain
with the indices
\begin{eqnarray}
\alpha_k = T_1 T_2 ... T_k
\label{alphaprod}
\end{eqnarray}

The maximal weight $\vert \psi_{T_1,...,T_N} \vert^2 $ corresponds to 
the case where each bond $[k,k+1]$ is in its biggest Schmidt value $\lambda^{+}_{k,k+1} $
corresponding to $T_k=+1$ for all $k$ (i.e. the state of Eq. \ref{meanfield})
\begin{eqnarray}
\vert\psi_{max} \vert^2=\vert\psi_{+++...+++} \vert^2
= \prod_{k=1}^{N-1} \vert\lambda^{+}_{k,k+1} \vert^2
= \prod_{k=1}^{N-1} \cos^2 \left( \frac{\theta_{k+\frac{1}{2}}}{2} \right)
=\prod_{k=1}^{N-1} \left[ \frac{1+ \cos \left(\theta_{k+\frac{1}{2}} \right)  }{2} \right]
\label{psimax}
\end{eqnarray}

On the contrary, the minimal weight $\vert \psi_{T_1,...,T_N} \vert^2 $ corresponds to 
the case where each bond $[k,k+1]$ is in its lowest Schmidt value $\lambda^{-}_{k,k+1} $ :
the two possibilities are thus $T_k=(-1)^k$ , or $T_k=-(-1)^k$ for $N$ even
\begin{eqnarray}
\vert\psi_{min} \vert^2=\vert\psi_{+-+-+-...} \vert^2
= \vert\psi_{-+-+-+....} \vert^2= \prod_{k=1}^{N-1} \vert\lambda^{-}_{k,k+1} \vert^2
= \prod_{k=1}^{N-1} \sin^2 \left( \frac{\theta_{k+\frac{1}{2}}}{2} \right)
=\prod_{k=1}^{N-1} \left[ \frac{1- \cos \left(\theta_{k+\frac{1}{2}} \right)  }{2} \right]
\label{psimin}
\end{eqnarray}

More generally, the statistics of all the weights $\vert \psi_{T_1,...,T_N} \vert^2 $
 normalized to unity can be analyzed via the Inverse Participation Ratios $Y_q(N) $
 where $q$ is a continuous parameter 
\begin{eqnarray}
Y_q(N) && \equiv  \sum_{T_1=\pm 1}\sum_{T_2=\pm 1} ... \sum_{T_N=\pm 1} 
\vert\psi_{T_1,...,T_N} \vert^2
\nonumber \\
&& = \sum_{\alpha_1=\pm 1}\sum_{\alpha_2=\pm 1} ... \sum_{\alpha_{N-1}=\pm 1} 
\prod_{k=1}^{N-1} \vert\lambda^{\alpha_k}_{k,k+1} \vert^{2q}
= \prod_{k=1}^{N-1} \left(  \vert\lambda^{+}_{k,k+1} \vert^{2q}
+\vert\lambda^{-}_{k,k+1} \vert^{2q}
\right)
\nonumber \\
&&= \prod_{k=1}^{N-1} \left(  \cos^{2q} \left( \frac{\theta_{k+\frac{1}{2}}}{2} \right)
+\sin^{2q} \left( \frac{\theta_{k+\frac{1}{2}}}{2} \right)
\right)
=\prod_{k=1}^{N-1} \left(  
\left[ \frac{1+ \cos \left(\theta_{k+\frac{1}{2}} \right)  }{2} \right]^q
+ \left[ \frac{1- \cos \left(\theta_{k+\frac{1}{2}} \right)  }{2} \right]^q
\right)
\label{yq}
\end{eqnarray}
or equivalently with the corresponding R\'enyi entropies
 \begin{eqnarray}
{\cal S}_q(N) \equiv  \frac{ \ln Y_q(N) }{1-q} 
= \frac{1}{1-q} \sum_{k=1}^{N-1} \ln \left(  
\left[ \frac{1+ \cos \left(\theta_{k+\frac{1}{2}} \right)  }{2} \right]^q
+ \left[ \frac{1- \cos \left(\theta_{k+\frac{1}{2}} \right)  }{2} \right]^q
\right)
\label{renyi}
\end{eqnarray}
The leading extensive behavior of the R\'enyi entropies define the generalized fractal dimensions $0 \leq {\cal D}_q \leq 1$
 \begin{eqnarray}
S_q(N)  \oppropto_{N \to +\infty} {\cal D}_q  (N \ln 2)
\label{renyid}
\end{eqnarray}
so here one obtains that the generalized fractal dimensions
 \begin{eqnarray}
 {\cal D}_q 
=\frac{1}{(1-q) \ln 2   } \ \  \lim_{N \to +\infty} \left[ \frac{1}{ N  } \sum_{k=1}^{N-1} \ln \left(  
\left[ \frac{1+ \cos \left(\theta_{k+\frac{1}{2}} \right)  }{2} \right]^q
+ \left[ \frac{1- \cos \left(\theta_{k+\frac{1}{2}} \right)  }{2} \right]^q
\right) \right]
\label{multifdim}
\end{eqnarray}
correspond to the spatial average over $k=1,..,N-1$
 of some q-dependent function of the bond angles $\theta_{k+\frac{1}{2}} $.

This specific example suggests that in other contexts,
it would be also interesting to analyze the multifractality in 
the basis that diagonalize the individual single-site reduced density matrices
in order to avoid the arbitrariness of the choice of the basis 
and in order to characterize directly the global entanglement properties
along the chain.


\section{ Reduced density matrix $\rho_{k,k+1,..,k+r}$ of $(r+1)$ consecutive sites in the bulk }

\label{sec_rhointerval}

\subsection{ Simultaneous Schmidt decomposition across the two distant bonds $(k-1,k)$ and $(k+r,k+r+1)$ }

After considering the Schmidt decomposition across a single bond (Eq. \ref{schmidtk})
and across two neighboring bonds (Eq. \ref{schmidtkm1}),
it is now interesting to focus on the simultaneous 
Schmidt decomposition across the two distant bonds $(k-1,k)$ and $(k+r,k+r+1)$
by rewriting Eq. \ref{mpsfinal} as
\begin{eqnarray}
\ket{\psi} && =  \sum_{\alpha_{L}=\pm} \sum_{\alpha_R=\pm}
\ket{\Phi^{[1,...,k-1]}_{\alpha_L } }
\lambda^{\alpha_L}_{k-1,k}
\ket{ I^{[k,..,k+r]}_{\alpha_L,\alpha_R}  }
\lambda^{\alpha_R}_{k+r,k+r+1}
\ket{\Phi^{[k+r+1,...,N]}_{\alpha_R} } 
\label{schmidtinterval}
\end{eqnarray}
where the four ket $\ket{ I^{[k,..,k+r]}_{\alpha_L=\pm,\alpha_R=\pm}  } $ associated to the interval $[k,..,k+r ]$
read
\begin{eqnarray}
\ket{ I^{[k,..,k+r]}_{\alpha_L,\alpha_R}  } && \equiv 
\sum_{\alpha_k=\pm} ... 
\sum_{\alpha_{k+r-1}=\pm}
\left[ \prod_{n=k}^{k+r-1}\lambda^{\alpha_n}_{n,n+1}  \right]
\ket{\tau^z_k= \alpha_L \alpha_k}
\left[ \prod_{n=k+1}^{k+r-1} \ket{\tau^z_n= \alpha_{n-1} \alpha_n} \right]
\ket{\tau^z_{k+r}= \alpha_{k+r-1} \alpha_R}
\label{ketinterval}
\end{eqnarray}
Their scalar products can be computed using Eq. \ref{lambdarhofinal}
\begin{eqnarray}
&& \braket{ I^{[k,..,k+r]}_{\beta_L,\beta_R}   \vert I^{[k,..,k+r]}_{\alpha_L,\alpha_R}  } 
 = 
\delta_{\beta_L\beta_R,  \alpha_L\alpha_R }
\prod_{n=k}^{k+r-1}
\left[ \sum_{\alpha_n=\pm} \lambda^{\alpha_n}_{n,n+1} \overline{ \lambda^{(\beta_L \alpha_L) \alpha_n}_{n,n+1} } \right] 
= \delta_{\beta_L\beta_R,  \alpha_L\alpha_R }
\prod_{n=k}^{k+r-1}
\left[ \sum_{\alpha_n=\pm} \Lambda^{\alpha_n,(\beta_L \alpha_L) \alpha_n}_{n,n+1}  \right]
\nonumber \\ &&
= \delta_{\beta_L\beta_R,  \alpha_L\alpha_R }
\nonumber \\ && 
\prod_{n=k}^{k+r-1}
\left[ \sum_{\alpha_n=\pm} 
\left(\delta_{\alpha_L,\beta_L } \left( \frac{1+\alpha_n\cos \left( \theta_{n+\frac{1}{2}} \right)  }{2}\right)
+ \delta_{\alpha_L,-\beta_L } \left(\frac{\sin \left( \theta_{n+\frac{1}{2}} \right) 
\left[   \cos \left( \phi_{n+\frac{1}{2}}  \right) -i \alpha_n\cos \left( \phi_{n+\frac{1}{2}}  \right)
\right]
 }{2}\right)
\right)
\right]
\nonumber \\ &&
= \delta_{\beta_L\beta_R,  \alpha_L\alpha_R }
\prod_{n=k}^{k+r-1}
\left[  \delta_{\alpha_L,\beta_L } 
+ \delta_{\alpha_L,-\beta_L } \sin \left( \theta_{n+\frac{1}{2}} \right)   \cos \left( \phi_{n+\frac{1}{2}}  \right) 
\right]
 \nonumber \\ &&
= \delta_{\beta_L\beta_R,  \alpha_L\alpha_R }
\left[  \delta_{\alpha_L,\beta_L } 
+ \delta_{\alpha_L,-\beta_L } \omega_{k,k+r} 
\right]
\label{ketintervalscalar}
\end{eqnarray}
where we have introduced the notation
\begin{eqnarray}
\omega_{k,k+r} \equiv 
&& 
= \prod_{n=k}^{k+r-1} \left[  \sin \left( \theta_{n+\frac{1}{2} } \right) \cos \left( \phi_{n+\frac{1}{2} } \right) \right]
\label{overlap}
\end{eqnarray}
This parameter will appear in various observables that involve the correlations between 
the sites $k$ and $k+r$ separated by the distance $r$.

Eq. \ref{ketintervalscalar} means that these four ket are normalized
\begin{eqnarray}
\braket{ I^{[k,..,k+r]}_{\alpha_L,\alpha_R}   \vert I^{[k,..,k+r]}_{\alpha_L,\alpha_R}  }=1
\label{ketintervalnorma}
\end{eqnarray}
and can be classified  into the two following orthogonal subspaces :

(i) the subspace $\alpha_L \alpha_R=1$
is generated by the two ket
$\ket{ I^{[k,..,k+r]}_{+,+}  } $ and $\ket{ I^{[k,..,k+r]}_{-,-}  } $
whose overlap is given by Eq \ref{overlap} 
\begin{eqnarray}
\braket{ I^{[k,..,k+r]}_{-,-}   \vert I^{[k,..,k+r]}_{+,+}  } = \omega_{k,k+r} 
\label{overlapi}
\end{eqnarray}

(ii) the subspace $\alpha_L \alpha_R=-1$
is generated by the two ket
$\ket{ I^{[k,..,k+r]}_{+,-}  } $ and $\ket{ I^{[k,..,k+r]}_{-,+}  } $ 
whose overlap is also given by Eq. \ref{overlap}
\begin{eqnarray}
 \braket{ I^{[k,..,k+r]}_{-,+}   \vert I^{[k,..,k+r]}_{+,-}  }= \omega_{k,k+r}
\label{overlapibis}
\end{eqnarray}

The comparison of Eq. \ref{schmidtinterval} with the Schmidt decomposition across the single bond $(k-1,k)$
yields the following recursions for the Schmidt eigenvectors of the Right part
\begin{eqnarray}
\ket{\Phi^{[k,...,N]}_{\alpha} } && =   \sum_{\alpha_R=\pm}
\ket{ I^{[k,..,k+r]}_{\alpha,\alpha_R}  }
\lambda^{\alpha_R}_{k+r,k+r+1}
\ket{\Phi^{[k+r+1,...,N]}_{\alpha_R} } 
\label{recintervalright}
\end{eqnarray}
while the comparison with the Schmidt decomposition across the single bond $(k+r,k+r+1)$
yields the following recursion for the Schmidt eigenvectors of the Right part
\begin{eqnarray}
\ket{\Phi^{[1,...,k+r]}_{\alpha } } && =  \sum_{\alpha_{L}=\pm} 
\ket{\Phi^{[1,...,k-1]}_{\alpha_L } }
\lambda^{\alpha_L}_{k-1,k}
\ket{ I^{[k,..,k+r]}_{\alpha_L,\alpha}  }
\label{recintervalleft}
\end{eqnarray}
that generalizes the recurrences of Eqs \ref{receigenvectors} concerning a difference of a single site.

The interval ket of Eq. \ref{ketinterval} satisfy the following recurrences
with respect to their leftmost spin $k$
or with respect to the rightmost spin $(k+r)$
\begin{eqnarray}
\ket{ I^{[k,..,k+r]}_{\alpha_L,\alpha_R}  } 
&& = 
\sum_{\alpha=\pm} 
\ket{\tau^z_k= \alpha_L \alpha}
\lambda^{\alpha}_{k,k+1}
\ket{ I^{[k+1,..,k+r]}_{\alpha,\alpha_R}  } 
\nonumber \\
\ket{ I^{[k,..,k+r]}_{\alpha_L,\alpha_R}  }
&& =
\sum_{\alpha=\pm} \ket{ I^{[k,..,k+r-1]}_{\alpha_L,\alpha}  }
\lambda^{\alpha}_{k+r-1,k+r}
\ket{\tau^z_{k+r}= \alpha \alpha_R}
\label{ketintervalrec}
\end{eqnarray}
and satisfy more generally the following decomposition with respect to any internal bond $(n,n+1)$
\begin{eqnarray}
\ket{ I^{[k,..,k+r]}_{\alpha_L,\alpha_R}  } 
&& = 
\sum_{\alpha=\pm} 
\ket{ I^{[k,..,n]}_{\alpha_L,\alpha}  }
\lambda^{\alpha}_{n,n+1}
\ket{ I^{[n+1,..,k+r]}_{\alpha,\alpha_R}  } 
\label{ketintervalrecitself}
\end{eqnarray}

The Schmidt decomposition of Eq. \ref{schmidtinterval} for the MPS ket
translates into the following decomposition for
the full density matrix 
\begin{eqnarray}
 \rho  
&& = \ket{\psi} \bra{\psi} 
=
\sum_{ \substack{\alpha_L=\pm \\ \beta_L=\pm} }
\sum_{ \substack{\alpha_R=\pm \\ \beta_R=\pm} }
 \nonumber \\&& 
\left(
\ket{\Phi^{[1,...,k-1]}_{\alpha_L} }
\bra{\Phi^{[1,...,k-1]}_{\beta_L} }
\right)
\Lambda^{\alpha_L,\beta_L}_{k-1,k} 
\left( \ket{ I^{[k,..,k+r]}_{\alpha_L,\alpha_R}  } \bra{ I^{[k,..,k+r]}_{\beta_L,\beta_R}  } \right)
\Lambda_{k+r,k+r+1}^{\alpha_R,\beta_R}
\left( 
\ket{\Phi^{[k+r+1,...,N]}_{\alpha_R} } 
\bra{\Phi^{[k+r+1,...,N]}_{\beta_R} } 
\right)
\label{rho2cut}
\end{eqnarray}

Depending on the applications, it will be convenient to keep 
the interval operators $\left( \ket{ I^{[k,..,k+r]}_{\alpha_L,\alpha_R}  } \bra{ I^{[k,..,k+r]}_{\beta_L,\beta_R}  } \right) $
in their global form in terms of the interval ket,
or it will be more convenient to write their Matrix-Product-Operator form analogous to Eq. \ref{mporho}
\begin{eqnarray}
 \ket{ I^{[k,..,k+r]}_{\alpha_L,\alpha_R}  } \bra{ I^{[k,..,k+r]}_{\beta_L,\beta_R}  } 
 && = 
\sum_{ \substack{\alpha_k=\pm \\ \beta_k=\pm} }
...
\sum_{ \substack{\alpha_{k+r-1}=\pm \\ \beta_{k+r-1}=\pm} }
\left[ \prod_{n=k}^{k+r-1} \Lambda_{n,n+1}^{\alpha_n,\beta_n}  \right]
 \nonumber \\ &&
O_k^{(\alpha_L \alpha_k),(\beta_L \beta_k)}
 \left[ \prod_{n=k+1}^{k+r-1}  O_n^{(\alpha_{n-1} \alpha_n),(\beta_{n-1} \beta_n)}\right]
O_{k+r}^{(\alpha_{k+r-1} \alpha_R),(\beta_{k+r-1} \beta_R)}
\label{mpomiddle}
\end{eqnarray}


\subsection{ Global properties of
the reduced density matrix $\rho_{k,k+1,..,k+r} $ of the interval $[k,..,k+r]$   }

The reduced density matrix $\rho_{k,k+1,..,k+r} $ of the interval $[k,..,k+r]$
can be computed from the full density matrix of Eq. \ref{rho2cut}
by taking the trace over the Left part $[1,..,k-1]$ that imposes $\beta_L=\alpha_L$
and by taking the trace over the Right part $[k+r+1,..,N]$ that imposes $\beta_R=\alpha_R$
\begin{eqnarray}
 \rho_{k,k+1,..,k+r}  
\equiv {\rm Tr}_{\{[1,..,k-1],[k+r+1,..,N]\}} \left( \rho \right) 
 =\sum_{\alpha_L=\pm} 
\sum_{\alpha_R=\pm} 
\Lambda^{\alpha_L,\alpha_L}_{k-1,k} 
\left( \ket{ I^{[k,..,k+r]}_{\alpha_L,\alpha_R}  } \bra{ I^{[k,..,k+r]}_{\alpha_L,\alpha_R}  } \right)
\Lambda_{k+r,k+r+1}^{\alpha_R,\alpha_R}
 \label{rhointerval} 
\end{eqnarray}
At the global level of the interval, this corresponds 
to the following weighted sum over the four projectors associated to
the four ket of Eq. \ref{ketinterval}
\begin{eqnarray}
 \rho_{k,k+1,..,k+r}  
&&  =\Lambda^{+,+}_{k-1,k} 
\ket{ I^{[k,..,k+r]}_{+,+}  } \bra{ I^{[k,..,k+r]}_{+,+}  } 
\Lambda_{k+r,k+r+1}^{+,+}
+ \Lambda^{-,-}_{k-1,k} 
\ket{ I^{[k,..,k+r]}_{-,-}  } \bra{ I^{[k,..,k+r]}_{-,-}  } 
\Lambda_{k+r,k+r+1}^{-,-}
\nonumber \\
&& +\Lambda^{+,+}_{k-1,k} 
\ket{ I^{[k,..,k+r]}_{+,-}  } \bra{ I^{[k,..,k+r]}_{+,-}  } 
\Lambda_{k+r,k+r+1}^{-,-}
+ \Lambda^{-,-}_{k-1,k} 
\ket{ I^{[k,..,k+r]}_{-,+}  } \bra{ I^{[k,..,k+r]}_{-,+}  } 
\Lambda_{k+r,k+r+1}^{+,+}
 \label{rhointervalglobal} 
\end{eqnarray}
Note that this is not the spectral decomposition of the reduced density matrix $\rho_{k,k+1,..,k+r}   $
 as a consequence as the non-vanishing overlaps of Eq. \ref{overlapi} and of Eq. \ref{overlapibis},
but the diagonal form of $\rho_{k,k+1,..,k+r}   $
is given in Appendix \ref{app_diago}.


\subsection{ Matrix-Product-Operator form of 
the reduced density matrix $\rho_{k,k+1,..,k+r} $ of the Interval $[k,..,k+r]$   }

Plugging the MPO form of Eq. \ref{mpomiddle}
for the interval operators with the special values $\beta_L=\alpha_L$ and $\beta_R=\alpha_R$
\begin{eqnarray}
 \ket{ I^{[k,..,k+r]}_{\alpha_L,\alpha_R}  } \bra{ I^{[k,..,k+r]}_{\alpha_L,\alpha_R}  } 
&& = \sum_{ \substack{\alpha_k=\pm \\ \beta_k=\pm} }
...
\sum_{ \substack{\alpha_{k+r-1}=\pm \\ \beta_{k+r-1}=\pm} }
\left[ \prod_{n=k}^{k+r-1} \Lambda_{n,n+1}^{\alpha_n,\beta_n}  \right]
 \nonumber \\ &&
O_k^{(\alpha_L \alpha_k),(\alpha_L \beta_k)}
 \left[ \prod_{n=k+1}^{k+r-1}  O_n^{(\alpha_{n-1} \alpha_n),(\beta_{n-1} \beta_n)}\right]
O_{k+r}^{(\alpha_{k+r-1} \alpha_R),(\beta_{k+r-1} \alpha_R)}
\label{mpomiddlediag}
\end{eqnarray}
into  Eq. \ref{rhointerval} 
leads to the MPO form
\begin{eqnarray}
 \rho_{k,k+1,..,k+r}  
&& 
 =\sum_{ \substack{\alpha_k=\pm \\ \beta_k=\pm} }
...
\sum_{ \substack{\alpha_{k+r-1}=\pm \\ \beta_{k+r-1}=\pm} }
\left[ \prod_{n=k}^{k+r-1} \Lambda_{n,n+1}^{\alpha_n,\beta_n}  \right]
W_{[k}^{ \alpha_k, \beta_k}
 \left[ \prod_{n=k+1}^{k+r-1}  O_n^{(\alpha_{n-1} \alpha_n),(\beta_{n-1} \beta_n)}\right]
W_{k+r]}^{\alpha_{k+r-1} ,\beta_{k+r-1} }
 \label{rhointervalmpo} 
\end{eqnarray}
with the two modified operators for the spins $k$ and $(k+r)$ at the boundaries of the interval $[k,..,k+r]$
\begin{eqnarray}
W_{[k}^{\alpha_k,\beta_k} && \equiv 
 \sum_{\alpha_L=\pm} 
\Lambda^{\alpha_L,\alpha_L}_{k-1,k} 
\  O_k^{(\alpha_L \alpha_k),(\alpha_L \beta_k)}
 \nonumber \\ 
W_{k+r]}^{\alpha_{k+r-1},\beta_{k+r-1}} 
&& \equiv  
\sum_{\alpha_R=\pm} 
\  O_{k+r}^{(\alpha_{k+r-1} \alpha_R),(\beta_{k+r-1} \alpha_R)}  
\Lambda_{k+r,k+r+1}^{\alpha_R,\alpha_R}
 \label{operatorsW} 
\end{eqnarray}

Using Eqs \ref{lambdarhofinal}
and \ref{operatorfinal}, 
one obtains their explicit expressions in terms of the Pauli operators and the angles
\begin{eqnarray}
W_{[k}^{\alpha_k,\beta_k} && 
= \sum_{\alpha_L=\pm} 
\left( \frac{1+\alpha_L\cos \left( \theta_{k-\frac{1}{2}} \right)  }{2}\right)
 \left[ \delta_{ \alpha_k, \beta_k} \left( \frac{1+\alpha_L \alpha_k \tau_k^z}{2}\right)
+ \delta_{ \alpha_k,- \beta_k} \left(\frac{ \tau^x_k +i \alpha_L \alpha_k \tau^y_k }{2} \right)
\right]
\nonumber \\
&& =
\delta_{ \alpha_k, \beta_k} \left( \frac{1+\cos \left( \theta_{k-\frac{1}{2} }\right) \alpha_k \tau_k^z}{2}\right)
+ \delta_{ \alpha_k,- \beta_k} 
\left( \frac{ \tau^x_k +i \cos \left( \theta_{k-\frac{1}{2} }\right) \alpha_k \tau^y_k }{2} \right)
 \label{Wleft} 
\end{eqnarray}
and similarly
\begin{eqnarray}
&& W_{k+r]}^{\alpha_{k+r-1},\beta_{k+r-1}} 
\label{Wright}
\\
&&  = \delta_{\alpha_{k+r-1} ,\beta_{k+r-1}} 
\left( \frac{  1
+   \cos \left( \theta_{k+r+\frac{1}{2} }\right) \alpha_{k+r-1}\tau_{k+r}^z
}{2}\right)
+ \delta_{\alpha_{k+r-1} ,-\beta_{k+r-1} }  
\left( \frac{ 
 \tau^x_{k+r} 
+i  \cos \left( \theta_{k+r+\frac{1}{2} }\right) \alpha_{k+r-1} \tau^y_{k+r}
 }{2} \right)
\nonumber 
\end{eqnarray}

These results will be used in the section \ref{sec_energy} 
concerning the energy of two-body Hamiltonians
where one needs the MPO form of $\rho_{k,k+1}$,
and will be used in the section \ref{sec_corre} 
concerning the correlations between two spins at distance $r$.


\section{ Optimization of the MPS to approximate ground states  }

\label{sec_energy}

The MPS $\ket{ \psi}$ or equivalently the corresponding full density matrix $\rho= \ket{\psi} \bra{\psi}$
allows to compute the energy associated to an Hamiltonian $H$ via
\begin{eqnarray}
E = \bra{\psi} H \ket{\psi}  = {\rm Tr} (H \rho) 
\label{energie}
\end{eqnarray}
For a given local Hamiltonian, the goal is then to find the MPS parameters
that minimize the energy in order to obtain the best approximation of the ground-state
within the MPS manifold.

\subsection { General Hamiltonian with one-body and two-body terms   }

Let us consider a general Hamiltonian with one-body and two-body terms
\begin{eqnarray}
H && = \sum_{k=1}^N h_k +  \sum_{k=1}^{N-1} h_{k,k+1}
\label{hgene}
\end{eqnarray}
The one-body Hamiltonian $h_k$ acting on the spin $k$ can be expanded in the Pauli basis $\sigma_k^{x,y,z}$
with three coefficients $B_k^{x,y,z}$ that play the role of the three components of the local magnetic field
\begin{eqnarray}
h_k && =  \sum_{a=x,y,z} B_k^a \sigma_k^a = B_k^x \sigma_k^x+B_k^y \sigma_k^y+B_k^z \sigma_k^z
\label{hk}
\end{eqnarray}
while the two-body Hamiltonian $h_{k,k+1} $ acting on the two consecutive spins $k$ and $(k+1)$
can be expanded in the Pauli basis of these two spins with up to nine couplings $J_{k+\frac{1}{2} }^{ab} $
\begin{eqnarray}
h_{k,k+1} && \equiv  \sum_{a=x,y,z} \sum_{b=x,y,z}  J_{k+\frac{1}{2} }^{ab} \sigma_k^a \sigma_{k+1}^b
\label{hkk}
\end{eqnarray}

As a consequence, the energy of Eq. \ref{energie} can be decomposed into 
\begin{eqnarray}
E && =  \sum_{k=1}^N e_k +  \sum_{k=1}^{N-1} e_{k,k+1} 
\label{energiedecomposition}
\end{eqnarray}
where the elementary contributions associated to $h_k$ and $h_{k,k+1}$
involve only the reduced density matrix of a single spin $\rho_k$
and the reduced density matrix of two neighboring spins $\rho_{k,k+1}$
\begin{eqnarray}
 e_k && =   {\rm Tr}(h_k \rho) 
={\rm Tr}_{\{k\}}(h_k \rho_k) 
\nonumber \\
 e_{k,k+1}  && =    {\rm Tr}( h_{k,k+1}\rho) =
{\rm Tr}_{\{k,k+1\}}( h_{k,k+1}\rho_{k,k+1})
\label{elementary}
\end{eqnarray}

\subsection { Contribution $e_k$ of $h_k$  }

Since the reduced density matrix $\rho_k$ of Eq. \ref{rhokbulkz}
only contains the operator $\tau_k^z$
\begin{eqnarray}
 \rho_k  =\frac{1 }{2} \left[1+\cos \left( \theta_{k-\frac{1}{2}} \right) \cos \left( \theta_{k+\frac{1}{2}} \right) \tau_k^z \right] 
\label{rhokall}
\end{eqnarray}
with the convention $\cos \left( \theta_{\frac{1}{2}} \right)=1=\cos \left( \theta_{N+\frac{1}{2}} \right) $
to describe also the two boundary spins (Eqs
\ref{rho1boundary} and \ref{rhoNboundary}), it is convenient to translate the one-body Hamiltonian $h_k$ of
Eq. \ref{hk} into the $\tau$ basis using Eq. \ref{rotationab}
\begin{eqnarray}
h_k && =  \sum_{a=x,y,z} B_k^a \sigma_k^a = \sum_{a=x,y,z} {\tilde B}_k^a\tau_k^a
\label{hktau}
\end{eqnarray}

The contribution $e_k$ of Eq. \ref{elementary} actually involves only the component 
\begin{eqnarray}
{\tilde B}_k^z && 
= \cos \theta_k B_k^z + \sin \theta_k \left( \cos \phi_k B_k^x +  \sin \phi_k B_k^y  \right)
\label{Btildez}
\end{eqnarray}
as a consequence of the specific form of Eq. \ref{rhokall}
\begin{eqnarray}
 e_k &&  ={\rm Tr}_{\{k\}}(h_k \rho_k) = \cos \left( \theta_{k-\frac{1}{2}} \right) \cos \left( \theta_{k+\frac{1}{2}} \right){\tilde B}_k^z 
\label{e1}
\end{eqnarray}

If one wishes to see the role of each initial magnetic field component $B_k^a $ in the initial basis $\sigma$,
one can decompose Eq. \ref{e1} 
\begin{eqnarray}
e_k  =  \sum_{a=x,y,z} B_k^a e_k^a
\label{e1decompose}
\end{eqnarray}
into the following contributions
\begin{eqnarray}
e_k^z && = \cos \left( \theta_{k-\frac{1}{2}} \right)  
\left[\cos \theta_k  \right]\cos \left( \theta_{k+\frac{1}{2}} \right)
\nonumber \\
e_k^x &&= \cos \left( \theta_{k-\frac{1}{2}} \right)  
\left[  \sin \theta_k \cos \phi_k   \right]\cos \left( \theta_{k+\frac{1}{2}} \right)
\nonumber \\
e_k^y && =\cos \left( \theta_{k-\frac{1}{2}} \right)  
\left[  \sin \theta_k  \sin \phi_k \right]\cos \left( \theta_{k+\frac{1}{2}} \right)
\label{e1a}
\end{eqnarray}


\subsection{ Matrix-Product-Operator form of the reduced density matrix $\rho_{k,k+1}   $ of two consecutive sites $(k,k+1)$ }

\label{sec_rhotwoconsecutive}

The Matrix-Product-Operator form of the reduced density matrix $\rho_{k,k+1}   $ of two consecutive sites $(k,k+1)$
corresponds to the special case $r=1$ in Eq. \ref{rhointervalmpo}.
Using the explicit forms of Eqs \ref{lambdarhofinal} , \ref{Wleft} and \ref{Wright}, 
one obtains the MPO form
\begin{eqnarray}
 \rho_{k,k+1}  
&&  =  \sum_{\alpha=\pm}  \sum_{\beta=\pm} 
 \Lambda_{k,k+1}^{\alpha,\beta}  
W_{[k}^{ \alpha, \beta}
W_{k+1]}^{\alpha ,\beta }
\label{rho2sitesmpo}
 \\
&& = \sum_{\alpha=\pm}  \sum_{\beta=\pm} 
\left[ \delta_{\alpha,\beta } \left( \frac{1+\alpha\cos \left( \theta_{k+\frac{1}{2}} \right)  }{2}\right)
+ \delta_{\alpha,-\beta } \left(\frac{\sin \left( \theta_{k+\frac{1}{2}} \right)  }{2}e^{-i \alpha \phi_{k+\frac{1}{2}} }\right)
\right]
\nonumber \\
&&\left[\delta_{ \alpha, \beta} \left( \frac{1+\cos \left( \theta_{k-\frac{1}{2} }\right) \alpha \tau_k^z}{2}\right)
+ \delta_{ \alpha,- \beta} 
\left( \frac{ \tau^x_k +i \cos \left( \theta_{k-\frac{1}{2} }\right) \alpha \tau^y_k }{2} \right)
\right]
\nonumber \\
&& \left[
\delta_{\alpha ,\beta} 
\left( \frac{  1
+   \cos \left( \theta_{k+\frac{3}{2} }\right) \alpha\tau_{k+1}^z
}{2}\right)
+ \delta_{\alpha ,-\beta }  
\left( \frac{ 
 \tau^x_{k+1} 
+i  \cos \left( \theta_{k+\frac{3}{2} }\right) \alpha \tau^y_{k+1}
 }{2} \right)
\right]
\nonumber \\
&& = \sum_{\alpha=\pm} 
 \left( \frac{1+\alpha\cos \left( \theta_{k+\frac{1}{2}} \right)  }{2}\right)
\left( \frac{1+\alpha \cos \left( \theta_{k-\frac{1}{2} }\right)  \tau_k^z}{2}\right)
\left( \frac{  1
+  \alpha \cos \left( \theta_{k+\frac{3}{2} }\right) \tau_{k+1}^z}{2}\right)
\nonumber \\
&& + \frac{\sin \left( \theta_{k+\frac{1}{2}} \right)  }{2} 
\sum_{\alpha=\pm}  
  \left(  \cos(\phi_{k+\frac{1}{2}}  )  -i \alpha \sin( \phi_{k+\frac{1}{2}} )  \right)
\left( \frac{ \tau^x_k +i \alpha\cos \left( \theta_{k-\frac{1}{2} }\right)  \tau^y_k }{2} \right)
\left( \frac{ \tau^x_{k+1} +i \alpha \cos \left( \theta_{k+\frac{3}{2} }\right)  \tau^y_{k+1} }{2} \right)
 \nonumber
\end{eqnarray}
This factorized form with respect to the Pauli operators of the two spins
can be expanded in order to perform the summation over $\alpha=\pm$ 
and one obtains the final result
\begin{eqnarray}
  \rho_{k,k+1}  
 && =\frac{1}{4}
+\cos \left( \theta_{k+\frac{1}{2}} \right) \left[
 \cos \left( \theta_{k-\frac{1}{2} }\right)  \frac{\tau_k^z}{4}
+  \cos \left( \theta_{k+\frac{3}{2} }\right)  \frac{\tau_{k+1}^z}{4} \right]
+\cos \left( \theta_{k-\frac{1}{2} }\right)  \cos \left( \theta_{k+\frac{3}{2} }\right) \frac{\tau_k^z \tau_{k+1}^z}{4}
\nonumber \\
&& + 
 \sin \left( \theta_{k+\frac{1}{2}} \right)   \cos(\phi_{k+\frac{1}{2}}  ) 
\left[ \frac{\tau^x_k\tau^x_{k+1} }{4}
-
\cos \left( \theta_{k-\frac{1}{2} }\right) 
\cos \left( \theta_{k+\frac{3}{2} }\right)  
\frac{\tau^y_k \tau^y_{k+1} }{4} \right]
\nonumber \\
&& + 
 \sin \left( \theta_{k+\frac{1}{2}} \right) \sin( \phi_{k+\frac{1}{2}} ) 
\left[ 
 \cos \left( \theta_{k+\frac{3}{2} }\right)   \frac{ \tau^x_k\tau^y_{k+1} }{4}
+  \cos \left( \theta_{k-\frac{1}{2} }\right)  
\frac{ \tau^y_k \tau^x_{k+1} }{4} \right]
 \label{rho2sitesdv}
\end{eqnarray}
Again as in Eq. \ref{rho12}, it involves only the 8 operators that commute with $\tau^z_1\tau^z_2  $
among the 16 operators of the tensor basis $\tau_1^a \otimes \tau_2^b$ with
$a=0,x,y,z$ and $b=0,x,y,z$.

\subsection { Contribution $e_{k,k+1}$ of $h_{k,k+1}$  }

Again it is convenient to translate the two-body Hamiltonian $h_{k,k+1}$ of
Eq. \ref{hkk} into the $\tau$ basis using Eq. \ref{rotationab}
\begin{eqnarray}
h_{k,k+1} && \equiv  \sum_{a=x,y,z} \sum_{b=x,y,z}  J_{k+\frac{1}{2} }^{ab} \sigma_k^a \sigma_{k+1}^b
=  \sum_{a=x,y,z} \sum_{b=x,y,z}  {\tilde J}_{k+\frac{1}{2} }^{ab} \tau_k^a \tau_{k+1}^b
\label{hkktilde}
\end{eqnarray}
with the effective couplings in the $\tau$ basis
\begin{eqnarray}
 {\tilde J}_{k+\frac{1}{2} }^{ab} = \sum_{a'=x,y,z} \sum_{b'=x,y,z}  J_{k+\frac{1}{2} }^{a'b'} R_k^{a' a} R_{k+1}^{b' b} 
\label{jtilde}
\end{eqnarray}

Since the reduced density matrix of Eq. \ref{rho2sitesdv} involves only five terms containing two Pauli matrices,
the contribution $e_{k,k+1}$ of Eq. \ref{elementary}
involve the corresponding five effective couplings
\begin{eqnarray}
  e_{k,k+1}  
 && =
 \cos \left( \theta_{k-\frac{1}{2} }\right)  \cos \left( \theta_{k+\frac{3}{2} }\right) {\tilde J}_{k+\frac{1}{2} }^{zz}
 + 
 \sin \left( \theta_{k+\frac{1}{2}} \right)   \cos(\phi_{k+\frac{1}{2}}  )
\left[ {\tilde J}_{k+\frac{1}{2} }^{xx}
-
\cos \left( \theta_{k-\frac{1}{2} }\right) 
\cos \left( \theta_{k+\frac{3}{2} }\right)  {\tilde J}_{k+\frac{1}{2} }^{yy} \right]
\nonumber \\
&& + 
 \sin \left( \theta_{k+\frac{1}{2}} \right) \sin( \phi_{k+\frac{1}{2}} ) 
\left[ 
 \cos \left( \theta_{k+\frac{3}{2} }\right)  
 {\tilde J}_{k+\frac{1}{2} }^{xy}
+  \cos \left( \theta_{k-\frac{1}{2} }\right) 
 {\tilde J}_{k+\frac{1}{2} }^{yx} \right]
 \label{e2}
\end{eqnarray}

If one wishes to see the role of each initial coupling $J_{k+\frac{1}{2} }^{ab}  $ in the initial basis $\sigma$,
one can decompose Eq. \ref{e2} 
\begin{eqnarray}
e_{k,k+1}  =  \sum_{a=x,y,z} \sum_{b=x,y,z} J_{k+\frac{1}{2} }^{ab}  e^{a b}_{k,k+1} 
\label{e2decompose}
\end{eqnarray}
into the corresponding contributions
\begin{eqnarray}
  e^{a b}_{k,k+1}  
 && =
 \cos \left( \theta_{k-\frac{1}{2} }\right)  \cos \left( \theta_{k+\frac{3}{2} }\right) R_k^{az}R_{k+1}^{bz}
 + 
 \sin \left( \theta_{k+\frac{1}{2}} \right)   \cos(\phi_{k+\frac{1}{2}}  )
\left[ R_k^{ax}R_{k+1}^{bx}
-
\cos \left( \theta_{k-\frac{1}{2} }\right) 
\cos \left( \theta_{k+\frac{3}{2} }\right) R_k^{ay}R_{k+1}^{by}
  \right]
\nonumber \\
&& + 
 \sin \left( \theta_{k+\frac{1}{2}} \right) \sin( \phi_{k+\frac{1}{2}} ) 
\left[ 
 \cos \left( \theta_{k+\frac{3}{2} }\right)  
 R_k^{ax}R_{k+1}^{by}
+  \cos \left( \theta_{k-\frac{1}{2} }\right) R_k^{ay}R_{k+1}^{bx}
 \right]
 \label{e2ab}
\end{eqnarray}
where the rotation coefficients $R_k^{ab}$ involve the angles $(\theta_k,\phi_k)$ of the site $k$ (Eq. \ref{operatorsigmatau})
while the rotation coefficients $R_{k+1}^{ab}$ involve the angles $(\theta_{k+1},\phi_{k+1})$ of the site $(k+1)$.

As examples, let us mention the three contributions $a=b$
that correspond to the three standard couplings $J_{k+1/2}^{zz}$, $J_{k+1/2}^{xx}$ and $J_{k+1/2}^{yy}$.
The $zz$ contribution involves only two terms
\begin{eqnarray}
 e^{zz}_{k,k+1} 
&&=
\cos \left( \theta_{k-\frac{1}{2} }\right) 
\cos \left( \theta_{k+\frac{3}{2} }\right)  \cos \theta_k    \cos \theta_{k+1} 
+ 
 \sin \left( \theta_{k+\frac{1}{2} }\right)
\cos \left( \phi_{k+\frac{1}{2} }\right) 
\sin \theta_k \sin \theta_{k+1} 
\label{e2singlezz}
\end{eqnarray}
while the $xx$ and $yy$ contributions involve the five terms of Eq. \ref{e2ab}
\begin{eqnarray}
  e^{xx}_{k,k+1} 
&&=
\cos \left( \theta_{k-\frac{1}{2} }\right) 
\cos \left( \theta_{k+\frac{3}{2} }\right) 
\left[ \sin \theta_k \cos \phi_k \right]
\left[ \sin \theta_{k+1} \cos \phi_{k+1} \right]
\nonumber \\
&& 
+
 \sin \left( \theta_{k+\frac{1}{2} }\right)
\cos \left( \phi_{k+\frac{1}{2} }\right) 
\left(
\left[ \cos \theta_k \cos \phi_k  \right]
\left[ \cos \theta_{k+1} \cos \phi_{k+1}\right]
-  \cos \left( \theta_{k-\frac{1}{2} }\right) 
\cos \left( \theta_{k+\frac{3}{2} }\right) 
\sin \phi_k
\sin \phi_{k+1} \right)
\nonumber \\
&& 
-\sin \left( \theta_{k+\frac{1}{2} }\right)
\sin \left( \phi_{k+\frac{1}{2} }\right)
\left(  \cos \left( \theta_{k+\frac{3}{2} }\right) 
\left[  \cos \theta_k \cos \phi_k \right]
\sin \phi_{k+1}
+  \cos \left( \theta_{k-\frac{1}{2} }\right) 
\sin \phi_k
\left[  \cos \theta_{k+1} \cos \phi_{k+1}\right]
\right)
\label{e2singlexx}
\end{eqnarray}
and

\begin{eqnarray}
 e^{yy}_{k,k+1}  
&&=
\cos \left( \theta_{k-\frac{1}{2} }\right) 
\cos \left( \theta_{k+\frac{3}{2} }\right) 
\left[\sin \theta_k \sin \phi_k \right]
\left[\sin \theta_{k+1} \sin \phi_{k+1} \right]
\nonumber \\
&& 
+\sin \left( \theta_{k+\frac{1}{2} }\right)
\cos \left( \phi_{k+\frac{1}{2} }\right)
\left(
\left[ \cos \theta_k \sin \phi_k \right]
\left[ \cos \theta_{k+1} \sin \phi_{k+1}  \right]
 -  \cos \left( \theta_{k-\frac{1}{2} }\right) 
\cos \left( \theta_{k+\frac{3}{2} }\right) 
 \cos \phi_k 
\cos \phi_{k+1}
\right)
\nonumber \\
&& 
+\sin \left( \theta_{k+\frac{1}{2} }\right)
\sin \left( \phi_{k+\frac{1}{2} }\right) 
\left( \cos \left( \theta_{k+\frac{3}{2} }\right) 
\left[ \cos \theta_k \sin \phi_k \right]
  \cos \phi_{k+1} 
+  \cos \left( \theta_{k-\frac{1}{2} }\right) 
 \cos \phi_k 
\left[ \cos \theta_{k+1} \sin \phi_{k+1} \right]
\right)
\label{e2singleyy}
\end{eqnarray}


\subsection{ Optimization of the MPS parameters to minimize the energy }

As explained above, the energy can be computed in terms of the $(4N-2)$ angles
via the elementary contributions of Eq. \ref{energiedecomposition},
where $e_k^a$ was found in Eq. \ref{e1a} to depend only 
on the four angles $[\theta_{k-\frac{1}{2}}; \theta_k , \phi_k ; \theta_{k+\frac{1}{2}}  ]$,
while  $e^{a b}_{k,k+1} $ was found in Eq. \ref{e2ab} to depend 
on the eight angles $[\theta_{k-\frac{1}{2}}; \theta_k , \phi_k ; \theta_{k+\frac{1}{2}},\phi_{k+\frac{1}{2}} ;
 \theta_{k+1} , \phi_{k+1} ; \theta_{k+\frac{3}{2}} ]$.
It is thus convenient to rewrite Eq. \ref{energiedecomposition}
with these explicit dependences 
\begin{eqnarray}
E && =  \sum_{k=1}^N \sum_{a=x,y,z} 
B_k^a \  e^a_k \left( \theta_{k-\frac{1}{2}}; \theta_k , \phi_k ; \theta_{k+\frac{1}{2}}  \right)
\nonumber \\
&& +  \sum_{k=1}^{N-1} \sum_{a=x,y,z} \sum_{b=x,y,z}  J_{k+\frac{1}{2} }^{ab} 
\  e^{a b}_{k,k+1} 
\left( \theta_{k-\frac{1}{2}}; \theta_k , \phi_k ; \theta_{k+\frac{1}{2}},\phi_{k+\frac{1}{2}} ;
 \theta_{k+1} , \phi_{k+1} ; \theta_{k+\frac{3}{2}} \right)
\label{energiedecompositionbis}
\end{eqnarray}

For $k=1,..,N$, the optimizations with respect to the sites angles $(\theta_k,\phi_k)$ for $k=1,..,N$ 
only involve the terms related to $(e^{a b}_{k-1,k},e^a_k,e^{a b}_{k,k+1} )$ as expected
\begin{eqnarray}
 0 && = \frac{ \partial E }{ \partial \theta_k}
 =\sum_{a=x,y,z} B_k^a 
 \frac{ \partial e^a_k  }{ \partial \theta_k} 
 + \sum_{a=x,y,z} \sum_{b=x,y,z}  
\left[ J_{k-\frac{1}{2} }^{ab} 
 \frac{ \partial e^{a b}_{k-1,k}  }{ \partial \theta_k}
+J_{k+\frac{1}{2} }^{ab} 
 \frac{ \partial   e^{a b}_{k,k+1}  }{ \partial \theta_k}
\right]
\nonumber \\
 0 && = \frac{ \partial E }{ \partial \phi_k}
 =\sum_{a=x,y,z} B_k^a 
 \frac{ \partial e^a_k  }{ \partial \phi_k} 
 + \sum_{a=x,y,z} \sum_{b=x,y,z}  
\left[ J_{k-\frac{1}{2} }^{ab} 
 \frac{ \partial e^{a b}_{k-1,k}  }{ \partial \phi_k}
+J_{k+\frac{1}{2} }^{ab} 
 \frac{ \partial   e^{a b}_{k,k+1}  }{ \partial \phi_k}
\right]
\label{opti2}
\end{eqnarray}

For $k=1,..,N-1$, the optimizations with respect to the bond angles $\phi_{k+\frac{1}{2}}$
only involve the term related to $(e^{a b}_{k,k+1} )$
\begin{eqnarray}
 0 = \frac{ \partial E }{ \partial \phi_{k+\frac{1}{2}}}
 = \sum_{a=x,y,z} \sum_{b=x,y,z}  J_{k+\frac{1}{2} }^{ab} 
\frac{ \partial  e^{a b}_{k,k+1}  }{ \partial \phi_{k+\frac{1}{2}}}
\label{opti4}
\end{eqnarray}
while the optimizations with respect to the bond angles $\theta_{k+\frac{1}{2}}$
involve the terms related to $(e^{a b}_{k-1,k},e^a_k,e^{a b}_{k,k+1},e^a_{k+1},e^{a b}_{k+1,k+2} )$
\begin{eqnarray}
 0 = \frac{ \partial E }{ \partial \theta_{k+\frac{1}{2}}}
&& = 
\sum_{a=x,y,z} 
\left[ B_k^a     \frac{ \partial e^a_k  }{ \partial \theta_{k+\frac{1}{2}}}
+
B_{k+1}^a     \frac{ \partial e^a_{k+1} }{ \partial \theta_{k+\frac{1}{2}}}
\right]
 \nonumber \\&&
+ \sum_{a=x,y,z} \sum_{b=x,y,z}  
\left[ J_{k-\frac{1}{2} }^{ab}  
 \frac{ \partial e^{a b}_{k-1,k}  }{ \partial \theta_{k+\frac{1}{2}}}
+J_{k+\frac{1}{2} }^{ab} 
  e^{a b}_{k,k+1} 
 \frac{ \partial   e^{a b}_{k,k+1}  }{ \partial \theta_{k+\frac{1}{2}}}
+J_{k+\frac{3}{2} }^{ab} 
  e^{a b}_{k+1,k+2} 
 \frac{ \partial  e^{a b}_{k+1,k+2}  }{ \partial \theta_{k+\frac{1}{2}}}
\right]
\label{opti3}
\end{eqnarray}


\section{  Correlations between the two spins $(k,k+r)$ at distance $r$ }

\label{sec_corre}

To compute any correlations between the two spins $k$ and $(k+r)$ at distance $r$,
one needs the reduced density matrix $\rho_{k,k+r}$ of these two spins.

\subsection{ Reduced density matrix $\rho_{k,k+r}$ of the two spins $(k,k+r)$ at distance $r$ }

We have already computed $\rho_{k,k+1} $ for two consecutive spins in section \ref{sec_rhotwoconsecutive},
so in this section we focus only on the cases $r>1$.
It is convenient to use the MPO form of Eq \ref{rhointervalmpo} 
for the reduced density matrix $\rho_{k,k+1,..,k+r}   $
of the whole interval $(k,k+1,..,k+r-1,k+r)$,
and to compute the trace over all the interval spins $n=k+1,..,k+r-1$ using Eq. \ref{operatorfinal}
\begin{eqnarray}
&& \rho_{k,k+r}  = {\rm Tr}_{\{k+1,..,k+r-1\}} \left(\rho_{k,k+1,..,k+r}  \right)
 \label{rhokr} 
 \\
&& =
\sum_{ \substack{\alpha_k=\pm \\ \beta_k=\pm} }
...
\sum_{ \substack{\alpha_{k+r-1}=\pm \\ \beta_{k+r-1}=\pm} }
\left[ \prod_{n=k}^{k+r-1} \Lambda_{n,n+1}^{\alpha_n,\beta_n}  \right]
W_{[k}^{ \alpha_k, \beta_k}
 \left[ \prod_{n=k+1}^{k+r-1}  {\rm Tr}_{\{n\}} \left(O_n^{(\alpha_{n-1} \alpha_n),(\beta_{n-1} \beta_n)} \right) \right]
W_{k+r]}^{\alpha_{k+r-1} ,\beta_{k+r-1} }
\nonumber \\
&& =
\sum_{ \substack{\alpha_k=\pm \\ \beta_k=\pm} }
...
\sum_{ \substack{\alpha_{k+r-1}=\pm \\ \beta_{k+r-1}=\pm} }
\left[ \prod_{n=k}^{k+r-1} \Lambda_{n,n+1}^{\alpha_n,\beta_n}  \right]
W_{[k}^{ \alpha_k, \beta_k}
 \left[ \prod_{n=k+1}^{k+r-1}  \delta_{\beta_n,(\alpha_{n-1}\beta_{n-1}) \alpha_n)}  \right]
W_{k+r]}^{\alpha_{k+r-1} ,\beta_{k+r-1} }
\nonumber \\
&& =\sum_{ \substack{\alpha_k=\pm \\ \beta_k=\pm} }
\sum_{\alpha_{k+r-1}=\pm} 
W_{[k}^{ \alpha_k, \beta_k}
W_{k+r]}^{\alpha_{k+r-1} ,(\alpha_k\beta_k)\alpha_{k+r-1} }
\Lambda_{k,k+1}^{\alpha_k,\beta_k}\Lambda_{k+r-1,k+r}^{\alpha_{k+r-1},(\alpha_k\beta_k)\alpha_{k+r-1} }
\left[ \prod_{n=k+1}^{k+r-2} \left( \sum_{\alpha_{n}=\pm}\Lambda_{n,n+1}^{\alpha_n,(\alpha_k\beta_k)\alpha_n} \right) \right]
 \nonumber
\end{eqnarray}

Replacing the boundary operators by their explicit forms of
Eqs  \ref{Wleft}, \ref{Wright} 
\begin{eqnarray}
 W_k^{\alpha_k,\beta_k}  
&& 
 =
 \delta_{ \alpha_k, \beta_k} \left( \frac{1+\cos \left( \theta_{k-\frac{1}{2} }\right) \alpha_k \tau_k^z}{2}\right)
+ \delta_{ \alpha_k,- \beta_k} 
\left( \frac{ \tau^x_k +i \cos \left( \theta_{k-\frac{1}{2} }\right) \alpha_k \tau^y_k }{2} \right) 
 \label{ww} \\
W_{k+r}^{\alpha_{k+r-1},(\alpha_k \beta_k)\alpha_{k+r-1}} 
&& = \delta_{\alpha_k , \beta_k} 
\left( \frac{  1+   \cos \left( \theta_{k+r+\frac{1}{2} }\right) \alpha_{k+r-1}\tau_{k+r}^z}{2}\right)
+ \delta_{\alpha_k ,- \beta_k}  
\left( \frac{  \tau^x_{k+r} +i  \cos \left( \theta_{k+r+\frac{1}{2} }\right) \alpha_{k+r-1} \tau^y_{k+r} }{2} \right)
\nonumber 
\end{eqnarray}
replacing the bond variables by their explicit forms of Eq. \ref{lambdarhofinal}
\begin{eqnarray}
\Lambda_{k,k+1}^{\alpha_k,\beta_k}
&& = \delta_{\alpha_k,\beta_k } \left( \frac{1+\alpha_k\cos \left( \theta_{k+\frac{1}{2}} \right)  }{2}\right)
+ \delta_{\alpha_k,-\beta_k } \left(\frac{\sin \left( \theta_{k+\frac{1}{2}} \right)  }{2}e^{-i \alpha_k \phi_{k+\frac{1}{2}} }\right)
\nonumber \\
\Lambda_{k+r-1,k+r}^{\alpha_{k+r-1},(\alpha_k\beta_k)\alpha_{k+r-1} }
&& 
= \delta_{\alpha_k,\beta_k } \left( \frac{1+\alpha_{k+r-1}\cos \left( \theta_{k+r-\frac{1}{2}} \right)  }{2}\right)
+ \delta_{\alpha_k,-\beta_k }
 \left(\frac{\sin \left( \theta_{k+r-\frac{1}{2}} \right)  }{2}e^{-i \alpha_{k+r-1} \phi_{k+r-\frac{1}{2}} }\right)
\label{lambdarhofinalinter}
\end{eqnarray}
and replacing the last factor by its explicit form in terms of the overlap introduced in Eq. \ref{overlapi}
\begin{eqnarray}
&& \prod_{n=k+1}^{k+r-2} 
\left( \sum_{\alpha_{n}=\pm}\Lambda_{n,n+1}^{\alpha_n,(\alpha_k\beta_k)\alpha_n} \right) =
 \prod_{n=k+1}^{k+r-2} \left( 
\Lambda_{n,n+1}^{+,(\alpha_k\beta_k)}  + \Lambda_{n,n+1}^{-,-(\alpha_k\beta_k)} \right)
 = \delta_{\alpha_k,\beta_k }
+ \delta_{\alpha_k,-\beta_k }\omega_{k+1,k+r-1} 
\label{lastfactor}
\end{eqnarray}
with the convention
\begin{eqnarray}
\omega_{k+1,k+1} 
&& = 1
\label{omeganeg}
\end{eqnarray}
in order to describe also the special case $r=2$ within the same formula,
Eq \ref{rhokr} reduces to

\begin{eqnarray}
 \rho_{k,k+r}  
&& =
\left[  \sum_{\alpha_k=\pm} 
\left( \frac{1+\alpha_k \cos \left( \theta_{k-\frac{1}{2} }\right) \tau_k^z}{2}\right)
\left( \frac{1+\alpha_k\cos \left( \theta_{k+\frac{1}{2}} \right)  }{2}\right)
\right]
\nonumber \\
&& 
\left[
\sum_{\alpha_{k+r-1}=\pm} 
\left( \frac{  1+ \alpha_{k+r-1}  \cos \left( \theta_{k+r+\frac{1}{2} }\right) \tau_{k+r}^z}{2}\right)
\left( \frac{1+\alpha_{k+r-1}\cos \left( \theta_{k+r-\frac{1}{2}} \right)  }{2}\right)
\right]
 \nonumber \\
&&
+\sin \left( \theta_{k+\frac{1}{2}} \right) \omega_{k+1,k+r-1} \sin \left( \theta_{k+r-\frac{1}{2}} \right) 
 \left[\sum_{\alpha_k=\pm} 
\left( \frac{ \tau^x_k +i \alpha_k \cos \left( \theta_{k-\frac{1}{2} }\right)  \tau^y_k }{2} \right) 
\left(\frac{\cos \left( \phi_{k+\frac{1}{2}}\right) -i \alpha_k \sin \left( \phi_{k+\frac{1}{2}}\right)  }{2}\right)
\right]
\nonumber \\
&& 
\left[ 
\sum_{\alpha_{k+r-1}=\pm} 
\left( \frac{  \tau^x_{k+r} +i \alpha_{k+r-1} \cos \left( \theta_{k+r+\frac{1}{2} }\right)  \tau^y_{k+r} }{2} \right)
 \left(\frac{\cos \left( \phi_{k+r-\frac{1}{2}} \right) -i \alpha_{k+r-1} \sin \left( \phi_{k+r-\frac{1}{2}} \right) }{2}\right)
\right]
\nonumber \\
&& =
\left( \frac{1+ \cos \left( \theta_{k-\frac{1}{2} }\right) \tau_k^z\cos \left( \theta_{k+\frac{1}{2}} \right)  }{2}\right)
\left( \frac{  1+   \cos \left( \theta_{k+r-\frac{1}{2} }\right) \tau_{k+r}^z \cos \left( \theta_{k+r+\frac{1}{2}} \right)  }{2}\right)
 \nonumber \\
&&
+\omega_{k,k+r} 
\left( \frac{ \tau^x_k + \cos \left( \theta_{k-\frac{1}{2} }\right)  \tau^y_k \tan \left( \phi_{k+\frac{1}{2}}\right)  }{2}\right)
\left( \frac{  \tau^x_{k+r} 
+ \tan \left( \phi_{k+r-\frac{1}{2}} \right)\tau^y_{k+r} 
\cos \left( \theta_{k+r+\frac{1}{2} }\right)  
  }{2}\right)
\label{rhokrfactor}
\end{eqnarray}
One could of course expand to obtain the form analogous to Eq \ref{rho2sitesdv}
for $\rho_{k,k+1}$, but the factorized form of Eq. \ref{rhokrfactor}
has a very direct physical meaning in terms of connected correlations.

\subsection{  Connected reduced density matrix $\rho^{connected}_{k,k+r} $ and physical meaning of $\omega_{k,k+r}   $ }

Since the first contribution in the final result of Eq. \ref{rhokrfactor}
corresponds to the product of the one-point reduced density matrices
$\rho_k$ and $\rho_{k+r}$ of Eq \ref{rhokbulkz},
it is useful to rewrite Eq. \ref{rhokrfactor} with the difference
\begin{eqnarray}
\rho^{connected}_{k,k+r} && \equiv \rho_{k,k+r}   - \rho_k  \rho_{k+r} 
\nonumber \\
&& 
=\omega_{k,k+r} 
\left( \frac{ \tau^x_k + \cos \left( \theta_{k-\frac{1}{2} }\right)  \tau^y_k \tan \left( \phi_{k+\frac{1}{2}}\right)  }{2}\right)
\left( \frac{  \tau^x_{k+r} + \tan \left( \phi_{k+r-\frac{1}{2}} \right)\tau^y_{k+r} \cos \left( \theta_{k+r+\frac{1}{2} }\right)    }{2}\right)
\label{rhoconnected}
\end{eqnarray}
that shows clearly the physical meaning of the prefactor $\omega_{k,k+r}  $
that governs the decay with the distance $r$ via the multiplicative structure of Eq. \ref{overlap}.


\subsection{  Computation of the two-point connected correlations between the two sites $(k,k+r)$ at distance $r$ }

Any type of connected correlations with $a=x,y,z$ and $b=x,y,z$ 
between the two sites $k$ and $(k+r)$ can be now computed from the reduced density matrix $\rho^{connected}_{k,k+r}$
of Eq. \ref{rhoconnected}
via
\begin{eqnarray}
C^{ab}_{k,k+r} && \equiv \bra{\psi} \sigma_k^a \sigma_{k+r}^b \ket{\psi} 
 - \bra{\psi} \sigma_k^a  \ket{\psi} \bra{\psi}  \sigma_{k+r}^b \ket{\psi} 
=
 {\rm Tr}_{\{k,k+r\}}( \sigma_k^a \sigma_{k+r}^b\rho_{k,k+r})- 
 \left({\rm Tr}_{\{k\}}(  \sigma_k^a  \rho_k)  \right)  \left({\rm Tr}_{\{k+r\}}(  \sigma_{k+r}^b \rho_{k+r})  \right)  
\nonumber \\
&&=
{\rm Tr}_{\{k,k+r\}}( \sigma_k^a \sigma_{k+r}^b\rho^{connected}_{k,k+r})
\label{corre}
\end{eqnarray}

The explicit form of 
Eq \ref{rhoconnected} and the rotation of Eq. \ref{rotationab}
then yields
\begin{eqnarray}
 C^{ab}_{k,k+r} 
&& =\omega_{k,k+r} {\rm Tr}_{\{k \}}\left[ 
\sigma_k^a
\left( \frac{ \tau^x_k + \cos \left( \theta_{k-\frac{1}{2} }\right)  \tau^y_k \tan \left( \phi_{k+\frac{1}{2}}\right)  }{2}\right)
\right]
 \nonumber \\ && 
{\rm Tr}_{\{k+1\}} \left[
\sigma_{k+r}^b\left( \frac{  \tau^x_{k+r} + \tan \left( \phi_{k+r-\frac{1}{2}} \right)\tau^y_{k+r} \cos \left( \theta_{k+r+\frac{1}{2} }\right)    }{2}\right)
\right]
\nonumber \\
&& = \omega_{k,k+r} 
\left( \frac{ R_k^{ax} + \cos \left( \theta_{k-\frac{1}{2} }\right)  \tan \left( \phi_{k+\frac{1}{2}}\right) R_k^{ay} }{2}\right)
\left( \frac{  R^{bx}_{k+r} + \tan \left( \phi_{k+r-\frac{1}{2}} \right)  \cos \left( \theta_{k+r+\frac{1}{2} }\right) R^{by}_{k+r}   }{2}\right)
\label{cabconnected}
\end{eqnarray}

For instance, the diagonal $a=b$ connected correlations read
\begin{eqnarray}
 C^{zz}_{k,k+r} 
&& =\omega_{k,k+r} \sin \theta_k  \sin \theta_{k+r}
\nonumber \\
 C^{xx}_{k,k+r} 
&& =\omega_{k,k+r} 
\left[ - \cos \theta_k \cos \phi_k
+
 \cos \left( \theta_{k-\frac{1}{2} }\right)    \tan \left( \phi_{k+\frac{1}{2} }\right) \sin \phi_k
\right]
 \nonumber \\  && 
\left[ - \cos \theta_{k+r} \cos \phi_{k+r}
+ \tan \left( \phi_{k+r-\frac{1}{2} }\right) \cos \left( \theta_{k+r+\frac{1}{2} }\right) \sin \phi_{k+r}
\right]
\nonumber \\
 C^{yy}_{k,k+r} 
&& =\omega_{k,k+r} 
\left[ 
  \cos \theta_k \sin \phi_k
+ \cos \left( \theta_{k-\frac{1}{2} }\right)   \tan \left( \phi_{k+\frac{1}{2} } \right)  \cos \phi_k 
\right]
 \nonumber \\  && 
 \left[
 \cos \theta_{k+r} \sin \phi_{k+r}
+ \tan \left( \phi_{k+r-\frac{1}{2} }\right)  \cos \left( \theta_{k+r+\frac{1}{2} } \right) \cos \phi_{k+r}
\right]
\label{caaconnected}
\end{eqnarray}


\section{ Riemann metric for the MPS manifold parametrized by the $(4N-2)$ angles  }

\label{sec_riemann}

The geometry of quantum states is a fascinating field (see the book \cite{book_geom} and references therein).
The Riemann Fubini-Study metric for the tangent space can be introduced 
either for the ket $\ket{\psi}$ or for the density matrix $\rho=\ket{\psi}\bra{\psi}$ 
with the same output for the metric \cite{book_geom}, as it is well known
for the Bloch sphere geometry of a single spin.
The tangent space of MPS kets has been much studied,
both for the detailed analysis of their geometric properties \cite{geomMPS,tangentspace}
and for the applications to dynamical algorithms \cite{haeg2011,haeg2012,haeg2013,haeg2016,ulrich_time,timetree}.
Here we have chosen instead the formulation in terms of the density matrix $\rho$
in order to use the properties of the Hilbert-Schmidt inner product for operators
that simplify some calculations, since our goal in this section is to compute explicitly
the metric for the MPS manifold parametrized by the $(4N-2)$ angles.

\subsection{ Reminder on the Hilbert-Schmidt inner product for operators } 

The Hilbert-Schmidt inner product between two operators $X$ and $Y$
\begin{eqnarray}
(X \vert Y)_{HS} \equiv {\rm Tr} ( X^{\dagger} Y)
\label{hsproduct}
\end{eqnarray}
allows to define the squared norm of an operator $X$
\begin{eqnarray}
\vert \vert X \vert \vert^2_{HS} \equiv (X \vert X)_{HS} = {\rm Tr} ( X^{\dagger} X)
\label{hsnorm}
\end{eqnarray}
and the squared distance between two operators $X$ and $Y$
\begin{eqnarray}
\vert \vert X -Y \vert \vert^2_{HS} \equiv (X-Y \vert X-Y)_{HS} = {\rm Tr} ( (X-Y)^{\dagger} (X-Y) )
\label{hsdistance}
\end{eqnarray}

The full density matrix $\rho=\ket{ \psi} \bra{\psi}$ 
satisfies
\begin{eqnarray}
\rho^{\dagger} && =  \rho
\nonumber \\
\rho^2 && = \rho
\nonumber \\
{ \rm Tr} (\rho) && =1
\label{rhopure}
\end{eqnarray}
so its Hilbert-Schmidt squared norm is unity
\begin{eqnarray}
\vert \vert \rho \vert \vert^2_{HS} = {\rm Tr} ( \rho^{\dagger} \rho) = { \rm Tr} (\rho^2)= { \rm Tr} (\rho)  =1
\label{hsnormrho}
\end{eqnarray}

\subsection{ Notion of Tangent Space around the density matrix $\rho$ } 

The MPS density matrix $\rho$ 
is a function of the $(2N-2)$ bond angles $( \theta_{k+\frac{1}{2}} ,\phi_{k+\frac{1}{2}})$ with $k=1,..,N-1$
and of the $(2N)$ site angles $(\theta_k,\phi_k)$ with $k=1,..,N$.
It will be momentarily convenient
to relabel them by $(4N-2)$ coordinates $x^{\mu}$ with $\mu=1,..,4N-2$,
for instance according to their appearance along the chain
\begin{eqnarray}
\theta_k && \equiv x^{4k-3} \ \ {\rm for } \ k=1,2,..,N
\nonumber \\
\phi_k && \equiv x^{4k-2} \ \ {\rm for } \ k=1,2,..,N
\nonumber \\
\theta_{k+\frac{1}{2}} && \equiv x^{4k-1} \ \ {\rm for } \ k=1,2,..,N-1
\nonumber \\
\phi_{k+\frac{1}{2}} && \equiv x^{4k} \ \ {\rm for } \ k=1,2,..,N-1
\label{xmu}
\end{eqnarray}

Let us now consider how the full density matrix $\rho$ changes
when all the parameters are changed from $x^{\mu}$ to $x^{\mu}+dx^{\mu}$ 
\begin{eqnarray}
d \rho = 
\sum_{k=1}^{N-1} \left[ 
d \theta_{k+\frac{1}{2}} V_{\theta_{k+\frac{1}{2}}}
+ d \phi_{k+\frac{1}{2}} V_{\phi_{k+\frac{1}{2}}}
\right]
+
\sum_{k=1}^{N-1} \left[ 
d \theta_k V_{\theta_k} 
+ d \phi_k V_{\phi_k} 
\right] \equiv \sum_{\mu=1}^{4N-2} dx^{\mu} V_{x^{\mu} }
\label{tangentrho}
\end{eqnarray}
The Tangent space around $\rho$ is thus generated by
the $(4N-2)$ tangent operators
\begin{eqnarray}
V_{x^{\mu} } \equiv   \frac{ \partial \rho} {\partial x^{\mu}  } 
\label{tangentrhomu}
\end{eqnarray}
The properties of the density matrix $\rho$ summarized in Eq \ref{rhopure}
yields that each tangent operator $V_{x^{\mu} } $ in Eq. \ref{tangentrho} is hermitian
\begin{eqnarray}
V_{x^{\mu} }^{\dagger} && =V_{x^{\mu} }
\label{daggervmu}
\end{eqnarray}
has a vanishing trace
\begin{eqnarray}
{ \rm Tr} (V_{x^{\mu} } ) && =0
\label{tracevmu}
\end{eqnarray}
and is orthogonal to the density matrix $\rho$ for the Hilbert-Schmidt inner-product
\begin{eqnarray}
0= { \rm Tr} ( \rho V_{x^{\mu} } )  = (\rho \vert  V_{x^{\mu} })_{HS}
\label{tracerhovmu}
\end{eqnarray}

\subsection{ Reminder on the Riemann Fubini-Study metric  }

The Riemann Fubini-Study metric \cite{book_geom}
can be defined in terms of the Hilbert-Schmidt distance of Eq. \ref{hsdistance}
between $\rho$ and $\rho+d\rho$ (Eq .\ref{tangentrho})
\begin{eqnarray}
ds^2 = \vert \vert d \rho \vert \vert^2_{HS}
 = {\rm Tr} \left( d \rho  d \rho\right)
 =\sum_{\mu=1}^{4N-2} \sum_{\nu=1}^{4N-2}  g_{(x^{\mu} ,x^{\nu})}  dx^{\mu}dx^{\nu} 
\label{metricdef}
\end{eqnarray}
where the metric coefficient $g_{(x^{\mu} ,x^{\nu})} $ between the coordinates $x^{\mu}$ and $x^{\nu}$
corresponds to the Hilbert-Schmidt inner product between their associated tangent operators $V_{x^{\mu} } $ and 
$V_{x^{\nu} } $
\begin{eqnarray}
g_{(x^{\mu} ,x^{\nu})} =  {\rm Tr} \left(  V_{x^{\mu} }  V_{x^{\nu} } \right) = ( V_{x^{\mu} } \vert V_{x^{\nu}}  )_{HS}
\label{metricg}
\end{eqnarray}
In particular, the diagonal coefficients correspond to the squared norms of the tangent operators
\begin{eqnarray}
g_{(x^{\mu} ,x^{\mu})} =  {\rm Tr} \left(  V_{x^{\mu} }  V_{x^{\mu} } \right) =\vert \vert  V_{x^{\mu} } \vert \vert^2_{HS}
\label{metricgdiag}
\end{eqnarray}
while the off-diagonal coefficients satisfy the symmetry $g_{(x^{\mu} ,x^{\nu})}=g_{(x^{\nu} ,x^{\mu})}$.
In the following, our goal is thus to write explicitly the tangent operators 
associated to bonds angles and to sites angles and to compute the metric.

\subsection{ Tangent operators associated to the bonds angles $( \theta_{k+\frac{1}{2}} ,\phi_{k+\frac{1}{2}})$  }

In the Schmidt decomposition of Eq. \ref{rho1bond}
\begin{eqnarray}
 \rho 
&& = \sum_{ \substack{\alpha =\pm \\ \beta=\pm} } 
\left( \ket{\Phi^{[1,...,k]}_{\alpha} } \bra{\Phi^{[1,...,k]}_{\beta} } \right)
\Lambda_{k,k+1}^{\alpha,\beta}
\left( \ket{\Phi^{[k+1,...,N]}_{\alpha} } \bra{\Phi^{[k+1,...,N]}_{\beta} } \right)
\label{rho1bondbis}
\end{eqnarray}
the only dependence with respect to the 
the two bond angles $( \theta_{k+\frac{1}{2}} ,\phi_{k+\frac{1}{2}})$
is contained in the bond variables $\Lambda_{k,k+1}^{\alpha,\beta} $ of Eq. \ref{lambdarhofinal}.
So the tangent operators associated to these two angles read 
\begin{eqnarray}
 V_{\theta_{ k+\frac{1}{2}}} && \equiv \frac{ \partial \rho} {\partial \theta_{k+\frac{1}{2}}  } 
 = \sum_{ \substack{\alpha =\pm \\ \beta=\pm} }
\left( \ket{\Phi^{[1,...,k]}_{\alpha} } \bra{\Phi^{[1,...,k]}_{\beta} } \right)
\left[  \frac{ \partial \Lambda_{k,k+1}^{\alpha,\beta}} {\partial \theta_{k+\frac{1}{2}}  } \right]
\left( \ket{\Phi^{[k+1,...,N]}_{\alpha} } \bra{\Phi^{[k+1,...,N]}_{\beta} } \right)
\nonumber \\
V_{\phi_{ k+\frac{1}{2}}} 
&& \equiv \frac{ \partial \rho} {\partial \phi_{k+\frac{1}{2}}  } 
=\sum_{ \substack{\alpha =\pm \\ \beta=\pm} }
\left( \ket{\Phi^{[1,...,k]}_{\alpha} } \bra{\Phi^{[1,...,k]}_{\beta} } \right)
\left[  \frac{ \partial \Lambda_{k,k+1}^{\alpha,\beta}} {\partial \phi_{k+\frac{1}{2}}  } \right]
\left( \ket{\Phi^{[k+1,...,N]}_{\alpha} } \bra{\Phi^{[k+1,...,N]}_{\beta} } \right)
\label{tangentbond}
\end{eqnarray}
where the derivatives of the bond variables $\Lambda_{k,k+1}^{\alpha,\beta} $ of Eq. \ref{lambdarhofinal} 
\begin{eqnarray}
\Lambda_{k,k+1}^{\alpha,\beta} 
= \delta_{\alpha,\beta } \left( \frac{1+\alpha\cos \left( \theta_{k+\frac{1}{2}} \right)  }{2}\right)
+ \delta_{\alpha,-\beta } \left(\frac{\sin \left( \theta_{k+\frac{1}{2}} \right) 
 }{2}e^{-i \alpha \phi_{k+\frac{1}{2}} }\right)
\label{lambdarhofinalbis}
\end{eqnarray}

read
\begin{eqnarray}
\frac{ \partial \Lambda_{k,k+1}^{\alpha,\beta}} {\partial \theta_{k+\frac{1}{2}}  } 
&& =  \delta_{\alpha,\beta } \left( \frac{-\alpha\sin \left( \theta_{k+\frac{1}{2}} \right)  }{2}\right)
+ \delta_{\alpha,-\beta } \left(\frac{\cos \left( \theta_{k+\frac{1}{2}} \right)  }{2}e^{-i \alpha \phi_{k+\frac{1}{2}} }\right)
\nonumber \\
\frac{ \partial \Lambda_{k,k+1}^{\alpha,\beta}} {\partial \phi_{k+\frac{1}{2}}  } 
&& = -i \alpha \delta_{\alpha,-\beta } \left(\frac{\sin \left( \theta_{k+\frac{1}{2}} \right)  }{2}
e^{-i \alpha \phi_{k+\frac{1}{2}} }
\right)
\label{lambdaderi}
\end{eqnarray}

Using the orthonormality of the Schmidt eigenvectors, one obtains that
the conditions of vanishing trace (Eq \ref{tracevmu}) and orthogonality with 
the density matrix $\rho$ (Eq. \ref{tracerhovmu}) correspond to the following local properties for the variables $\Lambda_{k,k+1}^{\alpha,\beta} $ of a single bond $(k,k+1)$
\begin{eqnarray}
0&& ={ \rm Tr} (V_{\theta_{k+\frac{1}{2}}  } ) = \sum_{\alpha=\pm}   
\left[  \frac{ \partial \Lambda_{k,k+1}^{\alpha,\alpha}} {\partial \theta_{k+\frac{1}{2}}  } \right]
\nonumber \\
0&& ={ \rm Tr} (V_{\phi_{k+\frac{1}{2}}  } ) = \sum_{\alpha=\pm}   
\left[  \frac{ \partial \Lambda_{k,k+1}^{\alpha,\alpha}} {\partial \phi_{k+\frac{1}{2}}  } \right]
\nonumber \\
0&& ={ \rm Tr} ( \rho V_{\theta_{k+\frac{1}{2}}  }) =
\sum_{ \substack{\alpha =\pm \\ \beta=\pm} } 
\left[  \frac{ \partial \Lambda_{k,k+1}^{\alpha,\beta}} {\partial \theta_{k+\frac{1}{2}}  } \right]
\Lambda_{k,k+1}^{\beta,\alpha}
\nonumber \\
0&& ={ \rm Tr} ( \rho V_{\phi_{k+\frac{1}{2}}  }) =
\sum_{ \substack{\alpha =\pm \\ \beta=\pm} } 
\left[  \frac{ \partial \Lambda_{k,k+1}^{\alpha,\beta}} {\partial \phi_{k+\frac{1}{2}}  } \right]
\Lambda_{k,k+1}^{\beta,\alpha}
\label{trvmubond}
\end{eqnarray}


\subsection{  Tangent operators associated to the site angles $(\theta_k,\phi_k)$  }

In the Schmidt decomposition of Eq. \ref{schmidtkmrho}
\begin{eqnarray}
 \rho  = 
\sum_{ \substack{\alpha_L=\pm \\ \beta_L=\pm} }
\sum_{ \substack{\alpha_R=\pm \\ \beta_R=\pm} }
\left(\ket{\Phi^{[1,...,k-1]}_{\alpha_L} }\bra{\Phi^{[1,...,k-1]}_{\beta_L} }\right)
\Lambda_{k-1,k}^{\alpha_L,\beta_L}
O_k^{\alpha_L \alpha_R,\beta_L \beta_R } 
\Lambda_{k,k+1}^{\alpha_R,\beta_R}
\left( \ket{\Phi^{[k+1,...,N]}_{\alpha_R} } \bra{\Phi^{[k+1,...,N]}_{\beta_R} } \right)
\label{rho2bondnext}
\end{eqnarray}
 the only dependence with respect to the 
the two sites angles $(\theta_k,\phi_k)$
is contained in the operators $O_k^{T,T' }  $ of Eq \ref{operatorfinal} 
via the Pauli operators $\tau_k^{x,y,x}$ of Eq. \ref{taukzxy}.
The derivatives 
of the Pauli operators $\tau_k^{x,y,x}$ with respect to the two angles $(\theta_k,\phi_k)$
can be obtained from their expressions 
in the fixed $\sigma$ basis (Eq. \ref{taukzxy})
and can be translated then back in the $\tau$ basis via Eq \ref{operatorsigmatau}
in order to obtain the derivatives with respect to $\theta_k$
\begin{eqnarray}
\frac{ \partial \tau_k^z} {\partial \theta_k  }   && 
=-   \tau_k^x  
\nonumber \\
\frac{ \partial \tau_k^x} {\partial \theta_k  }     && 
= \tau_k^z 
\nonumber \\
\frac{ \partial \tau_k^y} {\partial \theta_k  }     && 
=  0
\label{taukderitheta}
\end{eqnarray}
and with respect to $\phi_k$
\begin{eqnarray}
\frac{ \partial \tau_k^z} {\partial \phi_k  }   && 
=-  \sin \theta_k  \tau_k^y
\nonumber \\
\frac{ \partial \tau_k^x} {\partial \phi_k  }    && 
= \cos \theta_k  \tau_k^y 
\nonumber \\
\frac{ \partial \tau_k^y} {\partial \phi_k  }    && 
=   \sin \theta_k\tau_k^z - \cos \theta_k \tau_k^x 
\label{taukderiphi}
\end{eqnarray}

The
derivatives of the operators Eq \ref{operatorfinal} 
\begin{eqnarray}
O_k^{T,T'}  = \ket{\tau^z_k=T }\bra{\tau^z_k=T' } 
= \delta_{T,T'} \left( \frac{1+T \tau_k^z}{2}\right)
+ \delta_{T,-T'} \left( \frac{ \tau^x_k +i T \tau^y_k }{2}\right)
\label{operatorfinalbis}
\end{eqnarray}
then read
\begin{eqnarray}
 \frac{ \partial O_k^{T,T' } } {\partial \theta_k  } 
&& = \delta_{T,T'} \left( - \frac{T \tau_k^x}{2}\right)
+ \delta_{T,-T'} \left( \frac{ \tau^z_k  }{2}\right)
\nonumber \\
\frac{ \partial O_k^{T,T' } } {\partial \phi_k  } 
&& =\delta_{T,T'} \left( - \frac{T \sin \theta_k  \tau_k^y}{2}\right)
+ \delta_{T,-T'} \left( \frac{ \cos \theta_k  \tau_k^y  +i T\left[ \sin \theta_k\tau_k^z - \cos \theta_k \tau_k^x \right] }{2}\right)
\label{operatorderi}
\end{eqnarray}
and appear in
the tangent operators associated to the two sites angles $(\theta_k,\phi_k)$
\begin{eqnarray}
 V_{\theta_k} && \equiv \frac{ \partial \rho} {\partial \theta_k  }  
=  \sum_{ \substack{\alpha_L=\pm \\ \beta_L=\pm} }
\sum_{ \substack{\alpha_R=\pm \\ \beta_R=\pm} }
\left(\ket{\Phi^{[1,...,k-1]}_{\alpha_L} }\bra{\Phi^{[1,...,k-1]}_{\beta_L} }\right)
\Lambda_{k-1,k}^{\alpha_L,\beta_L}
\left[ \frac{ \partial O_k^{\alpha_L \alpha_R,\beta_L \beta_R } } {\partial \theta_k  }  \right]
\Lambda_{k,k+1}^{\alpha_R,\beta_R}
\left( \ket{\Phi^{[k+1,...,N]}_{\alpha_R} } \bra{\Phi^{[k+1,...,N]}_{\beta_R} } \right)
\nonumber \\
 V_{\phi_k} 
&&  \equiv \frac{ \partial \rho} {\partial \phi_k  } 
=  \sum_{ \substack{\alpha_L=\pm \\ \beta_L=\pm} }
\sum_{ \substack{\alpha_R=\pm \\ \beta_R=\pm} }
 \left(\ket{\Phi^{[1,...,k-1]}_{\alpha_L} }\bra{\Phi^{[1,...,k-1]}_{\beta_L} }\right)
\Lambda_{k-1,k}^{\alpha_L,\beta_L}
\left[ \frac{ \partial O_k^{\alpha_L \alpha_R,\beta_L \beta_R } } {\partial \phi_k  }  \right]
\Lambda_{k,k+1}^{\alpha_R,\beta_R}
\left( \ket{\Phi^{[k+1,...,N]}_{\alpha_R} } \bra{\Phi^{[k+1,...,N]}_{\beta_R} } \right)
\nonumber \\
\label{tangentsite}
\end{eqnarray}

Here the conditions of vanishing trace (Eq \ref{tracevmu}) and orthogonality with 
the density matrix $\rho$ (Eq. \ref{tracerhovmu})
translate into the local conditions
\begin{eqnarray}
0&& ={ \rm Tr} (V_{\theta_k  } ) =
\sum_{\alpha_L=\pm} \sum_{\alpha_R=\pm}
\Lambda_{k-1,k}^{\alpha_L,\alpha_L}
\left( {\rm Tr}_{\{k\}}\left[ \frac{ \partial O_k^{\alpha_L \alpha_R,\alpha_L \alpha_R } } {\partial \theta_k  }  \right] \right)
\Lambda_{k,k+1}^{\alpha_R,\alpha_R}
\label{trvmusite}
\\
0&& ={ \rm Tr} (V_{\phi_k  } ) = 
\sum_{\alpha_L=\pm} \sum_{\alpha_R=\pm}
\Lambda_{k-1,k}^{\alpha_L,\alpha_L}
\left( {\rm Tr}_{\{k\}}\left[ \frac{ \partial O_k^{\alpha_L \alpha_R,\alpha_L \alpha_R } } {\partial \phi_k  }  \right] \right)
\Lambda_{k,k+1}^{\alpha_R,\alpha_R}
\nonumber \\
0&& ={ \rm Tr} ( \rho V_{\theta_k }) =
 \sum_{ \substack{\alpha_L=\pm \\ \beta_L=\pm} }
\sum_{ \substack{\alpha_R=\pm \\ \beta_R=\pm} }
\Lambda_{k-1,k}^{\alpha_L,\beta_L}\Lambda_{k-1,k}^{\beta_L,\alpha_L}
\left( {\rm Tr}_{\{k\}}
\left[ \frac{ \partial O_k^{\alpha_L \alpha_R,\beta_L \beta_R } } {\partial \theta_k  } 
O_k^{\beta_L \beta_R,\alpha_L \alpha_R }
 \right]
\right)
\Lambda_{k,k+1}^{\alpha_R,\beta_R}
\Lambda_{k,k+1}^{\beta_R,\alpha_R}
\nonumber \\
0&& ={ \rm Tr} ( \rho V_{\phi_k }) =
\sum_{ \substack{\alpha_L=\pm \\ \beta_L=\pm} }
\sum_{ \substack{\alpha_R=\pm \\ \beta_R=\pm} }
\Lambda_{k-1,k}^{\alpha_L,\beta_L}\Lambda_{k-1,k}^{\beta_L,\alpha_L}
\left( {\rm Tr}_{\{k\}}
\left[ \frac{ \partial O_k^{\alpha_L \alpha_R,\beta_L \beta_R } } {\partial \phi_k  } 
O_k^{\beta_L \beta_R,\alpha_L \alpha_R }
 \right]
\right)
\Lambda_{k,k+1}^{\alpha_R,\beta_R}
\Lambda_{k,k+1}^{\beta_R,\alpha_R}
\nonumber 
\end{eqnarray}
The two first conditions are trivially satisfied because the derivatives of the operators of Eq. \ref{operatorderi}
have zero trace
\begin{eqnarray}
 {\rm Tr}_{\{k\}}\left[ \frac{ \partial O_k^{T,T' } } {\partial \theta_k  } \right]
&& = 0
\nonumber \\
 {\rm Tr}_{\{k\}}\left[ \frac{ \partial O_k^{T,T' } } {\partial \phi_k  } \right]
&& =0
\label{operatorderitr}
\end{eqnarray}
The third condition is also straightforward because the following trace over $k$ vanishes
\begin{eqnarray}
{\rm Tr}_{\{k\}}\left[ \frac{ \partial O_k^{T,T' } } {\partial \theta_k  } O_k^{T',T }\right]=0
\label{trvmusitetheta}
\end{eqnarray}
The trace appearing in the fourth condition does not vanish but reduces to
\begin{eqnarray}
{\rm Tr}_{\{k\}}\left[ \frac{ \partial O_k^{T,T' } } {\partial \phi_k  } O_k^{T',T }\right]=
-i T (\cos \theta_k ) \delta_{T,-T'} 
\label{trvmusitephi}
\end{eqnarray}
so the fourth condition vanishes only as a consequence of the remaining
sum over the indices $(\alpha_L,\alpha_R,\beta_L,\beta_R)$.


\subsection{  Metric within the sector of bonds angles $( \theta_{k+\frac{1}{2}} ,\phi_{k+\frac{1}{2}})$ with $k=1,..,N-1$  }

\subsubsection{  Metric coefficients between the two angles $( \theta_{k+\frac{1}{2}} ,\phi_{k+\frac{1}{2}})$ 
associated to the same bond $(k,k+1)$  }

The tangent operators of Eq. \ref{tangentbond} yields with Eq \ref{lambdaderi}
that the two diagonal elements read
\begin{eqnarray}
&&  g_{(\theta_{ k+\frac{1}{2}} ,\theta_{ k+\frac{1}{2}})} 
= {\rm Tr} \left(  V_{\theta_{ k+\frac{1}{2}} }  V_{\theta_{ k+\frac{1}{2}} } \right)
=\sum_{ \substack{\alpha =\pm \\ \beta=\pm} }
\left[  \frac{ \partial \Lambda_{k,k+1}^{\alpha,\beta}} {\partial \theta_{k+\frac{1}{2}}  } \right]
\left[  \frac{ \partial \Lambda_{k,k+1}^{\beta,\alpha}} {\partial \theta_{k+\frac{1}{2}}  } \right]
= \frac{1}{2}
\nonumber \\
&&  g_{(\phi_{ k+\frac{1}{2}} ,\phi_{ k+\frac{1}{2}})} 
= {\rm Tr} \left(  V_{\phi_{ k+\frac{1}{2}} }  V_{\phi_{ k+\frac{1}{2}} } \right)
=\sum_{ \substack{\alpha =\pm \\ \beta=\pm} }
\left[  \frac{ \partial \Lambda_{k,k+1}^{\alpha,\beta}} {\partial \phi_{k+\frac{1}{2}}  } \right]
\left[  \frac{ \partial \Lambda_{k,k+1}^{\beta,\alpha}} {\partial \phi_{k+\frac{1}{2}}  } \right]
=\frac{\sin^2 \left( \theta_{k+\frac{1}{2}} \right)  }{2}
\label{metricbonddiago}
\end{eqnarray}
while the off-diagonal element vanishes
\begin{eqnarray}
&&  g_{(\theta_{ k+\frac{1}{2}} ,\phi_{ k+\frac{1}{2}})} 
= {\rm Tr} \left(  V_{\theta_{ k+\frac{1}{2}} }  V_{\phi_{ k+\frac{1}{2}} } \right)
= \sum_{ \substack{\alpha =\pm \\ \beta=\pm} }
\left[  \frac{ \partial \Lambda_{k,k+1}^{\alpha,\beta}} {\partial \theta_{k+\frac{1}{2}}  } \right]
\left[  \frac{ \partial \Lambda_{k,k+1}^{\beta,\alpha}} {\partial \phi_{k+\frac{1}{2}}  } \right]
=0
\label{metricgttbond}
\end{eqnarray}


\subsubsection{  Metric coefficients between the angles 
associated to the two bonds $(k,k+1)$ and $(k+r,k+r+1)$ with $r \geq 1$  }

In order to compare the tangent vectors of Eq. \ref{tangentbond}
associated to the bonds $(k,k+1)$ and $(k+r,k+r+1)$,
it is convenient to rewrite them using the interval ket introduced in Eq. \ref{ketinterval}
as
\begin{eqnarray}
 && V_{\theta_{ k+\frac{1}{2}}}  
 = \sum_{ \substack{\alpha_L=\pm \\ \beta_L=\pm} }
\sum_{ \substack{\alpha_R=\pm \\ \beta_R=\pm} }
\nonumber \\ &&
\left( \ket{\Phi^{[1,...,k]}_{\alpha_L} } \bra{\Phi^{[1,...,k]}_{\beta_L} } \right)
\left[  \frac{ \partial \Lambda_{k,k+1}^{\alpha_L,\beta_L}} {\partial \theta_{k+\frac{1}{2}}  } \right]
\left( \ket{ I^{[k+1,..,k+r]}_{\alpha_L,\alpha_R}  }
\bra{ I^{[k+1,..,k+r]}_{\beta_L,\beta_R}  } \right)
\Lambda^{\alpha_R,\beta_R}_{k+r,k+r+1}
\left( \ket{\Phi^{[k+r+1,...,N]}_{\alpha_R} } 
\bra{\Phi^{[k+r+1,...,N]}_{\beta_R} } \right)
\nonumber \\
&& V_{\theta_{ k+r+\frac{1}{2}}}  
 =  \sum_{ \substack{\alpha_L=\pm \\ \beta_L=\pm} }
\sum_{ \substack{\alpha_R=\pm \\ \beta_R=\pm} }
\nonumber \\
&&
\left( \ket{\Phi^{[1,...,k]}_{\alpha_L} } \bra{\Phi^{[1,...,k]}_{\beta_L} } \right)
\Lambda^{\alpha_L,\beta_L}_{k,k+1}
\left( \ket{ I^{[k+1,..,k+r]}_{\alpha_L,\alpha_R}  }
\bra{ I^{[k+1,..,k+r]}_{\beta_L,\beta_R}  } \right)
\left[  \frac{ \partial \Lambda_{k+r,k+r+1}^{\alpha_R,\beta_R}} {\partial \theta_{k+r+\frac{1}{2}}  } \right]
\left( \ket{\Phi^{[k+r+1,...,N]}_{\alpha_R} } \bra{\Phi^{[k+r+1,...,N]}_{\beta_R} } \right)
\label{comparetwobonds}
\end{eqnarray}

Using the normalization of the interval ket (Eq \ref{ketintervalnorma}), one obtains that all the off-diagonal elements vanish as a consequence of Eq. \ref{trvmubond}
\begin{eqnarray}
  g_{(\theta_{ k+\frac{1}{2}} ,\theta_{ k+r+\frac{1}{2}})} 
 &&  
=0
\nonumber \\
 g_{(\phi_{ k+\frac{1}{2}} ,\phi_{ k+r+\frac{1}{2}})} 
&&
 =0
\nonumber \\
 g_{(\theta_{ k+\frac{1}{2}} ,\phi_{ k+r+\frac{1}{2}})} &&=  0
\nonumber \\
 g_{(\phi_{ k+\frac{1}{2}} ,\theta_{ k+r+\frac{1}{2}})} &&
 =0
\label{metrictwobonds}
\end{eqnarray}

So within the sector of bond angles, the metric has the nice property to be diagonal.


\subsection{  Metric within the sector of site angles $(\theta_k,\phi_k)$ with $k=1,..,N$ }

\subsubsection{  Metric coefficients between the two angles $(\theta_k,\phi_k)$
associated to the same site $(k)$  }

The tangent operators of Eq. \ref{tangentsite} yields with Eqs \ref{operatorderi}
that the two diagonal elements read

\begin{eqnarray}
 g_{(\theta_k ,\theta_k)} &&
= \sum_{ \substack{\alpha_L=\pm \\ \beta_L=\pm} }
\sum_{ \substack{\alpha_R=\pm \\ \beta_R=\pm} }
\Lambda_{k-1,k}^{\alpha_L,\beta_L}\Lambda_{k-1,k}^{\beta_L,\alpha_L}
\Lambda_{k,k+1}^{\alpha_R,\beta_R}\Lambda_{k,k+1}^{\beta_R,\alpha_R}
{\rm Tr}_{\{k\}} \left( \left[ \frac{ \partial O_k^{\alpha_L \alpha_R,\beta_L \beta_R } } {\partial \theta_k  }  \right]
\left[ \frac{ \partial O_k^{\beta_L \beta_R,\alpha_L \alpha_R } } {\partial \theta_k  }  \right]
\right) = \frac{1}{2}
\nonumber \\
 g_{(\phi_k ,\phi_k)} 
&& =\sum_{ \substack{\alpha_L=\pm \\ \beta_L=\pm} }
\sum_{ \substack{\alpha_R=\pm \\ \beta_R=\pm} }
\Lambda_{k-1,k}^{\alpha_L,\beta_L}\Lambda_{k-1,k}^{\beta_L,\alpha_L}
\Lambda_{k,k+1}^{\alpha_R,\beta_R}\Lambda_{k,k+1}^{\beta_R,\alpha_R}
{\rm Tr}_{\{k\}} \left( \left[ \frac{ \partial O_k^{\alpha_L \alpha_R,\beta_L \beta_R } } {\partial \phi_k  }  \right]
\left[ \frac{ \partial O_k^{\beta_L \beta_R,\alpha_L \alpha_R } } {\partial \phi_k  }  \right]
\right)
\nonumber \\
&& = \frac{1-\cos^2 \left(\theta_{k-\frac{1}{2}} \right) \cos^2 \theta_k
\cos^2 \left(\theta_{k+\frac{1}{2} }\right)}{2}
\label{metricsitepp}
\end{eqnarray}

while the off-diagonal element vanishes

\begin{eqnarray}
 g_{(\theta_k ,\phi_k)} 
= \sum_{ \substack{\alpha_L=\pm \\ \beta_L=\pm} }
\sum_{ \substack{\alpha_R=\pm \\ \beta_R=\pm} }
\Lambda_{k-1,k}^{\alpha_L,\beta_L}\Lambda_{k-1,k}^{\beta_L,\alpha_L}
\Lambda_{k,k+1}^{\alpha_R,\beta_R}\Lambda_{k,k+1}^{\beta_R,\alpha_R}
{\rm Tr}_{\{k\}} \left( \left[ \frac{ \partial O_k^{\alpha_L \alpha_R,\beta_L \beta_R } } {\partial \theta_k  }  \right]
\left[ \frac{ \partial O_k^{\beta_L \beta_R,\alpha_L \alpha_R } } {\partial \phi_k  }  \right]
\right) =0
\nonumber
\end{eqnarray}

Here one sees how the two neighboring bond angles $\theta_{k\pm\frac{1}{2}} $
modify the Bloch sphere metric of $(\theta_k,\phi_k)$ that the spin $k$ would have if it were isolated.


\subsubsection{  Metric coefficients between the angles 
associated to the two neighboring sites $(k)$ and $(k+1)$   }

Here the appropriate common decomposition reads

\begin{eqnarray}
 V_{\theta_k} &&  
=  \sum_{ \substack{\alpha_L=\pm \\ \beta_L=\pm} }
\sum_{ \substack{\alpha =\pm \\ \beta=\pm} }
\sum_{ \substack{\alpha_R=\pm \\ \beta_R=\pm} }
\label{tangentsiteneigh}
 \\
&& \left(\ket{\Phi^{[1,...,k-1]}_{\alpha_L} }\bra{\Phi^{[1,...,k-1]}_{\beta_L} }\right)
\Lambda_{k-1,k}^{\alpha_L,\beta_L}
\left[ \frac{ \partial O_k^{\alpha_L \alpha,\beta_L \beta } } {\partial \theta_k  }  \right]
\Lambda_{k,k+1}^{\alpha,\beta}
O_{k+1}^{\alpha \alpha_R,\beta \beta_R } 
\Lambda^{\alpha_R,\beta_R}_{k+1,k+2}
\left( \ket{\Phi^{[k+2,...,N]}_{\alpha_R} } 
\bra{\Phi^{[k+2,...,N]}_{\beta_R} } \right)
\nonumber \\
 V_{\theta_{k+1}} && 
=     \sum_{ \substack{\alpha_L=\pm \\ \beta_L=\pm} }
\sum_{ \substack{\alpha =\pm \\ \beta=\pm} }
\sum_{ \substack{\alpha_R=\pm \\ \beta_R=\pm} }
\nonumber \\
&& \left(\ket{\Phi^{[1,...,k-1]}_{\alpha_L} }\bra{\Phi^{[1,...,k-1]}_{\beta_L} }\right)
\Lambda_{k-1,k}^{\alpha_L,\beta_L}
 O_k^{\alpha_L \alpha,\beta_L \beta } 
\Lambda_{k,k+1}^{\alpha,\beta}
\left[ \frac{ \partial O_{k+1}^{\alpha \alpha_R,\beta \beta_R }  } {\partial \theta_{k+1}  }  \right]
\Lambda^{\alpha_R,\beta_R}_{k+1,k+2}
\left( \ket{\Phi^{[k+2,...,N]}_{\alpha_R} } 
\bra{\Phi^{[k+2,...,N]}_{\beta_R} } \right)
\nonumber
\end{eqnarray}

leading to the metric elements
\begin{eqnarray}
  g_{(\theta_k ,\theta_{k+1})}  
&&  = - \frac{\cos \left( \theta_{k-\frac{1}{2} }\right) 
\left[ \sin \left( \theta_{k+\frac{1}{2} }\right)
\cos \left( \phi_{k+\frac{1}{2} }\right) \right]
\cos \left( \theta_{k+\frac{3}{2} }\right) }{2} 
\label{metricsitetp}
\\
  g_{(\theta_k ,\phi_{k+1})}  
&& = - \frac{\cos \left( \theta_{k-\frac{1}{2} }\right) 
 \left[ \sin \left( \theta_{k+\frac{1}{2} }\right)
\cos \left( \phi_{k+\frac{1}{2} }\right) \right]\sin \theta_{k+1} }{2} 
\nonumber \\
  g_{(\phi_k ,\theta_{k+1})}  
&& = - \frac{\sin \theta_k
 \left[ \sin \left( \theta_{k+\frac{1}{2} }\right)\cos \left( \phi_{k+\frac{1}{2} }\right) \right]
\cos \left( \theta_{k+\frac{3}{2} }\right) 
 }{2} 
\nonumber \\
  g_{(\phi_k ,\phi_{k+1})}  
&& =  \frac{\cos \left( \theta_{k-\frac{1}{2} }\right) 
\cos \theta_k
 \sin^2 \left( \theta_{k+\frac{1}{2} }\right)
\cos \theta_{k+1}
\cos \left( \theta_{k+\frac{3}{2} }\right) 
+ \sin \theta_k \left[ \sin \left( \theta_{k+\frac{1}{2} }\right)
\cos \left( \phi_{k+\frac{1}{2} }\right) \right]\sin \theta_{k+1} }{2} 
\nonumber 
\end{eqnarray}


\subsubsection{  Metric coefficients between the angles 
associated to the two sites $(k)$ and $(k+r)$ with $r > 1$  }

Here the appropriate common decomposition involves the interval ket of Eq. \ref{ketinterval}

\begin{eqnarray}
 V_{\theta_k} &&  
=  
\sum_{\substack{\alpha_L=\pm \\ \beta_L=\pm}} 
\sum_{ \substack{\alpha_1 =\pm \\ \beta_1=\pm} }
\sum_{ \substack{\alpha_2 =\pm \\ \beta_2=\pm} }
\sum_{ \substack{\alpha_R=\pm \\ \beta_R=\pm} }
\left(\ket{\Phi^{[1,...,k-1]}_{\alpha_L} }\bra{\Phi^{[1,...,k-1]}_{\beta_L} }\right)
\Lambda_{k-1,k}^{\alpha_L,\beta_L}
\left[ \frac{ \partial O_k^{\alpha_L \alpha_1,\beta_L \beta_1 } } {\partial \theta_k  }  \right]
\Lambda_{k,k+1}^{\alpha_1,\beta_1}
\nonumber \\ && 
\left( \ket{ I^{[k+1,..,k+r-1]}_{\alpha_1,\alpha_2}  }
\bra{ I^{[k+1,..,k+r-1]}_{\beta_1,\beta_2}  } \right)
\Lambda^{\alpha_2,\beta_2}_{k+r-1,k+r}
O_{k+r}^{\alpha_2 \alpha_R,\beta_2 \beta_R } 
\Lambda^{\alpha_R,\beta_R}_{k+r,k+r+1}
\left( \ket{\Phi^{[k+r+1,...,N]}_{\alpha_R} } 
\bra{\Phi^{[k+r+1,...,N]}_{\beta_R} } \right)
\label{tangentsitek}
\end{eqnarray}

\begin{eqnarray}
 V_{\theta_{k+r}} && 
=  \sum_{\substack{\alpha_L=\pm \\ \beta_L=\pm}} 
\sum_{ \substack{\alpha_1 =\pm \\ \beta_1=\pm} }
\sum_{ \substack{\alpha_2 =\pm \\ \beta_2=\pm} }
\sum_{ \substack{\alpha_R=\pm \\ \beta_R=\pm} }
\left(\ket{\Phi^{[1,...,k-1]}_{\alpha_L} }\bra{\Phi^{[1,...,k-1]}_{\beta_L} }\right)
\Lambda_{k-1,k}^{\alpha_L,\beta_L}
 O_k^{\alpha_L \alpha_1,\beta_L \beta_1 } 
\Lambda_{k,k+1}^{\alpha_1,\beta_1}
\label{tangentsitefar} \\ 
&& 
\left( \ket{ I^{[k+1,..,k+r-1]}_{\alpha_1,\alpha_2}  }
\bra{ I^{[k+1,..,k+r-1]}_{\beta_1,\beta_2}  } \right)
\Lambda^{\alpha_2,\beta_2}_{k+r-1,k+r}
\left[ \frac{ \partial O_{k+r}^{\alpha_2 \alpha_R,\beta_2 \beta_R } } {\partial \theta_{k+r}  }  \right]
\Lambda^{\alpha_R,\beta_R}_{k+r,k+r+1}
\left( \ket{\Phi^{[k+r+1,...,N]}_{\alpha_R} } 
\bra{\Phi^{[k+r+1,...,N]}_{\beta_R} } \right)
\nonumber
\end{eqnarray}

Using the scalar products of the interval ket (Eq \ref{ketintervalscalar}), one obtains
that the metric coefficients involve the overlap $\omega_{k+1,k+r-1} $ of Eq. \ref{overlap},
that contains all the bond variables of the interval
and that corresponds to the factor that governs the decay of correlations 
as explained in the previous section (Eq. \ref{rhoconnected} and Eq. \ref{corre})
\begin{eqnarray}
  g_{(\theta_k ,\theta_{k+r})}  && =  \frac{\cos \left( \theta_{k-\frac{1}{2} }\right) 
\left[ \sin \left( \theta_{k+\frac{1}{2} }\right)\sin \left( \phi_{k+\frac{1}{2} }\right) \right]
\omega_{k+1,k+r-1}
\left[ \sin \left( \theta_{k+r-\frac{1}{2} }\right)\sin \left( \phi_{k+ri\frac{1}{2} }\right) \right]
\cos \left( \theta_{k+r+\frac{1}{2} }\right) }{2} 
\nonumber \\
  g_{(\theta_k ,\phi_{k+r})} &&
=  - \frac{\cos \left( \theta_{k-\frac{1}{2} }\right) 
\left[ \sin \left( \theta_{k+\frac{1}{2} }\right)\sin \left( \phi_{k+\frac{1}{2} }\right) \right]
\omega_{k+1,k+r-1}
\left[ \sin \left( \theta_{k+r-\frac{1}{2} }\right)\cos \left( \phi_{k+ri\frac{1}{2} }\right) \right]
\sin \theta_{k+r} }{2} 
\nonumber \\
  g_{(\phi_k ,\theta_{k+r})} &&
=  - \frac{\sin \theta_{k}
\left[ \sin \left( \theta_{k+\frac{1}{2} }\right)\cos \left( \phi_{k+\frac{1}{2} }\right) \right]
\omega_{k+1,k+r-1}
\left[ \sin \left( \theta_{k+r-\frac{1}{2} }\right)\sin \left( \phi_{k+ri\frac{1}{2} }\right) \right]
\cos \left( \theta_{k+r+\frac{1}{2} }\right) 
 }{2} 
\nonumber \\
  g_{(\phi_k ,\theta_{k+r})} &&
=   \frac{\sin \theta_{k}
\left[ \sin \left( \theta_{k+\frac{1}{2} }\right)\cos \left( \phi_{k+\frac{1}{2} }\right) \right]
\omega_{k+1,k+r-1}
\left[ \sin \left( \theta_{k+r-\frac{1}{2} }\right)\cos \left( \phi_{k+r-\frac{1}{2} }\right) \right]
\sin \theta_{k+r} 
 }{2} 
\label{metricsitetpr}
\end{eqnarray}


\subsection{ Metric coefficients between bond angles and sites angles }

\subsubsection{ Metric coefficients between the angles of the site $k$ and the angles of the bond $(k,k+1)$ }

Here the appropriate common decomposition reads
\begin{eqnarray}
V_{\theta_k}  
&& =  \sum_{ \substack{\alpha_L=\pm \\ \beta_L=\pm} }
\sum_{ \substack{\alpha_R=\pm \\ \beta_R=\pm} }
\left(\ket{\Phi^{[1,...,k-1]}_{\alpha_L} }\bra{\Phi^{[1,...,k-1]}_{\beta_L} }\right)
\Lambda_{k-1,k}^{\alpha_L,\beta_L}
\left[ \frac{ \partial O_k^{\alpha_L \alpha_R,\beta_L \beta_R } } {\partial \theta_k  }  \right]
\Lambda_{k,k+1}^{\alpha_R,\beta_R}
\left( \ket{\Phi^{[k+1,...,N]}_{\alpha_R} } \bra{\Phi^{[k+1,...,N]}_{\beta_R} } \right)
\nonumber \\
 V_{\theta_{ k+\frac{1}{2}}} && 
= \sum_{ \substack{\alpha_L=\pm \\ \beta_L=\pm} }
\sum_{ \substack{\alpha_R=\pm \\ \beta_R=\pm} }
\left(\ket{\Phi^{[1,...,k-1]}_{\alpha_L} }\bra{\Phi^{[1,...,k-1]}_{\beta_L} }\right)
\Lambda_{k-1,k}^{\alpha_L,\beta_L}
  O_k^{\alpha_L \alpha_R,\beta_L \beta_R } 
\left[  \frac{ \partial
\Lambda_{k,k+1}^{\alpha_R,\beta_R}
} {\partial \theta_{k+\frac{1}{2}}  } \right]
\left( \ket{\Phi^{[k+1,...,N]}_{\alpha_R} } \bra{\Phi^{[k+1,...,N]}_{\beta_R} } \right)
\nonumber
\end{eqnarray}

Using Eq. \ref{trvmusitetheta} and \ref{trvmusitephi}
 one obtains 
\begin{eqnarray}
 g_{(\theta_k ,\theta_{ k+\frac{1}{2}})} &&=0
\nonumber \\
g_{(\theta_k ,\phi_{ k+\frac{1}{2}})} && =0
\nonumber \\
 g_{(\phi_k ,\theta_{ k+\frac{1}{2}})} &&  =0
\nonumber \\
g_{(\phi_k ,\phi_{ k+\frac{1}{2}})} && =\frac{\cos \left( \theta_{k-\frac{1}{2} }\right) 
\cos(\theta_k) 
 \sin^2 \left( \theta_{k+\frac{1}{2} }\right) }{2} 
\label{metricsiteppb}
\end{eqnarray}

Similarly, the metric elements between the bond $(k,k+1)$ and the site $(k+1)$ read
\begin{eqnarray}
 g_{(\theta_{ k+\frac{1}{2}},\theta_{k+1})} &&
=0
\nonumber \\
g_{(\phi_{ k+\frac{1}{2}},\theta_{k+1})} && =0
\nonumber \\
 g_{(\theta_{ k+\frac{1}{2}},\phi_{k+1})} && =0
\nonumber \\
g_{(\phi_{ k+\frac{1}{2}},\phi_{k+1})} && = 
\frac{\sin^2 \left( \theta_{k+\frac{1}{2} }\right)
\cos(\theta_{k+1}) 
\cos \left( \theta_{k+\frac{3}{2} }\right) 
  }{2} 
\label{metricsitepp2}
\end{eqnarray}

\subsubsection{ Metric between the angles of the site $k$ and the angles of the bond $(k+r,k+r+1)$ with $r>1$ }

Here the appropriate common decomposition involves the interval ket of Eq. \ref{ketinterval}
\begin{eqnarray}
V_{\theta_k}  
&& =   \sum_{ \substack{\alpha_L=\pm \\ \beta_L=\pm} }
\sum_{ \substack{\alpha =\pm \\ \beta=\pm} }
\sum_{ \substack{\alpha_R=\pm \\ \beta_R=\pm} }
\left(\ket{\Phi^{[1,...,k-1]}_{\alpha_L} }\bra{\Phi^{[1,...,k-1]}_{\beta_L} }\right)
\Lambda_{k-1,k}^{\alpha_L,\beta_L}
\nonumber \\ &&
\left[ \frac{ \partial O_k^{\alpha_L \alpha,\beta_L \beta } } {\partial \theta_k  }  \right]
\Lambda_{k,k+1}^{\alpha,\beta}
\left( \ket{ I^{[k+1,..,k+r]}_{\alpha,\alpha_R}  }
\bra{ I^{[k+1,..,k+r]}_{\beta,\beta_R}  } \right)
\Lambda_{k+r,k+r+1}^{\alpha_R,\beta_R} 
\left( \ket{\Phi^{[k+r+1,...,N]}_{\alpha_R} } \bra{\Phi^{[k+r+1,...,N]}_{\beta_R} } \right)
\nonumber \\
 V_{\theta_{ k+r+\frac{1}{2}}}  
&& = \sum_{ \substack{\alpha_L=\pm \\ \beta_L=\pm} }
\sum_{ \substack{\alpha =\pm \\ \beta=\pm} }
\sum_{ \substack{\alpha_R=\pm \\ \beta_R=\pm} }
\left(\ket{\Phi^{[1,...,k-1]}_{\alpha_L} }\bra{\Phi^{[1,...,k-1]}_{\beta_L} }\right)
\Lambda_{k-1,k}^{\alpha_L,\beta_L}
\nonumber \\ &&
  O_k^{\alpha_L \alpha,\beta_L \beta } 
\Lambda_{k,k+1}^{\alpha,\beta}
\left( \ket{ I^{[k+1,..,k+r]}_{\alpha,\alpha_R}  }
\bra{ I^{[k+1,..,k+r]}_{\beta,\beta_R}  } \right)
\left[  \frac{ \partial \Lambda_{k+r,k+r+1}^{\alpha_R,\beta_R}  } {\partial \theta_{k+r+\frac{1}{2}}  } \right]
\left( \ket{\Phi^{[k+r+1,...,N]}_{\alpha_R} } \bra{\Phi^{[k+r+1,...,N]}_{\beta_R} } \right)
\label{comparebondfarsite}
\end{eqnarray}

Using Eq \ref{ketintervalscalar}, one obtains that the following metric elements vanish
as a consequence of Eq. \ref{trvmubond}
\begin{eqnarray}
 g_{(\theta_k ,\theta_{ k+r+\frac{1}{2}})} &&=0
\nonumber \\
 g_{(\theta_k ,\phi_{ k+r+\frac{1}{2}})} &&= 0
\nonumber \\
 g_{(\phi_k ,\theta_{ k+r+\frac{1}{2}})} &&= 0
\nonumber \\
 g_{(\phi_k ,\phi_{ k+r+\frac{1}{2}})} &&= 0
\label{metricsiteppr0}
\end{eqnarray}

\subsection{ Summary of the metric coefficients for each variable }

Let us summarize the properties of this metric from the point of view of each variable :

(i) the bond angle $\theta_{k+\frac{1}{2}}$ is orthogonal to all the other angles.

(ii) the bond angle $\phi_{k+\frac{1}{2}}$ has non-vanishing off-diagonal metric coefficients with $\phi_k$ and $\phi_{k+1}$ only.

(iii) the site angle $\theta_{k}$ has non-vanishing off-diagonal metric coefficients with 
all the other site angles $(\theta_{k+r},\phi_{k+r})$ with $r\ne 0$,
where the decay with the distance is governed by the overlap $\omega_{k+1,k+r-1} $ of Eq. \ref{overlap}.

(iv)  the site angle $\phi_{k}$ has non-vanishing off-diagonal metric coefficients 
with the two bond angles $(\phi_{k-\frac{1}{2}},\phi_{k+\frac{1}{2}})$, and 
with all the other site variables $(\theta_{k+r},\phi_{k+r})$ with $r \ne 0$,
where the decay with the distance is again governed by the overlap $\omega_{k+1,k+r-1} $ of Eq. \ref{overlap}.

The in-depth analysis of all the properties of this metric clearly goes beyond the scope of the present work
and is left for the future studies.


\section{ Generalization of the simplest MPS to any structure without loops }

\label{sec_tree}

Since Matrix-Product-States for chains can be directly generalized to any tree-like structure without loops
\cite{vidal_tree}, it is interesting to describe in this section how the MPS of Eq. \ref{mpsfinal}
can be extended to such tree-like structures without loops.

\subsection{ Parametrization of the ket on an arbitrary tree-like structure without loops }

For each bond $b$ that cuts the tree-like structure into two independent parts $A$ and $B$,
the Schmidt decomposition across this bond $b$ keeps its meaning,
 and one can thus still introduce the two angles $\theta_{b} $ and $ \phi_{b} $
to parametrize the two complex Schmidt values as in Eq \ref{lambdathetaphi}
\begin{eqnarray}
\lambda_{b}^+ && \equiv \cos \left( \frac{\theta_{b}}{2} \right)
\nonumber \\
 \lambda_{b}^- && \equiv \sin \left( \frac{\theta_{b}}{2} \right) e^{i \phi_b }
\label{lambdatreeb}
\end{eqnarray}
For each site $k$, one can still introduce the two Bloch angles $\theta_k $ and $\phi_k $
that parametrize the new appropriate local basis $\ket{\tau^z_k=\pm } $.
With respect to the one-dimensional chain,
the novelty is that each site $k$ is connected to a certain number $c_k \geq 1$ of bonds 
that will be labelled by $b^{[k]}_1,...,b^{[k]}_{c_k}$.
The leaves of the tree correspond to the sites $k$ connected to a single link $c_k=1$.
The direct generalization of the MPS of Eq. \ref{mpsfinal}
reads
\begin{eqnarray}
\ket{\psi}   
&& = \left[ \prod_b  \left( \sum_{\alpha_b=\pm} \lambda_b^{\alpha_b} \right) \right]
\left[ \prod_{k} \ket{\tau^z_k= \prod_{j=1}^{c_k} \alpha_{b^{[k]}_j } } \right]
\label{mpswithoutloops}
\end{eqnarray}
The corresponding MPO generalization of the density matrix of Eq. \ref{mporho}
reads
\begin{eqnarray}
\rho = \ket{\psi} \bra{\psi}  = 
\left[ \prod_b  \left(\sum_{ \substack{\alpha_b=\pm \\ \beta_b=\pm} } \Lambda_b^{\alpha_b,\beta_b} \right) \right]
\left[ \prod_{k} O_k^{\displaystyle \prod_{j=1}^{c_k} \alpha_{b^{[k]}_j }  ,  \prod_{j=1}^{c_k} \beta_{b^{[k]}_j }   } \right]
\label{mporhowithoutloops}
\end{eqnarray}
and involves a number of parameters given by
\begin{eqnarray}
P^{Tree}_{N,N_b} = 2 N +2 N_b 
\label{ptree}
\end{eqnarray}
in terms of the number $N$ of spins and the number $N_b$ of bonds.

\subsection{ Reduced density matrix $\rho_k$ of the site $k$ alone }

Around the site $k$, the $c_k$ outgoing bonds labelled by $b^{[k]}_{j=1,..,c_k}$
corresponds to independent branches of the tree, and 
are thus associated to two orthonormal Schmidt eigenvectors
$\ket{\Phi^{b^{[k]}_{j}}_{\pm} }  $, so the MPS of Eq. \ref{mpswithoutloops}
can be rewritten as
\begin{eqnarray}
\ket{\psi}   
&& = \sum_{\alpha_1 =\pm} \sum_{\alpha_2 =\pm}  ...  \sum_{\alpha_{c_k} =\pm}
 \left[ \prod_{j=1}^{c_k}   \lambda_{b^{[k]}_j}^{\alpha_j} \ket{\Phi^{b^{[k]}_{j}}_{\alpha_j} }  
 \right]
 \ket{\tau^z_k= \prod_{j=1}^{c_k} \alpha_j } 
\label{mpswithoutloopsk}
\end{eqnarray}
with the corresponding full density matrix
\begin{eqnarray}
\rho = \ket{\psi}    \bra{\psi}   
&& =\sum_{ \substack{\alpha_1 =\pm \\ \beta_1=\pm} }
\sum_{ \substack{\alpha_2 =\pm \\ \beta_2=\pm} }
...
\sum_{ \substack{\alpha_{c_k} =\pm \\ \beta_{c_k}=\pm} }
 \left[ \prod_{j=1}^{c_k}   \Lambda_{b^{[k]}_j}^{\alpha_j,\beta_j} 
\left( \ket{\Phi^{b^{[k]}_{j}}_{\alpha_j} }  \bra{\Phi^{b^{[k]}_{j}}_{\beta_j} }  
\right)
 \right]
 O_k^{\displaystyle\prod_{j=1}^{c_k} \alpha_j , \prod_{j=1}^{c_k} \beta_j} 
\label{mporhowithoutloopskdebut}
\end{eqnarray}
The trace over the $c_k$ outgoing branches imposes $\beta_j=\alpha_j$ for $j=1,..,c_k$,
so the reduced density matrix of the site $k$ alone reads
using the explicit expressions of Eq. \ref{lambdarhofinal}
and Eq. \ref{operatorfinal}
\begin{eqnarray}
\rho_k 
&& =\sum_{\alpha_1 =\pm} \sum_{\alpha_2 =\pm}  ...  \sum_{\alpha_{c_k} =\pm}
 \left[ \prod_{j=1}^{c_k}   \Lambda_{b^{[k]}_j}^{\alpha_j,\alpha_j} 
 \right]
 O_k^{\displaystyle\prod_{j=1}^{c_k} \alpha_j , \prod_{j=1}^{c_k} \alpha_j} 
\nonumber \\
&& = \frac{1}{ 2^{1+c_k} } \sum_{\alpha_1 =\pm} \sum_{\alpha_2 =\pm}  ...  \sum_{\alpha_{c_k} =\pm}
 \left[ \prod_{j=1}^{c_k}  
\left( 1+\alpha_j \cos \left( \theta_{b^{[k]}_j } \right)  \right)
 \right]
\left( 1+\left( \prod_{j=1}^{c_k} \alpha_j\right)  \tau_k^z \right)
\nonumber \\
&& =\frac{1}{2}
+\frac{\tau_k^z}{2}\prod_{j=1}^{c_k} \cos \left( \theta_{b^{[k]}_j } \right)
\label{mpsrhowithoutloopsk}
\end{eqnarray}
So the two ket $ \ket{ \tau_k^z=\pm } $ are still the two eigenvectors 
of the reduced density matrix $\rho_k$,
and the corresponding eigenvalues
\begin{eqnarray}
 p_k^{\pm} 
&& 
 = \frac{1 }{2} \left[1\pm \prod_{j=1}^{c_k} \cos \left( \theta_{b^{[k]}_j } \right) \right]
\label{rhokbulkwtree}
\end{eqnarray}
now involves the $c_k$ angles $ \theta_{b^{[k]}_{j=1,..,c_k} }$ of the $c_k$ bonds connected to the site $k$,
instead of the two angles of Eq. \ref{rhokbulkw}
for the chain.

\subsection{ Reduced density matrix $\rho_{k_L,k_R}$ of two neighboring sites $(k_L,k_R)$ connected by the bond $b$ }

Let us now focus on two neighboring sites $(k_L,k_R)$ connected by the bond $b$ :
the site $k_L$ is connected to $(c_{k_L}-1)$ other bonds that will be relabeled by $ \theta_{b^{[k_L]}_{j_L=1,..,(c_{k_L}-1)} }$,
and the site $k_R$ is connected to $(c_{k_R}-1)$  other bonds  that will be relabeled by $ \theta_{b^{[k_R]}_{j_R=1,..,(c_{k_R}-1)} }$.
All these out-going branches are independent, so the appropriate decomposition
of the ket of Eq. \ref{mpswithoutloops} reads
\begin{eqnarray}
\ket{\psi}   
 = 
\sum_{\alpha=\pm}
 \left[ \prod_{j_L=1}^{c_{k_L}-1} \sum_{\alpha^L_{j_L}=\pm}  \lambda_{b^{[k_L]}_{j_L}}^{\alpha^L_{j_L}} \ket{\Phi^{b^{[k_L]}_{j_L}}_{\alpha^L_{j_L}}}
\right]
 \left[ \prod_{j_R=1}^{c_{k_R}-1} \sum_{\alpha^R_{j_R} =\pm}  \lambda_{b^{[k_R]}_{j_R}}^{\alpha^R_{j_R}} \ket{\Phi^{b^{[k_R]}_{j_R}}_{\alpha^R_{j_R}} }  
\right]
 \ket{\tau^z_{k_L}= \alpha \prod_{j_L=1}^{c_{k_L}-1 } \alpha^L_{j_L} } 
\lambda_b^{\alpha}
\ket{\tau^z_{k_R}= \alpha \prod_{j_R=1}^{c_{k_R}-1 } \alpha^R_{j_R} } 
\label{mpswithoutloopsklkr}
\end{eqnarray}
with the corresponding full density matrix
\begin{eqnarray}
\rho
&& = \sum_{ \substack{\alpha =\pm \\ \beta=\pm} }
 \left[ \prod_{j_L=1}^{c_{k_L}-1} 
\sum_{ \substack{\alpha^L_{j_L} =\pm \\ \beta^L_{j_L}=\pm} }
  \Lambda_{b^{[k_L]}_{j_L}}^{\alpha^L_{j_L}, \beta^L_{j_L}} 
\left( \ket{\Phi^{b^{[k_L]}_{j_L}}_{\alpha^L_{j_L}}} \bra{\Phi^{b^{[k_L]}_{j_L}}_{\beta^L_{j_L}}}
\right)
\right]
 \left[ \prod_{j_R=1}^{c_{k_R}-1} 
 \sum_{ \substack{\alpha^R_{j_R} =\pm \\ \beta^R_{j_R}=\pm} }
  \Lambda_{b^{[k_R]}_{j_R}}^{\alpha^R_{j_R},  \beta^R_{j_R} } 
\left(\ket{\Phi^{b^{[k_R]}_{j_R}}_{\alpha^R_{j_R}} }  \bra{\Phi^{b^{[k_R]}_{j_R}}_{\beta^R_{j_R}} }  
\right)
\right]
\nonumber \\
&& 
 O_{k_L}^{ \displaystyle\alpha \prod_{j_L=1}^{c_{k_L}-1 } \alpha^L_{j_L} ,\beta \prod_{j_L=1}^{c_{k_L}-1 } \beta^L_{j_L}  } 
\Lambda_b^{\alpha,\beta}
O_{k_R}^{ \displaystyle\alpha \prod_{j_R=1}^{c_{k_R}-1 } \alpha^R_{j_R} , \beta \prod_{j_R=1}^{c_{k_R}-1 } \beta^R_{j_R} } 
\label{mpsrhowithoutloopsklkr}
\end{eqnarray}
The trace over the $(c_{k_L}-1)+(c_{k_R}-1)$ outgoing branches imposes $\beta^L_{j_L}=\beta^L_{j_L}
$ for $j_L=1,..,c_{k_L}-1$ and $\beta^R_{j_R}=\beta^R_{j_R}
$ for $j_R=1,..,c_{k_R}-1$
so the reduced density matrix of the two neighboring sites $(k_L,k_R)$ reads
\begin{eqnarray}
\rho_{k_L,k_R}
&& = \sum_{ \substack{\alpha =\pm \\ \beta=\pm} }
 W_{[k_L}^{\alpha,\beta}
\Lambda_b^{\alpha,\beta}
W_{k_R]}^{ \alpha,\beta } 
\nonumber \\
&& = \sum_{ \substack{\alpha =\pm \\ \beta=\pm} }
 W_{[k_L}^{\alpha,\beta}
\left[ \delta_{\alpha,\beta } \left( \frac{1+\alpha\cos \theta_b  }{2}\right)
+ \delta_{\alpha,-\beta } \left(\frac{\sin \theta_b  
\left[   \cos  \phi_b   -i \alpha\cos \phi_b 
\right]
 }{2}\right)
\right]
W_{k_R]}^{ \alpha,\beta } 
\label{mporhowithoutloopsklkr}
\end{eqnarray}
in terms of the two modified operators that are the analog of Eqs \ref{operatorsW} \ref{Wleft} \ref{Wright}
\begin{eqnarray}
 W_{[k_L}^{\alpha,\beta} && = \left[ \prod_{j_L=1}^{c_{k_L}-1} 
\sum_{ \alpha^L_{j_L} =\pm  }
  \Lambda_{b^{[k_L]}_{j_L}}^{\alpha^L_{j_L}, \alpha^L_{j_L}} 
\right] 
 O_{k_L}^{ \displaystyle\alpha \prod_{j_L=1}^{c_{k_L}-1 } \alpha^L_{j_L} ,\beta \prod_{j_L=1}^{c_{k_L}-1 } \alpha^L_{j_L}  } 
 = \frac{\delta_{ \alpha, \beta}}{2}  \left( 1+\alpha\tau_{k_L}^z {\cal C}_{out}^{[k_L]}   \right)
+\frac{ \delta_{ \alpha,- \beta} }{2} \left( \tau^x_{k_L} +i  \alpha \tau^y_{k_L} {\cal C}_{out}^{[k_L]}\right)
\nonumber \\
W_{k_R]}^{ \alpha,\beta } && = \left[ \prod_{j_R=1}^{c_{k_R}-1} 
 \sum_{ \alpha^R_{j_R} =\pm  }
  \Lambda_{b^{[k_R]}_{j_R}}^{\alpha^R_{j_R},  \alpha^R_{j_R} } 
\right]
O_{k_R}^{ \displaystyle\alpha \prod_{j_R=1}^{c_{k_R}-1 } \alpha^R_{j_R} , \beta \prod_{j_R=1}^{c_{k_R}-1 } \alpha^R_{j_R} } 
 = \frac{\delta_{ \alpha, \beta}}{2}  \left( 1+\alpha\tau_{k_R}^z {\cal C}_{out}^{[k_R]}   \right)
+\frac{ \delta_{ \alpha,- \beta} }{2} \left( \tau^x_{k_R} +i  \alpha \tau^y_{k_R} {\cal C}_{out}^{[k_R]}\right)
\nonumber \\
\label{wrighttree}
\end{eqnarray}
where the two coefficients involve all the angles of the outgoing links
\begin{eqnarray}
{\cal C}_{out}^{[k_L]} && \equiv 
\prod_{j_L=1}^{c_{k_L}-1 }\cos \left( \theta_{b^{[k_L]}_{j_L} } \right) 
\nonumber \\
{\cal C}_{out}^{[k_R]} && = 
\prod_{j_R=1}^{c_{k_R}-1 }\cos \left( \theta_{b^{[k_R]}_{j_R} } \right) 
\label{coefstree}
\end{eqnarray}

The final result
\begin{eqnarray}
\rho_{k_L,k_R}
&& = \frac{1}{8} \sum_{ \alpha=\pm 1  }
 \left( 1+\alpha\cos \theta_b \right)
\left( 1+\alpha\tau_{k_L}^z {\cal C}_{out}^{[k_L]}   \right)
\left( 1+\alpha\tau_{k_R}^z {\cal C}_{out}^{[k_R]}   \right)
\nonumber \\
&& +\frac{\sin \theta_b }{8}
\sum_{ \alpha=\pm 1  } \left(  \cos  \phi_b   -i \alpha\cos \phi_b  \right)
\left( \tau^x_{k_L} +i  \alpha \tau^y_{k_L} {\cal C}_{out}^{[k_L]}\right)
\left( \tau^x_{k_R} +i  \alpha \tau^y_{k_R} {\cal C}_{out}^{[k_R]}\right)
\nonumber \\
&&=\frac{1}{4}
+\cos \theta_b
 \left[
   \frac{\tau_{k_L}^z}{4} {\cal C}_{out}^{[k_L]}
+    \frac{\tau_{k_R}^z}{4} {\cal C}_{out}^{[k_R]} 
\right]
+  \frac{\tau_{k_L}^z \tau_{k_R}^z}{4}{\cal C}_{out}^{[k_L]}{\cal C}_{out}^{[k_R]} 
\nonumber \\
&& + 
 \sin \theta_b   \cos\phi_b   
\left[ \frac{\tau^x_{k_L}\tau^x_{k_R} }{4}
-
\frac{\tau^y_{k_L} \tau^y_{k_R} }{4} {\cal C}_{out}^{[k_L]}{\cal C}_{out}^{[k_R]} 
\right]
+ 
 \sin \theta_b \sin \phi_b  
\left[ 
    \frac{ \tau^x_{k_L}\tau^y_{k_R} }{4}{\cal C}_{out}^{[k_R]} 
+  
\frac{ \tau^y_{k_L} \tau^x_{k_R} }{4} {\cal C}_{out}^{[k_L]}
\right]
\label{mpswithoutloopsklkrfinal}
\end{eqnarray}
is thus a direct generalization of the corresponding result of Eq. \ref{rho2sitesdv}
 concerning the chain.

\subsection{ Discussion }

The reduced density matrices for a single site (Eq. \ref{rhokbulkwtree})
and for two neighboring sites (Eq. \ref{mpswithoutloopsklkrfinal})
can be then used to compute the energy of Hamiltonians containing only one-body and two-body
terms in order to optimize the MPS parameters as described in section \ref{sec_energy}.
Here we have chosen to remain very general with an arbitrary tree-like structure without loops,
but to analyze the more global properties of the MPS, one should specify the specific global geometry
of the tree structure one is interested in, so this is left for future studies.



\section{ Generalization of the simplest MPS to a periodic Ring of $N$ spins }

\label{sec_ring}

In the main text we have focused on the case of an open chain,
but since Matrix-Product-States are also much used in the presence of periodic boundary conditions,
it is interesting to discuss the changes that are needed for a ring of $N$ spins.

\subsection{ MPS ket on the ring with its normalization }

On a ring, the cut of a single bond does not give two independent parts,
so the bond variables $\lambda_{k,k+1}^{\alpha=\pm}$ unfortunately
loose their direct interpretation in terms of the Schmidt decomposition,
but one can nevertheless adapt the MPS ket of the open chain (Eq \ref{mpsfinal})
by adding on the bond $(N,N+1)=(N,1)$ that join the two boundaries
the following variables as in Eq. \ref{lambdacomplexk}
\begin{eqnarray}
\lambda_{N,1}^+ && \equiv \cos \left( \frac{\theta_{N+\frac{1}{2}}}{2} \right)
\nonumber \\
 \lambda_{N,1}^- && \equiv \sin \left( \frac{\theta_{N+\frac{1}{2}}}{2} \right) e^{i \phi_{N+\frac{1}{2}} }
\label{lambdacomplexkring}
\end{eqnarray}
and by adding some normalization $K_N$ that will be computed below
\begin{eqnarray}
\ket{\psi^{Ring}}  && = \frac{1}{ K_N}   \sum_{\alpha_1=\pm} ...  \sum_{\alpha_{N-1}=\pm}\sum_{\alpha_{N}=\pm}  
\left[ \prod_{k=1}^{N}\lambda^{\alpha_k}_{k,k+1}  \right]
\left[ \prod_{k=1}^N \ket{\tau^z_k= \alpha_{k-1} \alpha_k} \right]
\label{mpsring}
\end{eqnarray}

This can be rewritten in terms of the interval ket of Eq. \ref{ketinterval} as
\begin{eqnarray}
\ket{\psi^{Ring}}  && = \frac{1}{ K_N}   \sum_{\alpha=\pm}  
\lambda^{\alpha}_{N,1}  \ket{ I^{[1,..,N]}_{\alpha,\alpha}  }
\label{mpsringinterket}
\end{eqnarray}
so the normalization can be computed with the scalar product of Eq \ref{ketintervalscalar}
\begin{eqnarray}
1 =\braket{\psi^{Ring} \vert \psi^{Ring}}  && = \frac{1}{ K_N^2 }   \sum_{\alpha=\pm}   \sum_{\beta=\pm}  
\lambda^{\alpha}_{N,1} \overline{ \lambda^{\beta}_{N,1} } \braket{I^{[1,..,N]}_{\beta,\beta}  \vert I^{[1,..,N]}_{\alpha,\alpha}  }
 =\frac{1}{ K_N^2 }   \sum_{\alpha=\pm}   \sum_{\beta=\pm}  
\Lambda^{\alpha,\beta}_{N,1} 
\left[  \delta_{\alpha,\beta } 
+ \delta_{\alpha,-\beta } \omega_{1,N} 
\right]
\nonumber \\ &&
 =\frac{1}{ K_N^2 }   \sum_{\alpha=\pm}   
\left[   \Lambda^{\alpha,\alpha}_{N,1} 
+  \Lambda^{\alpha,-\alpha}_{N,1} \omega_{1,N} 
\right] = \frac{ 1+\omega^{Ring}_N   }{K_N^2} 
\label{mpsringinterketnorma}
\end{eqnarray}
where $\omega^{Ring}_N $ is the generalization of Eq \ref{overlap} for the total ring 
and thus involves the $(2N)$ bond angles of the whole ring
\begin{eqnarray}
\omega^{Ring}_N \equiv \omega_{1,N+1} \equiv 
&& 
= \prod_{n=1}^{N} \left[  \sin \left( \theta_{n+\frac{1}{2} } \right) \cos \left( \phi_{n+\frac{1}{2} } \right) \right]
\label{overlapring}
\end{eqnarray}
The normalization factor $K_N$ thus reads
\begin{eqnarray}
K_N = \sqrt{1+\omega^{Ring}_N } 
\label{normaring}
\end{eqnarray}

\subsection{ Reduced density matrix $\rho_k$ for the site $k$ alone }

In order to focus on the spin $k$, the appropriate decomposition of Eq. \ref{mpsring}
involves the interval ket $\ket{ I^{[k+1,..,N+k-1]}_{\alpha_R,\alpha_L}  } $
of the Ring without the site $k$ 
 (instead of Eq. \ref{schmidtkm1} for the open chain)
\begin{eqnarray}
\ket{\psi}  && = \frac{1}{ K_N}   \sum_{\alpha_L=\pm}  \sum_{\alpha_R=\pm} 
\lambda_{k-1,k}^{\alpha_L}
 \ket{ \tau_k^z=\alpha_L \alpha_R  }
\lambda_{k,k+1}^{\alpha_R}
\ket{ I^{[k+1,..,N+k-1]}_{\alpha_R,\alpha_L}  }
\label{mpsringk}
\end{eqnarray}
with the corresponding full density matrix (instead of Eq. \ref{schmidtkmrho} for the open chain)
\begin{eqnarray}
&& \rho  = \ket{\psi} \bra{\psi} =   \frac{1}{ K_N^2}  
 \sum_{\alpha_L=\pm} \sum_{\alpha_R=\pm}
\sum_{\beta_L=\pm} \sum_{\beta_R=\pm}
\nonumber \\
&&
\Lambda_{k-1,k}^{\alpha_L,\beta_L}
\left(  \ket{ \tau_k^z=\alpha_L \alpha_R  }
 \bra{ \tau_k^z=\beta_L \beta_R  }
\right)
\Lambda_{k,k+1}^{\alpha_R,\beta_R}
\left( 
\ket{I^{[k+1,...,N+k-1]}_{\alpha_R,\alpha_L} } 
\bra{I^{[k+1,...,N+k-1]}_{\beta_R,\beta_L} } 
\right)
\label{schmidtkmrhoring}
\end{eqnarray}
The trace over the other $(N-1)$ sites $(k+1,...,N+k-1)$ can be computed using Eq. \ref{ketintervalscalar}
\begin{eqnarray}
&& \rho_k  = {\rm Tr}_{\{k+1,...,N+k-1\}}  (\rho) =   \frac{1}{ K_N^2}  
 \sum_{\alpha_L=\pm} \sum_{\alpha_R=\pm}
\sum_{\beta_L=\pm} \sum_{\beta_R=\pm}
\Lambda_{k-1,k}^{\alpha_L,\beta_L}
O_k^{\alpha_L \alpha_R ,\beta_L \beta_R  }
\Lambda_{k,k+1}^{\alpha_R,\beta_R}
\braket{I^{[k+1,...,N+k-1]}_{\beta_R,\beta_L}  \vert  I^{[k+1,...,N+k-1]}_{\alpha_R,\alpha_L} } 
\nonumber \\ &&
=   \frac{1}{ K_N^2}  
 \sum_{\alpha_L=\pm} \sum_{\alpha_R=\pm}
\sum_{\beta_L=\pm} \sum_{\beta_R=\pm}
\Lambda_{k-1,k}^{\alpha_L,\beta_L}
O_k^{\alpha_L \alpha_R ,\beta_L \beta_R  }
\Lambda_{k,k+1}^{\alpha_R,\beta_R}
\delta_{\beta_L\beta_R,  \alpha_L\alpha_R }
\left[  \delta_{\alpha_L,\beta_L } 
+ \delta_{\alpha_L,-\beta_L } \omega_{k+1,N+k-1} 
\right]
\label{rhokring}
\end{eqnarray}
The explicit expressions of Eq. \ref{lambdarhofinal}
and of Eq. \ref{operatorfinal}
lead to the final result
\begin{eqnarray}
 \rho_k &&  
=\frac{1}{2} + \frac{\tau_k^z }{2} 
\left(\frac{ \cos \left( \theta_{k-\frac{1}{2}} \right)   \cos \left( \theta_{k+\frac{1}{2}} \right)
-\omega^{Ring}_N \tan \left( \phi_{k-\frac{1}{2}} \right)   \tan \left( \phi_{k+\frac{1}{2}} \right)
}{ 1+\omega^{Ring}_N  }
\right)
\label{rhokbulkzring}
\end{eqnarray}
So the two ket $ \ket{ \tau_k^z=\pm } $ are still the two eigenvectors 
of the reduced density matrix $\rho_k$ as for the open chain (Eq \ref{rhokbulkz}),
even if the corresponding eigenvalues have changed with respect to Eq. \ref{rhokbulkw},
and have lost their locality since $\omega^{Ring}_N $ contains the bond angles of the whole ring.

\subsection{ Reduced density matrix $\rho_{k,k+1}$ of two consecutive sites }

If one wishes to focus on the two consecutive spins $k$ and $(k+1)$,
 the appropriate decomposition of Eq. \ref{mpsring}
reads 
\begin{eqnarray}
\ket{\psi}  && = \frac{1}{ K_N}   \sum_{\alpha_L=\pm} \sum_{\alpha=\pm} \sum_{\alpha_R=\pm} 
\lambda_{k-1,k}^{\alpha_L}
 \ket{ \tau_k^z=\alpha_L \alpha  }
\lambda_{k,k+1}^{\alpha}
\ket{ \tau_{k+1}^z=\alpha \alpha_R  }
\lambda_{k+1,k+2}^{\alpha_R}
\ket{ I^{[k+2,..,N+k-1]}_{\alpha_R,\alpha_L}  }
\label{mpsring2sites}
\end{eqnarray}
with the corresponding full density matrix (instead of Eq. \ref{schmidtkmrho} for the open chain)
\begin{eqnarray}
&& \rho =   \frac{1}{ K_N^2}  
 \sum_{\alpha_L=\pm} \sum_{\alpha_R=\pm}\sum_{\alpha=\pm}
\sum_{\beta_L=\pm} \sum_{\beta_R=\pm}\sum_{\beta=\pm}
\nonumber \\
&&
\Lambda_{k-1,k}^{\alpha_L,\beta_L}
O_k^{\alpha_L \alpha ,\beta_L \beta  }
\Lambda_{k,k+1}^{\alpha,\beta}
O_k^{ \alpha\alpha_R , \beta \beta_R }
\Lambda_{k+1,k+2}^{\alpha_R,\beta_R}
\left( 
\ket{I^{[k+2,...,N+k-1]}_{\alpha_R,\alpha_L} } 
\bra{I^{[k+2,...,N+k-1]}_{\beta_R,\beta_L} } 
\right)
\label{schmidtkmrho2ring}
\end{eqnarray}
The trace over the other $(N-2)$ sites $(k+2,...,N+k-1)$ 
yields using Eq. \ref{ketintervalscalar}, Eq. \ref{lambdarhofinal}
and Eq. \ref{operatorfinal}

\begin{eqnarray}
&& \rho_{k,k+1}  = {\rm Tr}_{\{k+2,...,N+k-1\}} (\rho) =  
  \frac{1}{ K_N^2}   \sum_{\alpha_L=\pm} \sum_{\alpha_R=\pm}\sum_{\alpha=\pm}
\sum_{\beta_L=\pm} \sum_{\beta_R=\pm}\sum_{\beta=\pm}
\nonumber \\
&&
\Lambda_{k-1,k}^{\alpha_L,\beta_L}
O_k^{\alpha_L \alpha ,\beta_L \beta  }
\Lambda_{k,k+1}^{\alpha,\beta}
O_{k+1}^{ \alpha\alpha_R , \beta \beta_R }
\Lambda_{k+1,k+2}^{\alpha_R,\beta_R}
\delta_{\beta_L\beta_R,  \alpha_L\alpha_R }
\left[  \delta_{\alpha_L,\beta_L } 
+ \delta_{\alpha_L,-\beta_L } \omega_{k+2,N+k-1} 
\right]
\nonumber \\
&& =\frac{1}{4} 
+ \frac{\tau_k^z }{4} 
\left(\frac{ \cos \left( \theta_{k-\frac{1}{2}} \right)   \cos \left( \theta_{k+\frac{1}{2}} \right)
-\omega^{Ring}_N \tan \left( \phi_{k-\frac{1}{2}} \right)   \tan \left( \phi_{k+\frac{1}{2}} \right)
}{ 1+\omega^{Ring}_N  }
\right)
 \nonumber \\ &&
  + \frac{\tau_{k+1}^z }{4} 
\left(\frac{ \cos \left( \theta_{k+\frac{1}{2}} \right)   \cos \left( \theta_{k+\frac{3}{2}} \right)
-\omega^{Ring}_N \tan \left( \phi_{k+\frac{1}{2}} \right)   \tan \left( \phi_{k+\frac{3}{2}} \right)
}{ 1+\omega^{Ring}_N  }
\right)
\nonumber \\ &&
  + \frac{\tau_k^z\tau_{k+1}^z }{4} 
\left(\frac{ \cos \left( \theta_{k-\frac{1}{2}} \right)   \cos \left( \theta_{k+\frac{3}{2}} \right)
-\omega^{Ring}_N \tan \left( \phi_{k-\frac{1}{2}} \right)   \tan \left( \phi_{k+\frac{3}{2}} \right)
}{ 1+\omega^{Ring}_N  }
\right)
\nonumber \\ &&
  + \frac{\tau_k^x\tau_{k+1}^x }{4} 
\left(\frac{ \sin ( \theta_{k+\frac{1}{2}} )   \cos(\phi_{k+\frac{1}{2}}  ) 
+ \frac{\omega^{Ring}_N }{ \sin ( \theta_{k+\frac{1}{2}} )   \cos(\phi_{k+\frac{1}{2}}  ) }
}{ 1+\omega^{Ring}_N  }
\right)
\nonumber \\ &&
  - \frac{\tau_k^y\tau_{k+1}^y }{4} 
\left(\frac{ \cos \left( \theta_{k-\frac{1}{2}} \right)   \cos \left( \theta_{k+\frac{3}{2}} \right)\sin ( \theta_{k+\frac{1}{2}} )   \cos(\phi_{k+\frac{1}{2}}  ) 
- \frac{\omega^{Ring}_N \tan \left( \phi_{k-\frac{1}{2}} \right)   \tan \left( \phi_{k+\frac{3}{2}} \right)}{ \sin ( \theta_{k+\frac{1}{2}} )   \cos(\phi_{k+\frac{1}{2}}  ) }
}{ 1+\omega^{Ring}_N  }
\right)
\nonumber \\ &&
  + \frac{\tau_k^x\tau_{k+1}^y }{4} 
\left(\frac{ \cos \left( \theta_{k+\frac{3}{2}} \right)   \sin ( \theta_{k+\frac{1}{2}} )   \sin(\phi_{k+\frac{1}{2}}  ) 
+ \frac{\omega^{Ring}_N \tan \left( \phi_{k+\frac{3}{2}} \right)   }{ \tan ( \theta_{k+\frac{1}{2}} )   \cos(\phi_{k+\frac{1}{2}}  ) }
}{ 1+\omega^{Ring}_N  }
\right)
\nonumber \\ &&
  + \frac{\tau_k^y\tau_{k+1}^x }{4} 
\left(\frac{ \cos \left( \theta_{k-\frac{1}{2}} \right)   \sin ( \theta_{k+\frac{1}{2}} )   \sin(\phi_{k+\frac{1}{2}}  ) 
+ \frac{\omega^{Ring}_N \tan \left( \phi_{k-\frac{1}{2}} \right)   }{ \tan ( \theta_{k+\frac{1}{2}} )   \cos(\phi_{k+\frac{1}{2}}  ) }
}{ 1+\omega^{Ring}_N  }
\right)
\label{rhokring2sites}
\end{eqnarray}
So again the same eight Pauli operators appear as for the open chain (Eq. \ref{rho2sitesdv}) 
even the coefficients are more complicated.


\section{ Conclusion }

\label{sec_conclusion}

In this paper, we have focused on 
the simplest inhomogeneous Matrix-Product-State for an open chain of N quantum spins
that involves two angles per site and two angles per bond with a very clear physical meaning :
the two angles associated to the site $k$ are the two Bloch angles 
that parametrize the two orthonormal eigenvectors of the reduced density matrix $\rho_k$ of the spin $k$ alone, while the two angles associated to the bond $(k,k+1)$ parametrize the entanglement properties of the Schmidt decomposition across the bond $(k,k+1)$ within the gauge fixing described in section \ref{sec_gauge}.
We have then described how this simple structure allows to compute explicitly many observables,
including (i)  the reduced density matrix $\rho_{k,k+1}$ of two consecutive sites needed to evaluate the energy of two-body Hamiltonians (ii)  the reduced density matrix $\rho_{k,k+r}$ of two sites at distance $r$ needed to evaluate the spin-spin correlations at distance $r$ (iii) the Riemann metric of the MPS manifold as parametrized by these $(4N-2)$ angles. Finally, we have discussed the generalization to any tree-like structure without loops and to the chain with periodic boundary conditions.

Our main conclusion is thus that besides the outstanding achievements obtained by the Tensor Networks algorithms over the years, it is interesting to consider simple Tensor Networks of small dimension to compute explicitly their properties.
In the future, we hope to use the present framework to construct 
toy models of Many-Body-Localizion where all the eigenstates are Matrix-Product-States
\cite{c_mpsformbl}, to generalize the present approach to other entanglement architectures
like MERA in order to describe critical states of disordered spin chains \cite{c_future},
and to use the tangent space geometry to analyse dynamical problems \cite{c_future}.


 \appendix

\section{ Diagonal form of the reduced density matrix $\rho_{k,k+1,..,k+r}$ of $(r+1)$ consecutive sites in the bulk   }

\label{app_diago}

Many properties of the reduced density matrix $\rho_{k,k+1,..,k+r}$ of $(r+1)$ consecutive sites in the bulk
have already been discussed in section \ref{sec_rhointerval} in the text.
In this Appendix, we analyze its spectral decomposition.
But first it is instructive to analyze how the Left part $[1,..,k-1]$ and the Right part $[k+r+1,..,N]$ are entangled.

\subsection{ Diagonalization of the reduced density matrix $\rho_{[1,..,k-1],[k+r+1,..,N]} $ of the Left and Right parts together }

The reduced density matrix $\rho_{[1,..,k-1],[k+r+1,..,N]} $ of the Left and Right parts together
can be computed from the full density matrix of Eq. \ref{rho2cut}
by taking the trace over the interval spins $(k,..,k+r)$
that involve the scalar products of Eq. \ref{ketintervalscalar}
\begin{eqnarray}
{\rm Tr}_{\{k,..,k+r\}} \left( \left( \ket{ I^{[k,..,k+r]}_{\alpha_L,\alpha_R}  } \bra{ I^{[k,..,k+r]}_{\beta_L,\beta_R}  } \right) \right)
&& = \braket{ I^{[k,..,k+r]}_{\beta_L,\beta_R}   \vert I^{[k,..,k+r]}_{\alpha_L,\alpha_R}  }
= \delta_{\beta_L\beta_R,  \alpha_L\alpha_R }
\left[ \delta_{ \alpha_L , \beta_L }  
+ \omega_{k,k+r}  \delta_{ \alpha_L , -\beta_L }   \right]
\label{ketintervaltrace}
\end{eqnarray}
and one obtains the following form in terms of the two Schmidt eigenvectors $ \ket{\Phi^{[1,...,k]}_{\pm} }$
of the Left part alone and in terms of the two Schmidt eigenvectors $ \ket{\Phi^{[k+1,...,N]}_{\pm} }$
of the Right part alone
\begin{eqnarray}
 \rho_{[1,..,k-1],[k+r+1,..,N]}  
&& \equiv {\rm Tr}_{\{k,..,k+r\}} \left( \rho \right) 
\label{rhoLR}\nonumber \\
&&  =\sum_{\alpha_L=\pm} 
\sum_{\alpha_R=\pm} 
\sum_{\beta_L=\pm} 
\sum_{\beta_R=\pm} 
\rho^{(\alpha_L,\alpha_R), (\beta_L,\beta_R)}_{L,R}  
\left( \ket{\Phi^{[1,...,k]}_{\alpha_L} } \otimes \ket{\Phi^{[k+1,...,N]}_{\alpha_R} } \right)
\left( \bra{\Phi^{[1,...,k]}_{\beta_L} } \otimes \bra{\Phi^{[k+1,...,N]}_{\beta_R} } \right)
\nonumber
\end{eqnarray}
where the matrix elements in this basis of the Schmidt eigenvectors
can be computed using Eq. \ref{lambdarhofinal}
\begin{eqnarray}
&& \rho^{(\alpha_L,\alpha_R), (\beta_L,\beta_R)}_{L,R}  
=\delta_{\beta_L\beta_R,  \alpha_L\alpha_R }
\left[ \delta_{ \alpha_L , \beta_L }  
+ \omega_{k,k+r}  \delta_{ \alpha_L , -\beta_L }   \right]
\Lambda^{\alpha_L,\beta_L}_{k-1,k}
\Lambda^{\alpha_R,\beta_R}_{k+r,k+r+1}
\nonumber \\
&&
=\delta_{\beta_L\beta_R,  \alpha_L\alpha_R }
\left[ \delta_{ \alpha_L , \beta_L }  
+ \omega_{k,k+r}  \delta_{ \alpha_L , -\beta_L }   \right]
\left[
\delta_{\alpha_L,\beta_L } \left( \frac{1+\alpha_L\cos \left( \theta_{k-\frac{1}{2}} \right)  }{2}\right)
+ \delta_{\alpha_L,-\beta_L } \left(\frac{\sin \left( \theta_{k-\frac{1}{2}} \right)  }{2}e^{-i \alpha_L \phi_{k-\frac{1}{2}} }\right)
\right]
\nonumber \\
&&
\left[
 \delta_{\alpha_R,\beta_R } \left( \frac{1+\alpha_R\cos \left( \theta_{k+r+\frac{1}{2}} \right)  }{2}\right)
+ \delta_{\alpha_R,-\beta_R } \left(\frac{\sin \left( \theta_{k+r+\frac{1}{2}} \right)  }{2}e^{-i \alpha_R \phi_{k+r+\frac{1}{2}} }\right)
\right]
\nonumber \\
&&
= \delta_{ \alpha_L , \beta_L }  \delta_{\alpha_R,\beta_R }
\left( \frac{1+\alpha_L\cos \left( \theta_{k-\frac{1}{2}} \right)  }{2}\right)
\left( \frac{1+\alpha_R\cos \left( \theta_{k+r+\frac{1}{2}} \right)  }{2}\right)
\nonumber \\
&&+  \delta_{ \alpha_L , -\beta_L }  \delta_{\alpha_R,-\beta_R }
\omega_{k,k+r} \left(\frac{\sin \left( \theta_{k-\frac{1}{2}} \right)  }{2}e^{-i \alpha_L \phi_{k-\frac{1}{2}} }\right)
\left(\frac{\sin \left( \theta_{k+r+\frac{1}{2}} \right)  }{2}e^{-i \alpha_R \phi_{k+r+\frac{1}{2}} }\right)
\label{rhoLRmatrix}
\end{eqnarray}

This $4 \times4$ matrix is thus block-diagonal in the basis $(++,--,+-,-+)$
\begin{eqnarray}
\rho_{L,R} = \begin{pmatrix}
D^{++} & \Omega_F & 0 & 0     \\
 \overline{\Omega_F }  & D^{--} &  0 & 0        \\
0  & 0 & D^{+-}  & \Omega_A    \\
0  & 0 & \overline{\Omega_A } &   D^{-+}    \\
\end{pmatrix}
\label{psiSchmidt}
\end{eqnarray}
with the four diagonal elements 
\begin{eqnarray}
D^{ (\alpha_L,\alpha_R)} \equiv && \rho^{(\alpha_L,\alpha_R), (\alpha_L,\alpha_R)}_{L,R}  
= \left( \frac{1+ \alpha_L \cos \left( \theta_{k-\frac{1}{2} } \right) }{2} \right)
 \left( \frac{1+ \alpha_R \cos \left( \theta_{k+r+\frac{1}{2} } \right) }{2} \right)
\label{rhoLRdiag}
\end{eqnarray}
and the two off-diagonal elements
\begin{eqnarray}
\Omega_F \equiv  \rho^{(+,+), (-,-) }_{L,R}  
&& = \frac{\omega_{k,k+r}}{4} \sin \left( \theta_{k-\frac{1}{2} } \right)
\sin \left( \theta_{k+r+\frac{1}{2} } \right)
e^{-i \phi_{k-\frac{1}{2} } -i \phi_{k+r+\frac{1}{2} }  }
\nonumber \\
\Omega_A \equiv  \rho^{(+,-), (-,+ )}_{L,R}  
&&= \frac{\omega_{k,k+r}}{4} \sin \left( \theta_{k-\frac{1}{2} } \right)
\sin \left( \theta_{k+r+\frac{1}{2} } \right)
e^{-i \phi_{k-\frac{1}{2} } +i \phi_{k+r+\frac{1}{2} }  }
\nonumber
\end{eqnarray}
with their complex conjugates $\overline{ \Omega_F}$ and $\overline{ \Omega_A}$.
The overlap $\omega_{k,k+r} $ of Eq. \ref{overlap} appears in all 
the off-diagonal elements and will thus govern the magnitude of the entanglement 
between the Left part and the Right part separated by the distance $r$,
in agreement with the fact that the overlap $\omega_{k,k+r}  $ also governs
the correlations between the two spins $r$ and $k+r$ as described in section \ref{sec_corre}.

The traces of the two blocks of the matrix of Eq. \ref{psiSchmidt}
read respectively
\begin{eqnarray}
{\cal T}_F && \equiv D^{ (++)} + D^{ (--)} 
= \left( \frac{1+  \cos \left( \theta_{k-\frac{1}{2} } \right)\cos \left( \theta_{k+r+\frac{1}{2} } \right) }{2} \right)
\nonumber \\
{\cal T}_A && \equiv D^{ (+-)} + D^{ (-+)} 
= \left( \frac{1-  \cos \left( \theta_{k-\frac{1}{2} } \right)\cos \left( \theta_{k+r+\frac{1}{2} } \right) }{2} \right)
\label{rhoLRdiagtraces}
\end{eqnarray}
while they have the same determinant
\begin{eqnarray}
{\cal D} &&  \equiv D^{ (++)}  D^{ (--)} - \vert \Omega_F \vert^2 = D^{ (+-)}  D^{ (-+)} - \vert \Omega_A \vert^2
= \frac{ (1-\omega_{k,k+r}^2) \sin^2 \left( \theta_{k-\frac{1}{2} } \right)\sin^2 \left( \theta_{k+r+\frac{1}{2} } \right)}{16}
\label{rhoLRdet}
\end{eqnarray}
so their eigenvalues read respectively
\begin{eqnarray}
 p^F_{\pm}  && = \frac{ {\cal T}_F  \pm \sqrt{  {\cal T}_F^2 - 4  {\cal D}    } }{2}
\nonumber \\
&& = \frac{1+  \cos \left( \theta_{k-\frac{1}{2} } \right)\cos \left( \theta_{k+r+\frac{1}{2} } \right) 
\pm \sqrt{ \left[ \cos \left( \theta_{k-\frac{1}{2} } \right)  + \cos \left( \theta_{k+r+\frac{1}{2} } \right) \right]^2 
+\omega_{k,k+r}^2 \sin^2 \left( \theta_{k-\frac{1}{2} } \right)\sin^2 \left( \theta_{k+r+\frac{1}{2} } \right) }
}{4}
\nonumber \\
 p^A_{\pm}  && = \frac{ {\cal T}_A  \pm \sqrt{  {\cal T}_A^2 - 4  {\cal D}    } }{2}
\nonumber \\
&& = \frac{1 -   \cos \left( \theta_{k-\frac{1}{2} } \right)\cos \left( \theta_{k+r+\frac{1}{2} } \right) 
\pm \sqrt{ \left[ \cos \left( \theta_{k-\frac{1}{2} } \right)  - \cos \left( \theta_{k+r+\frac{1}{2} } \right) \right]^2 
+\omega_{k,k+r}^2 \sin^2 \left( \theta_{k-\frac{1}{2} } \right)\sin^2 \left( \theta_{k+r+\frac{1}{2} } \right) }
}{4}
\label{rhoLReigenFA}
\end{eqnarray}
The corresponding eigenvectors read
\begin{eqnarray}
 \ket{\Phi_{LR}^{F+}  } && = \cos \left( \frac{\theta_F}{2} \right) 
 \ket{\Phi^{[1,...,k-1]}_{+ } }  \ket{\Phi^{[k+r+1,...,N]}_{+} }
+ \sin \left( \frac{\theta_F}{2} \right) e^{i \phi_F} 
\ket{\Phi^{[1,...,k-1]}_{- } }  \ket{\Phi^{[k+r+1,...,N]}_{-} }
\nonumber \\
\ket{\Phi_{LR}^{F-}  } && = \sin \left( \frac{\theta_F}{2} \right) 
 \ket{\Phi^{[1,...,k-1]}_{+ } }  \ket{\Phi^{[k+r+1,...,N]}_{+} }
- \cos \left( \frac{\theta_F}{2} \right) e^{i \phi_F} 
\ket{\Phi^{[1,...,k-1]}_{- } }  \ket{\Phi^{[k+r+1,...,N]}_{-} }
\nonumber \\
 \ket{\Phi_{LR}^{A+}  } && = \cos \left( \frac{\theta_A}{2} \right) 
 \ket{\Phi^{[1,...,k-1]}_{+ } }  \ket{\Phi^{[k+r+1,...,N]}_{-} }
+ \sin \left( \frac{\theta_A}{2} \right) e^{i \phi_A} 
\ket{\Phi^{[1,...,k-1]}_{- } }  \ket{\Phi^{[k+r+1,...,N]}_{+} }
\nonumber \\
\ket{\Phi_{LR}^{A-}  } && = \sin \left( \frac{\theta_A}{2} \right) 
 \ket{\Phi^{[1,...,k-1]}_{+ } }  \ket{\Phi^{[k+r+1,...,N]}_{-} }
- \cos \left( \frac{\theta_A}{2} \right) e^{i \phi_A} 
\ket{\Phi^{[1,...,k-1]}_{- } }  \ket{\Phi^{[k+r+1,...,N]}_{+} }
\label{eigenvectorsFA}
\end{eqnarray}
in terms of the angles
\begin{eqnarray}
\phi_F \equiv && \phi_{k-\frac{1}{2} } + \phi_{k+r+\frac{1}{2} }  
\nonumber \\
\phi_A \equiv && \phi_{k-\frac{1}{2} } - \phi_{k+r+\frac{1}{2} }  
\nonumber
\\
\cos(\theta_F) \equiv && \frac{\cos \left( \theta_{k-\frac{1}{2} } \right)  + \cos \left( \theta_{k+r+\frac{1}{2} } \right) }
{ \left[ \cos \left( \theta_{k-\frac{1}{2} } \right)  + \cos \left( \theta_{k+r+\frac{1}{2} } \right) \right]^2 
+\omega_{k,k+r}^2 \sin^2 \left( \theta_{k-\frac{1}{2} } \right)\sin^2 \left( \theta_{k+r+\frac{1}{2} } \right) }
\nonumber
\\
\sin(\theta_F) \equiv && 
=  \frac{\omega_{k,k+r} \sin \left( \theta_{k-\frac{1}{2} } \right)\sin \left( \theta_{k+r+\frac{1}{2} } \right)}{
\left[ \cos \left( \theta_{k-\frac{1}{2} } \right)  + \cos \left( \theta_{k+r+\frac{1}{2} } \right) \right]^2 +\omega_{k,k+r}^2 \sin^2 \left( \theta_{k-\frac{1}{2} } \right)\sin^2 \left( \theta_{k+r+\frac{1}{2} } \right)
}
\nonumber
\\
\cos(\theta_A) \equiv && \frac{\cos \left( \theta_{k-\frac{1}{2} } \right)  - \cos \left( \theta_{k+r+\frac{1}{2} } \right) }
{ \left[ \cos \left( \theta_{k-\frac{1}{2} } \right)  - \cos \left( \theta_{k+r+\frac{1}{2} } \right) \right]^2 +\omega_{k,k+r}^2 \sin^2 \left( \theta_{k-\frac{1}{2} } \right)\sin^2 \left( \theta_{k+r+\frac{1}{2} } \right) }
\nonumber
\\
\sin(\theta_A) \equiv && 
=  \frac{\omega_{k,k+r} \sin \left( \theta_{k-\frac{1}{2} } \right)\sin \left( \theta_{k+r+\frac{1}{2} } \right)}{
\left[ \cos \left( \theta_{k-\frac{1}{2} } \right)  - \cos \left( \theta_{k+r+\frac{1}{2} } \right) \right]^2 +\omega_{k,k+r}^2 \sin^2 \left( \theta_{k-\frac{1}{2} } \right)\sin^2 \left( \theta_{k+r+\frac{1}{2} } \right)}
\label{FAangles}
\end{eqnarray}


\subsection{ Diagonal form of 
the reduced density matrix $\rho_{k,k+1,..,k+r} $ of the Interval $[k,..,k+r]$  }

\label{sec_rhointervaldiag}

The diagonal form of $\rho_{k,k+1,..,k+r}   $  can be directly obtained from
the eigenvectors found in Eq \ref{eigenvectorsFA} with the inversion formula
\begin{eqnarray}
\ket{\Phi^{[1,...,k-1]}_{+ } }  \ket{\Phi^{[k+r+1,...,N]}_{+} }&&
\cos \left( \frac{\theta_F}{2} \right)  \ket{\Phi_{LR}^{F+}  } 
+ \sin \left( \frac{\theta_F}{2} \right) \ket{\Phi_{LR}^{F-}  } 
\nonumber \\
\ket{\Phi^{[1,...,k-1]}_{- } }  \ket{\Phi^{[k+r+1,...,N]}_{-} }
&& = e^{-i \phi_F} \left[  
\sin \left( \frac{\theta_F}{2} \right)  \ket{\Phi_{LR}^{F+}  } 
- \cos \left( \frac{\theta_F}{2} \right) \ket{\Phi_{LR}^{F-}  } 
\right]
\nonumber \\
\ket{\Phi^{[1,...,k-1]}_{+ } }  \ket{\Phi^{[k+r+1,...,N]}_{-} }&&
\cos \left( \frac{\theta_A}{2} \right)  \ket{\Phi_{LR}^{A+}  } 
+ \sin \left( \frac{\theta_A}{2} \right) \ket{\Phi_{LR}^{A-}  } 
\nonumber \\
\ket{\Phi^{[1,...,k-1]}_{- } }  \ket{\Phi^{[k+r+1,...,N]}_{+} }
&& = e^{-i \phi_A} \left[  
\sin \left( \frac{\theta_A}{2} \right)  \ket{\Phi_{LR}^{A+}  } 
- \cos \left( \frac{\theta_A}{2} \right) \ket{\Phi_{LR}^{A-}  } 
\right]
\label{eigenvectorsFAinv}
\end{eqnarray}
that can be plugged into the initial full ket of Eq. \ref{schmidtinterval}
to obtain the Schmidt decomposition between the interval and the exterior (Left and Right together)
in the two sectors
\begin{eqnarray}
\ket{\psi^F} && \equiv 
\ket{\Phi^{[1,...,k-1]}_{+ } }\ket{\Phi^{[k+r+1,...,N]}_{+} } 
\lambda^{+}_{k-1,k}\lambda^{+}_{k+r,k+r+1}
\ket{ I^{[k,..,k+r]}_{+,+}  }
+\ket{\Phi^{[1,...,k-1]}_{- } }\ket{\Phi^{[k+r+1,...,N]}_{-} } 
\lambda^{-}_{k-1,k}\lambda^{-}_{k+r,k+r+1}
\ket{ I^{[k,..,k+r]}_{-,-}  }
\nonumber \\
&& = \sqrt{p_F^+}  \ket{\Phi_{LR}^{F+}  } \ket{I_F^+}
+  \sqrt{p_F^-}  \ket{\Phi_{LR}^{F-}  } \ket{I_F^-}
\nonumber \\
\ket{\psi^A} && \equiv 
\ket{\Phi^{[1,...,k-1]}_{+ } }\ket{\Phi^{[k+r+1,...,N]}_{-} } 
\lambda^{+}_{k-1,k}\lambda^{-}_{k+r,k+r+1}
\ket{ I^{[k,..,k+r]}_{+,+}  }
+\ket{\Phi^{[1,...,k-1]}_{+ } }\ket{\Phi^{[k+r+1,...,N]}_{-} } 
\lambda^{+}_{k-1,k}\lambda^{-}_{k+r,k+r+1}
\ket{ I^{[k,..,k+r]}_{+,-}  }
\nonumber \\
&& = e^{i \phi_R} \left[ \sqrt{p_A^+}  \ket{\Phi_{LR}^{A+}  } \ket{I_A^+}
+  \sqrt{p_A^-}  \ket{\Phi_{LR}^{A-}  } \ket{I_A^-} \right]
\label{schmidtintervaldiag}
\end{eqnarray}
where the four orthonormal ket concerning the interval $[k,..,k+r]$
\begin{eqnarray}
\ket{I_F^+} && = \frac{1}{\sqrt{p_F^+} }
\left[\cos \left( \frac{\theta_F}{2} \right)
\cos \left( \frac{\theta_{k-\frac{1}{2} } }{2} \right)  
 \cos \left( \frac{ \theta_{k+r+\frac{1}{2} } }{2}\right)
\ket{ I^{[k,..,k+r]}_{+,+}  }
+\sin \left( \frac{\theta_F}{2} \right)
\sin \left( \frac{\theta_{k-\frac{1}{2} } }{2} \right)  
 \sin \left( \frac{ \theta_{k+r+\frac{1}{2} } }{2}\right)
\ket{ I^{[k,..,k+r]}_{-,-}  }
\right]
\nonumber \\
\ket{I_F^-} && = \frac{1}{\sqrt{p_F^-} }
\left[\sin \left( \frac{\theta_F}{2} \right)
\cos \left( \frac{\theta_{k-\frac{1}{2} } }{2} \right)  
 \cos \left( \frac{ \theta_{k+r+\frac{1}{2} } }{2}\right)
\ket{ I^{[k,..,k+r]}_{+,+}  }
-\cos \left( \frac{\theta_F}{2} \right)
\sin \left( \frac{\theta_{k-\frac{1}{2} } }{2} \right)  
 \sin \left( \frac{ \theta_{k+r+\frac{1}{2} } }{2}\right)
\ket{ I^{[k,..,k+r]}_{-,-}  }
\right]
\nonumber \\
\ket{I_A^+} && = \frac{1}{\sqrt{p_A^+} }
\left[\cos \left( \frac{\theta_A}{2} \right)
\cos \left( \frac{\theta_{k-\frac{1}{2} } }{2} \right)  
 \sin \left( \frac{ \theta_{k+r+\frac{1}{2} } }{2}\right)
\ket{ I^{[k,..,k+r]}_{+,-}  }
+\sin \left( \frac{\theta_A}{2} \right)
\sin \left( \frac{\theta_{k-\frac{1}{2} } }{2} \right)  
 \cos \left( \frac{ \theta_{k+r+\frac{1}{2} } }{2}\right)
\ket{ I^{[k,..,k+r]}_{-,+}  }
\right]
\nonumber \\
\ket{I_A^-} && = \frac{1}{\sqrt{p_A^-} }
\left[\sin \left( \frac{\theta_A}{2} \right)
\cos \left( \frac{\theta_{k-\frac{1}{2} } }{2} \right)  
 \sin \left( \frac{ \theta_{k+r+\frac{1}{2} } }{2}\right)
\ket{ I^{[k,..,k+r]}_{+,-}  }
-\cos \left( \frac{\theta_A}{2} \right)
\sin \left( \frac{\theta_{k-\frac{1}{2} } }{2} \right)  
 \cos \left( \frac{ \theta_{k+r+\frac{1}{2} } }{2}\right)
\ket{ I^{[k,..,k+r]}_{-,+}  }
\right]
\nonumber
\end{eqnarray}
are the four eigenvectors of the reduced density matrix $\rho_{k,k+1,..,k+r} $ of the interval $[k,..,k+r]$,
associated to the four eigenvalues $(p_F^+,p_F^-,p_A^+,p_A^- )$
\begin{eqnarray}
\rho_{k,...,k+r} && = p_F^+\ket{ I_F^+  }  \bra{I_F^+  } + p_F^- \ket{I_F^-  }  \bra{I_F^-  } 
 + p_A^+\ket{ I_A^+  }  \bra{I_A^+  } + p_A^- \ket{I_A^-  }  \bra{I_A^-  } 
\label{spectralrhointerval}
\end{eqnarray}

\subsection{ Application to the construction of Parent Hamiltonians that have the MPS as exact ground-state }

Since the reduced density matrix $\rho_{k,k+1,..,k+r}   $ 
is of dimension $2^{r+1} \times 2^{r+1}$, and has only four non-trivial eigenvalues as discussed above,
this means that there are $n_0(r)=(2^{r+1}-4)$ vanishing eigenvalues.
For $r=1$ corresponding to an interval $[k,k+1]$ of two sites, there are no vanishing eigenvalue ($n_0(r=1)=0$), 
but for $r=2$ corresponding to an interval $[k,k+1,k+2]$ of three sites,
there are already $n_0(r=2)=4$ vanishing eigenvalues.
One may then introduce the projector on the support of $\rho_{k,k+1,k+2}$ of Eq. \ref{spectralrhointerval}
 \begin{eqnarray}
{\cal P}_{k,k+1,k+2} && \equiv \ket{ I_F^+  }  \bra{I_F^+  } +  \ket{I_F^-  }  \bra{I_F^-  } 
 + \ket{ I_A^+  }  \bra{I_A^+  } +  \ket{I_A^-  }  \bra{I_A^-  } 
\label{projecteursupport}
\end{eqnarray}
in order to construct the following Parent Hamiltonians \cite{wolf,qi}
with arbitrary positive couplings $ J_k >0 $
 \begin{eqnarray}
H^{Parent} && = \sum_k J_k \left( 1- {\cal P}_{k,k+1,k+2} \right)
\label{parent}
\end{eqnarray}
The energy of the MPS is zero by construction and the MPS is thus an exact ground state of these 
local Hamiltonians with three-body interactions.


\end{document}